\documentclass[%
superscriptaddress,
amsmath,amssymb,
aps,
floatfix,
]
{revtex4-2} 
\usepackage{graphicx}
\usepackage{dcolumn}
\usepackage{bm}
\usepackage{ulem}
\usepackage{xcolor}
\usepackage{float}
\usepackage{comment}
\usepackage{sidenotes}
\usepackage{booktabs}
\usepackage{graphicx}
\usepackage{epstopdf}
\usepackage[utf8]{inputenc}
\usepackage[T1]{fontenc}
\usepackage{mathtools}
\usepackage[thinc]{esdiff}
\usepackage{natbib}
\usepackage{subfigure}

\usepackage{hyperref}
\usepackage{inputenc}
\usepackage{subcaption}
\hypersetup{
    colorlinks=true,
    citecolor=blue,
    linkcolor=blue,
    filecolor=magenta,      
    urlcolor=blue,
    }

\DeclareMathOperator{\Tr}{Tr}

\begin{document}
\newcommand {\Wi}{ W\!i}

\title{Nature of continuous spectra in wall-bounded shearing flows of FENE-P fluids}
\author{Pratyush Kumar Mohanty}
\email{These authors contributed equally to this work}
\affiliation{Department of Chemical Engineering, Indian Institute of Technology, Kanpur 208016, India.}
\author{P.\,S.\,D.\,Surya Phani Tej}
\email{These authors contributed equally to this work}
\affiliation{Department of Chemical Engineering, Indian Institute of Technology, Kanpur 208016, India.}
\author{Ganesh Subramanian}
 \email{Contact author: sganesh@jncasr.ac.in}
\affiliation{Engineering Mechanics Unit, Jawaharlal Nehru Center for Advanced Scientific Research, Bangalore 560064, India.}
\author{V.\,Shankar}
 \email{Contact author: vshankar@iitk.ac.in}
\affiliation{Department of Chemical Engineering, Indian Institute of Technology, Kanpur 208016, India.}

\begin{abstract}
Owing to the spatially local nature of the constitutive equations typically used to model polymeric stresses, the differential operators governing the linearized dynamics of bounded viscoelastic shearing flows have singular points. As a result, 
the eigenspectra of such shearing flows contain, in addition to discrete eigenvalues,
continuous spectra (CS) comprising singular eigenfunctions.
On account of the singularity, numerical methods capture these spectra only as extended balloons that slowly converge to the theoretical loci. 
A clear understanding of the theoretical CS loci is crucial in discriminating physically genuine (discrete) eigenvalues from the poorly approximated numerical CS. This is especially important because, when a parameter such as the Weissenberg number ($\Wi$) is varied, discrete eigenvalues can emerge from or disappear into the CS, as the latter are often branch cuts in the complex spectral plane. For rectilinear shear flows of Oldroyd-B fluids, the CS are a pair of line segments, with lengths equal to the base-state range of velocities. 
In this study, we provide the first comprehensive account of the nature of the CS for both rectilinear and curvilinear shearing flows of the FENE-P fluid. The FENE-P model is one of the most widely used constitutive equations for dilute polymer solutions, and predicts both shear thinning of the polymer viscosity and the first normal stress coefficient. In stark contrast to the CS for the Oldroyd-B fluid mentioned above, we  show analytically that there are up to six distinct continuous spectra for shearing flows of FENE-P fluids. 
Further, owing to shear thinning, and due to the highly coupled nature of the linearized equations, these spectra possess nontrivial structures in the complex plane.
When the finite extensibility parameter $L > 50$, as appropriate for large molecular weight polymers used in experiments, three of the CS  are nearly identical, and independent of the solvent-to-solution viscosity ratio ($\beta$). The other three CS are $\beta$-dependent, with one of them being the analogue of the solvent (viscous) continuous spectrum in the Oldroyd-B fluid. The remaining two $\beta$-dependent CS are novel features of the FENE-P spectrum, and can have phase speeds outside the base range of velocities, including negative ones. 
For non-linear shearing flows subjected to streamwise invariant disturbances, the latter two CS have a novel  `wing'-like structure in the complex plane, symmetric about the imaginary axis. 
  All but two of the theoretical CS are shown to collapse onto master curves for different ($\Wi, L$) pairs,  for a fixed $\Wi/L$, a feature not shared by the discrete spectrum. The location and structure of the theoretical CS are broadly in accord with their numerical approximations obtained using the spectral method. The complexity of the CS predicted here for shearing flows of FENE-P fluids is expected to carry over to other nonlinear viscoelastic  models that exhibit a shear-thinning rheology.
 Our analytical predictions for the CS will therefore be helpful in identifying physically genuine, discrete eigenvalues in the numerically computed spectra.
\end{abstract}

\maketitle


\section{Introduction}

Both rectilinear and curvilinear shearing flows of viscoelastic fluids are known to exhibit novel hydrodynamic instabilities, often caused by the elastic nature of the fluid, and that are now well understood from both experimental and theoretical standpoints. A well-known example is the so-called `purely elastic' instability \cite{Shaqfeh1996,larson1992,muller_review}, which occurs even when inertial effects are negligible,  in curvilinear shear flows such as Taylor-Couette flow between concentric cylinders \cite{larson_shaqfeh_muller_1990,muller_etal_1989,shaqfeh_muller_larson_1992}, von-Karman swirling (torsional) flow between rotating discs \cite{PhanThien1983,Oztekin1993,byars1994}, and in the cone-and-plate configuration \cite{PhanThien1985,McKinley1991,McKinley1995}. Curvature of the base flow streamlines, coupled with the first-normal stress difference, drives this linear instability, as succinctly summarized by the Pakdel-McKinley criterion \cite{Pakdel_McKinley,PakdelMcKinley1996}.  More recently, even rectilinear shearing flows (such as plane and pipe Poiseuille flows) have been shown to be susceptible to a linear `center-mode'  instability at finite $Re$, with phase speeds of the disturbances approaching the base-state maximum velocity \cite{Garg_etal_2018,chaudharyetal_2021,khalid2021centermode,choueiri2021experimental,dong_zhang_2022}.
For pressure-driven channel flow, the center-mode instability has been shown to exist as a purely elastic instability even in the absence of inertia, despite the absence of a base-flow streamline curvature \cite{khalid_creepingflow_2021,page2020exact,buza_page_kerswell_2022,Kerswell2024asymptotics}. Thorough and up-to-date reviews of these instabilities can be found in Refs.\,\cite{Shaqfeh_Khomami_2021,Steinberg2021,Datta_etal2022,CastilloSanchez2022,Khomami_retrospective_2023}.

The Oldroyd-B constitutive equation \cite{Oldroyd1950} has often been the go-to model to analyze these instabilities, 
as it predicts a positive first-normal stress difference, one of the key ingredients of the purely elastic instabilities mentioned above. In contrast to its predictions of a constant shear viscosity and first normal stress coefficient, however, most polymer solutions and melts exhibit substantial shear thinning of these material functions, and it therefore becomes necessary to use more accurate constitutive models. In this regard, the FENE-P model \cite{Bird1980} has been one of the most widely used, especially in direct numerical simulations (DNS) of viscoelastic flows \cite{Alves2021}, both in the context of drag reduction \cite{Graham2014,Xi2019DRreview} as well as in simulations of elastoinertial and elastic turbulence \cite{Dubief_Hof_EIT_AFM_2023}. This is because the model predicts shear-rate dependence of both the viscosity and first normal stress coefficient, and has a closed-form (non-linear) relation between the stress and the conformation tensors, with the latter being governed by an evolution equation involving the upper-convected derivative and a nonlinear relaxation term (see Eqs.\,\ref{eq:tau_c_f}--\ref{eq:C_tau} below).
Surprisingly, and as we discuss below, despite these advantages, barring a few studies, the FENE-P model has not been extensively used in linear stability analyses.

One striking feature of stability analyses of bounded shear flows of viscoelastic fluids is the presence of continuous spectra (CS) \cite{Renardy2021}.  Such spectra arise when there is no stress diffusion in the constitutive equation, as is the case, for instance,  for the classical Oldroyd-B model. The eigenvalues belonging to the CS are associated with  singular points of the governing ordinary differential equation, implying that the associated eigenfunctions are singular. For this reason, the CS are not accurately captured by numerical methods which involve approximating the eigenfunctions in terms of a truncated series consisting of smooth basis functions.  
For the Oldroyd-B fluid, it is well known that the there are two CS \cite{Wilson1999}, which appear as distinct horizontal line segments in the complex $\omega$-plane (the complex frequency $\omega = \omega_r + i \omega_i$ being the eigenvalue),
with lengths dictated by $k \Delta V$, with $\Delta V$ being the base range of velocities. Numerical methods such as the spectral method capture these line segments only as extended balloons, with the vertical extent of these balloons (in the complex $\omega$-plane) shrinking slowly with increase in the numerical resolution. As a result, while carrying out stability analysis of non-diffusive constitutive models (be it Oldroyd-B or FENE-P), it becomes very difficult to distinguish between genuine (discrete) modes and the poorly-resolved eigenvalues belonging to the CS balloon; the latter can often straddle the unstable region of the complex plane, subsuming discrete eigenvalues in the process. The importance of this aspect has been discussed in the earlier work of  Arora \textit{et al.} \cite{arora2004hydrodynamic} in the context of the stability of plane Couette flow using the Pompom model.

 \begin{figure}
  \centering
  \subfigure[$Re = 0$]
  {\includegraphics[width=0.4\textwidth]{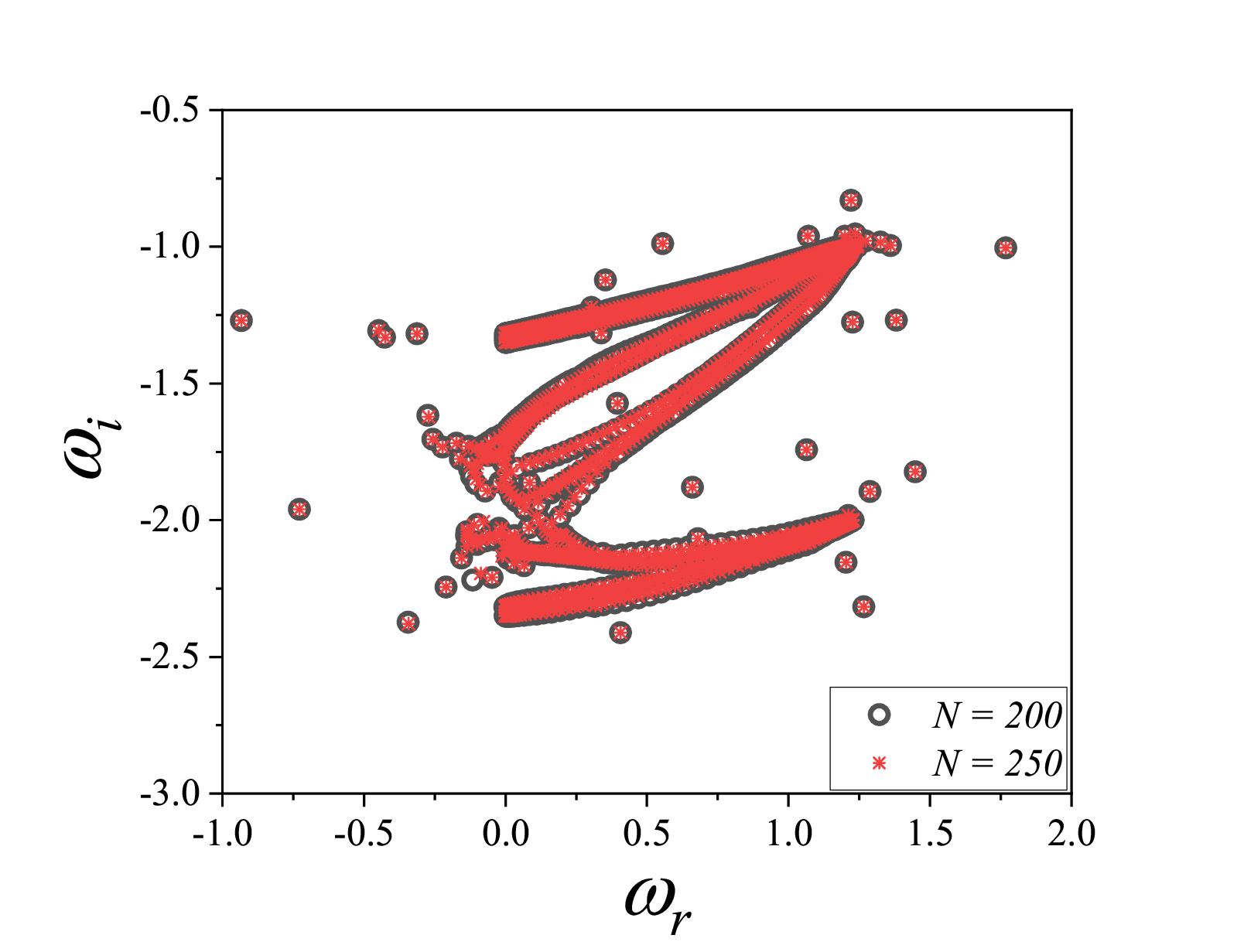}\label{fig:Re0_dean}}
    \subfigure[$Re = 10^4$]
  {\includegraphics[width=0.4\textwidth]{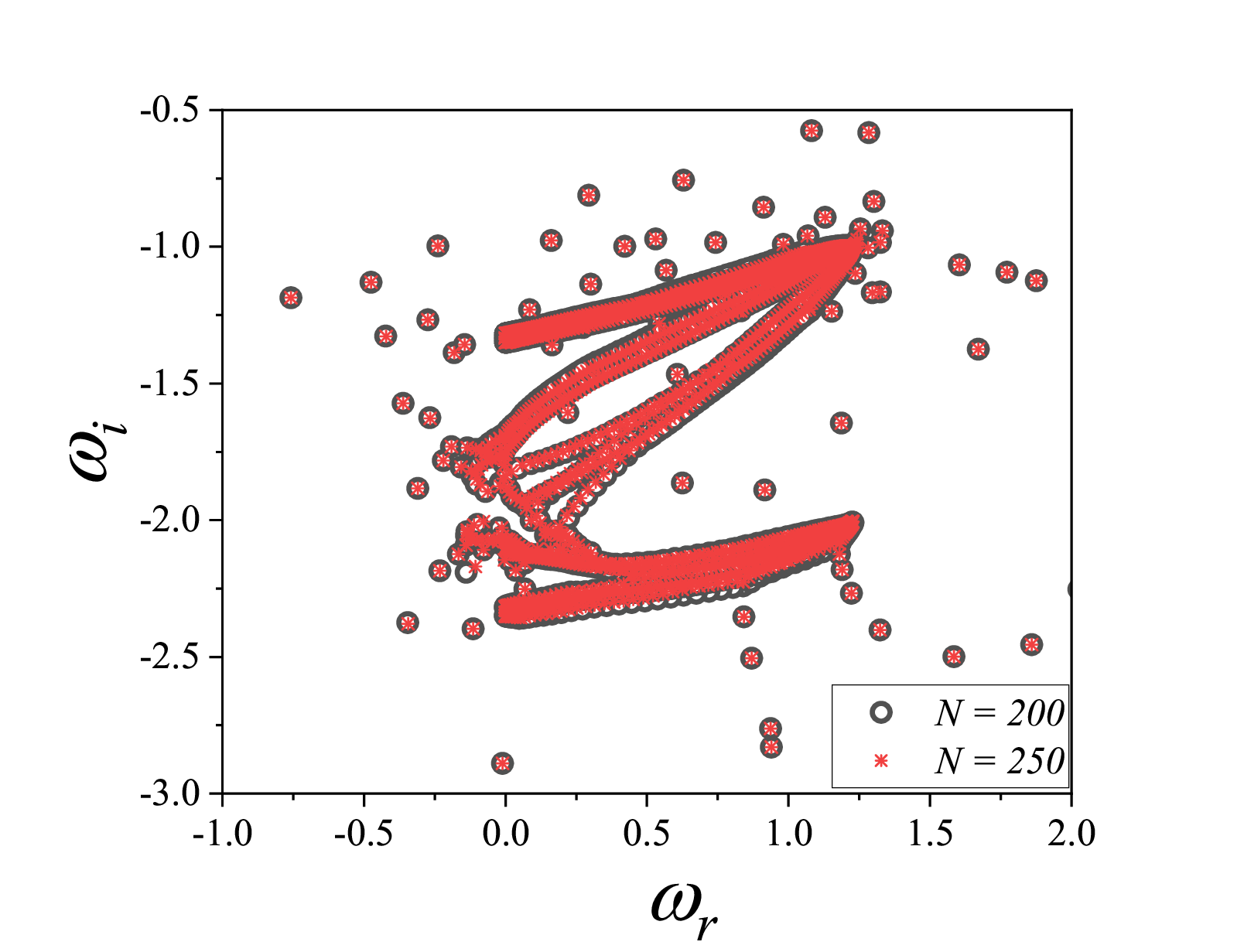}\label{fig:Re10000_dean}}
  \subfigure[$Re = 0$ (Oldroyd-B)]
  {\includegraphics[width=0.4\textwidth]{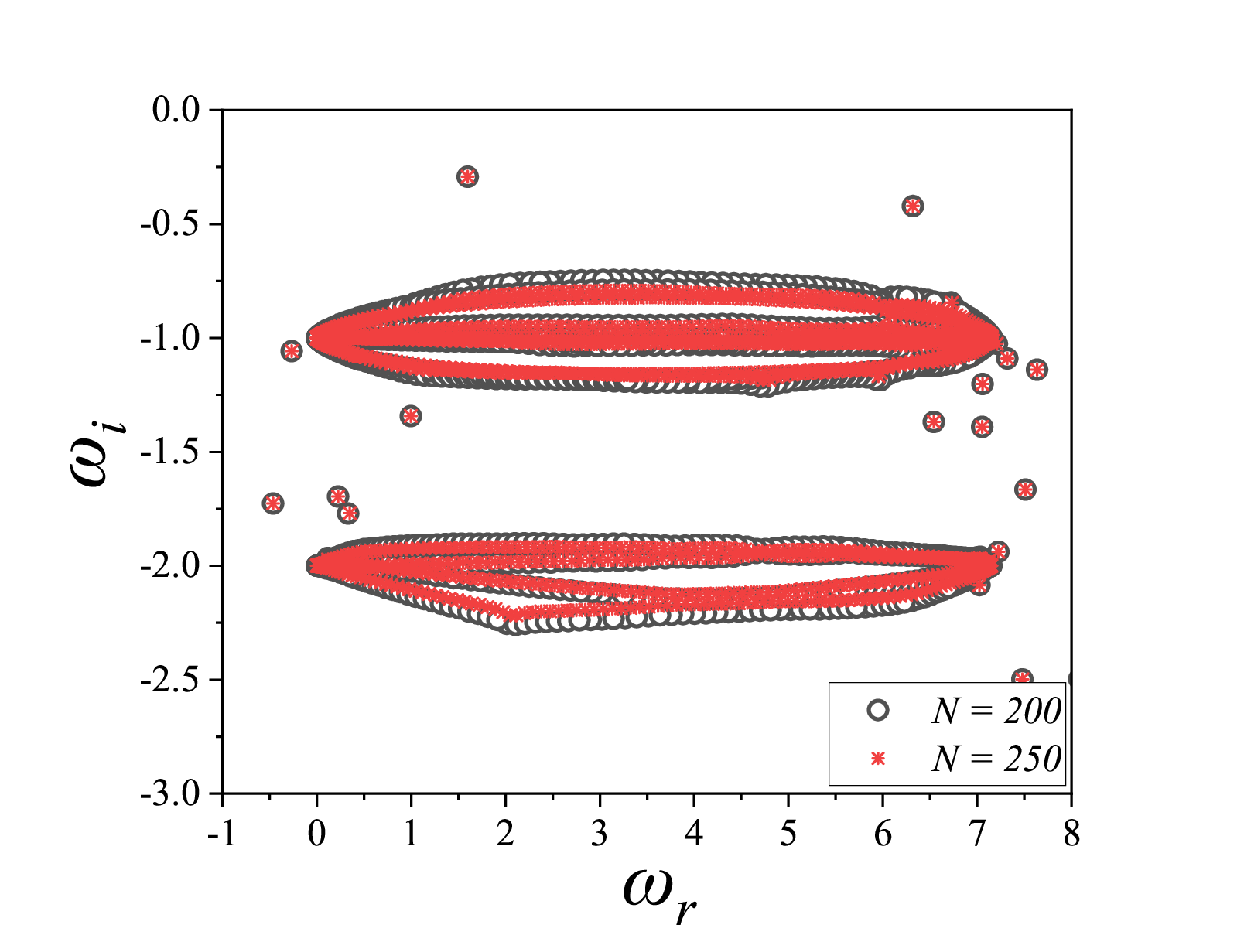}\label{fig:Re0_dean_OldroydB}}
    \subfigure[$Re = 10^4$ (Oldroyd-B)]
  {\includegraphics[width=0.4\textwidth]{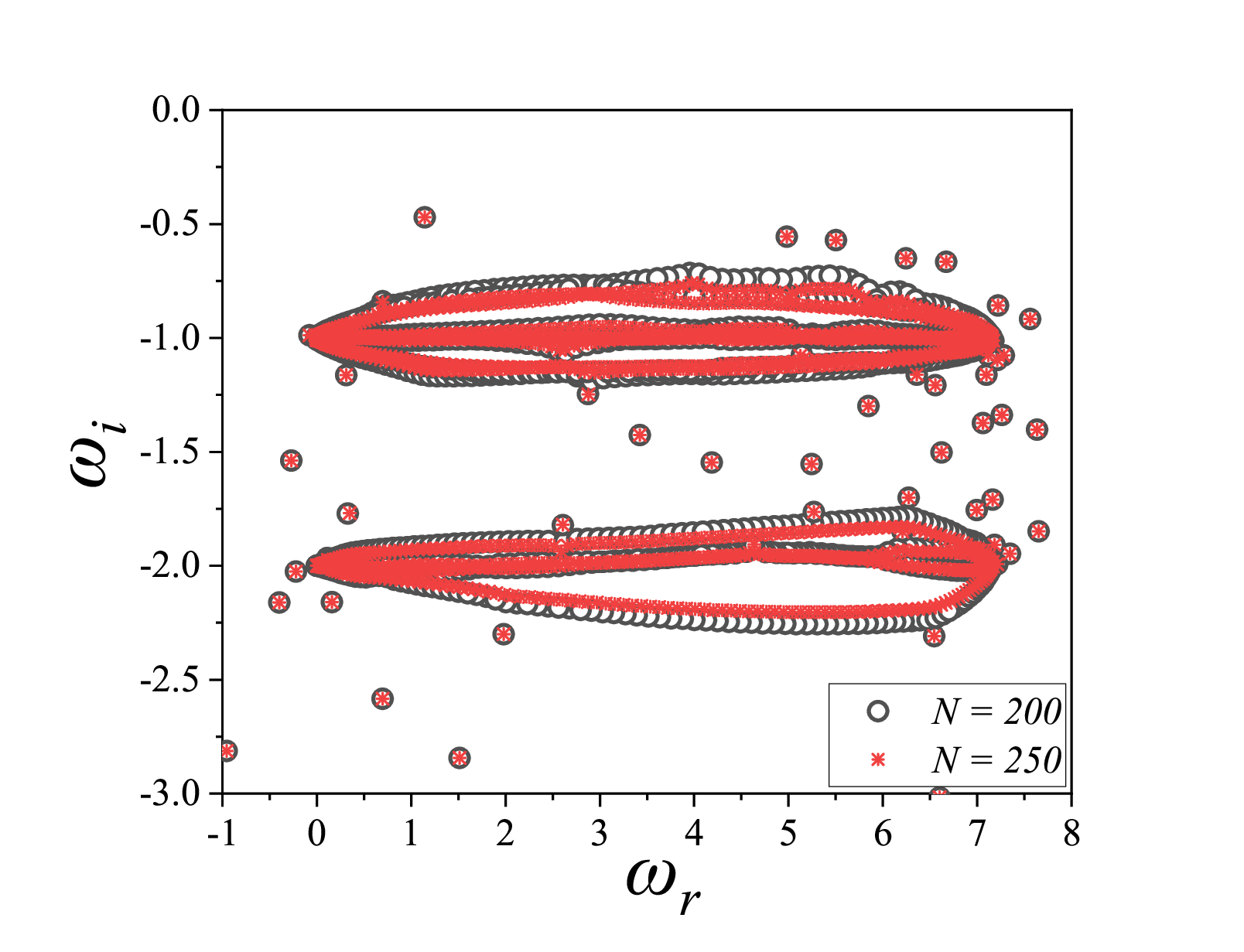}\label{fig:Re10000_dean_OldroydB}}
  \caption{Eigenspectra for viscoelastic Dean flow for $n = 1$, $\alpha = 7$, $\beta = 0.5$, $\Wi = 50, \epsilon = 0.1$. 
  Here, $\omega$ is the complex frequency made dimensionless using the relaxation time, and $\Wi$ is a suitably defined Weissenberg number discussed below in Sec.\,\ref{sec:probformulation}.
  The spectra in panels (a) and (b) are for the FENE-P fluid with $ L = 100$.
   The spectra in panels (c) and (d) are for the Oldroyd-B fluid for the same set of parameters. Note that the constitution of the densely filled curves, shown later to be the CS, appears to be independent of $Re$.}
  \label{fig:Effect_Re_Dean_CS}
\end{figure}

However, the nature of the CS for the FENE-P model has not been explored thus far in the literature. An exception is the effort of Arora and Khomami \cite{Arora_Khomami}, on the stability of plane Couette flow, which is discussed below. When one computes the eigenspectrum using the FENE-P model, for a curvilinear shearing flow such as viscoelastic Dean flow  (pressure-driven flow in a curved channel \cite{joo_shaqfeh_1991}), the structure of the spectrum turns out to be very complex, an example of which is shown in Fig.\,\ref{fig:Effect_Re_Dean_CS} (the symbols denoting the various parameters are explained in more detail in the next section). 
The corresponding Oldroyd-B spectra, also shown for the sake of comparison, exhibit a much simpler structure, with CS balloons around the aforementioned theoretical line segments with decay rates $-1/\lambda$ and $-1/(\beta \lambda)$; $\lambda$ being the polymer relaxation time, and $\beta = \eta_s/\eta_t$ being the ratio of solvent to (zero-shear) solution viscosity henceforth ($\eta_s$ and $\eta_t$ are the solvent and solution viscosities, respectively).
The motivation behind showing the FENE-P spectra here, in the Introduction, is to illustrate their nontrivial nature:  Ostensibly, the spectra do not seem to possess balloons as in the Oldroyd-B case, and 
therefore, it is not \textit{a priori} obvious, based on a visual inspection alone, as to which parts of the spectrum belong to the discrete eigenvalues, and which parts pertain to the CS. The usual way to numerically discriminate between the two is to compute the spectra for two different numbers of collocation points ($N$), and determine which set of eigenvalues converge with $N$ (hence, belonging to the discrete set) and which do not (hence, belonging to the CS); in addition, a study of the eigenfunctions of various velocity and stress fields can also reveal signatures of poorly resolved singular behavior, again serving to distinguish between the discrete and continuous spectra. In Fig.\,\ref{fig:Effect_Re_Dean_CS}, upon a cursory inspection,  the spectra for $N = 200$ and $N = 250$ appear to overlap well as a whole, making it necessary to zoom into various parts of the spectra to determine which eigenvalues are physically genuine. 
While this somewhat ad hoc procedure is useful, there is evidently an element of trial-and-error involved, rendered difficult  by the rather large spread of the numerical CS balloons and by the presence of fine-scale features (requiring magnification). Ideally, it would be desirable to have analytical predictions for the CS for such flows, in order for us to conclusively distinguish the discrete modes from those belonging to the poorly approximated CS. The objective of the present work is to address this issue for both rectilinear and curvilinear flows of FENE-P fluids. Below, we  provide a brief overview of the literature relevant to the present study, and thereby first set the context for the present work.

The Oldroyd-B model can be derived using a kinetic theory framework \cite{birdvol2} where the dilute polymer solution is modeled as a suspension of non-interacting bead-spring dumbbells, with a linear force-extension relation for the spring. The total stress in the polymer solution is given by a sum of the polymeric stress $\boldsymbol{\tau}$ and the (Newtonian) solvent contribution. The polymeric stress obeys the closed-form relation:
\begin{equation}	
\label{dim-Old-B-with-diffusion}
  \begin{aligned}
 \boldsymbol{\tau} + \lambda\left(\frac{\partial  \boldsymbol{\tau}}{\partial t} +  \boldsymbol{v.\nabla\tau} -  \boldsymbol{\nabla v}{^T}. \boldsymbol{\tau} - \boldsymbol{\tau}. \boldsymbol{\nabla v}\right) = \eta_{p} ( \boldsymbol{\nabla v}  \ +  \boldsymbol{\nabla v}{^T}) +  D \lambda \, \nabla^2 \boldsymbol{\tau} \, .
  \end{aligned}
\end{equation}
Here, $\lambda$ is the relaxation time, $\eta_p$ is the polymeric viscosity, and $D$ is the translational diffusivity of the dumbbell which is $O(10^{-12}\, \,\mathrm{m^2/s}
)$ for a typical high molecular weight polymer \cite{Dubief_Hof_EIT_AFM_2023}. On account of its negligibly small value, most theoretical treatments set $D = 0$, yielding the classical Oldroyd-B model which was proposed purely on continuum grounds by Oldroyd in 1950 \cite{Oldroyd1950}. The nature of the above constitutive relation changes from parabolic (for $D \neq 0$) to hyperbolic (for $D = 0$) -- as a result, the stress tensor is convected and its principal axes stretched and rotated along the fluid pathlines.
Physically, for a given velocity field, the constitutive relation becomes local in the absence of stress diffusion; this locality being, in fact, the reason underlying the viscoelastic continuous spectrum.

Gorodtsov and Leonov \cite{Gorodtsov1967} provided the first description of the spectrum of plane Couette flow of a UCM fluid (the limiting case of the Oldroyd-B fluid with no solvent) in the absence of fluid inertia (Reynolds number, $Re = 0$). They showed the existence of two stable discrete modes, and a (stable) continuous spectrum of modes with phase speeds spanning the entire base range of velocities, and with a common  decay rate that equals $-1/\lambda$. The location of the continuous spectrum is identified by recasting the governing linearized momentum and constitutive equations into a single fourth-order ordinary differential equation for the wall-normal velocity, in the wall-normal coordinate, and by setting the coefficient of the highest (fourth) derivative of the wall-normal velocity to zero. In other words, for eigenvalues belonging to the continuous spectrum, the governing fourth-order ODE exhibits a regular singular point. Renardy and Renardy \cite{renardy1986linear} were the first to report the spectrum for plane Couette flow of a UCM fluid at finite Reynolds number, and identified both the discrete spectrum and the numerical approximation of the elastic continuous spectrum. These authors pointed out that while the spectral method (used for the numerical calculation) represents the eigenfunctions as a sum of smooth basis functions, the eigenfunctions belonging to the CS are singular. As a result, they are typically not resolved accurately for a given number of collocation points that suffice for the discrete eigenmodes.
Graham \cite{Graham_1998} provided a more comprehensive analytical description of the eigenfunctions associated with the CS. 
Denoting the aforementioned regular singular point as $z = z_c$ ($z$ being the wall-normal coordinate, and $z_c$ being the wall-normal location where the base velocity equals the phase speed of the CS mode), Graham \cite{Graham_1998} showed that the perturbation stresses are proportional to $1/(z - z_c)^{m}$, where $m = 1, 2$ or $3$ for the $zz$, $xz$, and $xx$ stress components respectively.  The wall-normal velocity $v_z$ is a $C^1$ function, implying that its first derivative, $d v_z/dz$, exists and is continuous but not differentiable at all points in the domain. The second derivative, $d^2 v_z/dz^2$, can suffer a jump at $z = z_c$, as a result of which the third derivative of $v_z$ includes a contribution that is proportional to $\delta(z-z_c)$.  The linearized stress components are related to  the wall-normal velocity and its derivatives near $z_c$, and barring exceptional values of $z_c$ where one or more of these derivatives vanish, 
the stresses are singular at $z = z_c$. It is due to this singular behavior that the spectral method cannot capture the CS eigenvalues accurately, resulting in the aforementioned balloon-like structure in the complex plane.

Subsequently, Wilson \textit{et al.} \cite{Wilson1999} provided a detailed description of the spectrum for an Oldroyd-B fluid undergoing plane Couette and plane Poiseuille flow; the solvent to total solution viscosity ratio, $\beta$, now enters as an additional parameter.
These authors showed that, in addition to the elastic CS discussed above, there is another CS that arises due to solvent viscous effects, with decay rates proportional to $-1/(\beta \lambda)$, and which therefore recedes to $-\infty$ in the UCM limit ($\beta = 0$).  Physically, the timescale for the solvent viscous modes is proportional to $\eta_s/G$ ($G = \eta_p/\lambda$ being the elastic modulus), which vanishes for $\eta_s \rightarrow 0$, resulting in an infinitely large decay rate. From a mathematical standpoint, the highest derivative in the governing fourth-order equation has a term proportional to $\beta$, and the limit $\beta \rightarrow 0$ is therefore a singular one for the viscous CS above.
The Frobenius series solutions of the governing ODE near the regular singular point $z_c$ will, in general, comprise fractional powers or logarithms of $(z - z_c)$, implying multi-valuedness in the complex plane. Barring the case of plane Couette flow, the viscous CS is therefore a branch cut, and hence, the number of discrete eigenvalues can change due to eigenvalues crossing the branch cut, with changing values of one or more parameters ($\Wi$ or the streamwise wavenumber $k$). Thus, for plane Poiseuille flow, there are a maximum of six discrete eigenvalues in the inertialess limit, 
although this number decreases to five, upon change in the streamwise wavenumber, owing to one of the eigenvalues disappearing into the CS. 
Interestingly, the recently discovered unstable center mode \cite{Garg_etal_2018,chaudharyetal_2021,khalid2021centermode,khalid_creepingflow_2021}, alluded to above, and that is believed to underlie the transition to elastoinertial turbulence, also emerges from/disappears into the CS, upon variation of parameters such as $\Wi$.
 Renardy \cite{renardy2000location} has also provided a mathematical analysis of the location of CS in 2D flows of a UCM fluid without a stagnation point.

In contrast to the above comprehensive understanding of the CS for rectilinear flows of Oldroyd-B/UCM fluids,  an understanding of its nature in curvilinear flows is not fully mature yet, especially for the more realistic FENE-P model. The following efforts are exceptions, and worthy of mention.
Sureshkumar \cite{sureshkumar_2000} used the UCM model to examine the CS in journal bearing and eccentric Dean flows, where the flows are more complex, while still being shear-dominated.  Ghanbari and Khomami \cite{ghanbari2014onset} used the FENE-P model in the context of analyzing the role of thermal effects and the  gap-width ratio for Taylor-Couette flow.
Even for the simpler rectilinear case, the only effort to our knowledge,  that has examined the spectra (including the CS)
 for plane Couette flow of a FENE-P fluid, is that of Arora and Khomami \cite{Arora_Khomami}. They  showed that, as the finite extensibility parameter $L$ is decreased, the original CS with decay rate $-1/\lambda$ gives way to multiple distinct CS with different decay rates.  
Given that the decay rates of the CS are proportional to $1/\lambda$ (for $L \rightarrow \infty)$, and on account of
the shear-rate dependence of the relaxation time for the FENE-P model (finite extensibility leads to a decrease in relaxation time with increasing shear rates), one anticipates the CS for inhomogeneous shearing flows to be curves (not line segments) in the complex plane.   
There is also an added complexity in terms of the nontrivial coupling between different components of the conformation tensor, which we show, leads to up to six distinct CS for the FENE-P fluid, again in stark contrast to the Oldroyd-B model.

While the subject of purely elastic instabilities in curvilinear flows is quite well developed (see the reviews in Refs.\cite{Shaqfeh_Khomami_2021,Datta_etal2022,CastilloSanchez2022}), we are not aware of a comprehensive discussion on the CS in such flows when the FENE-P constitutive equation is used. 
The focus of the early experimental efforts was on Boger fluids, with an aim to minimize the effects of shear thinning, which perhaps explains the lack of theoretical analyses using the FENE-P model. However, 
there have been many experiments reported in the past decade in various curvilinear geometries where shear thinning is dominant \cite{dutcher2013effects,schaefer2018geometric,lacassagne2021shear,steinberg2022new,shakeri2022scaling,more2024elasto,bai2023viscoelastic,latrache2012transition}, and 
an accurate description of purely elastic instabilities in these experiments would entail use of the FENE-P model. Accordingly, 
in this work, we examine  both rectilinear (plane Couette and pressure-driven channel flows) and curvilinear (Dean and Taylor-Couette flows) configurations subjected to arbitrary three-dimensional disturbances,  and provide a comprehensive account of the nature of the CS in these flows using the FENE-P model.
The present study is closer in spirit to the earlier work of  Arora \textit{et al.} \cite{arora2004hydrodynamic}, who examined the structure of the spectra for plane Couette flow using the Pompom model (used to model melt rheology), and showed (numerically) that there are four CS in that model. Three of these are associated with the orientation tensor, and the fourth with the stretch equation.

An accurate knowledge of the CS is useful in many ways. From a purely numerical standpoint, this aids us in locating the true discrete modes and to separate them from the numerical approximations of the CS.
From a more fundamental perspective, we show that there are parameter regimes for the FENE-P fluid (for $L \lesssim 100$)
where there are no discrete modes in the spectrum, thereby pointing to the significance of the CS in the linearized dynamics. 
Interestingly, the CS is known to be especially significant in the classical inviscid context where the spectrum for the Rayleigh equation is purely continuous for non-inflexional velocity profiles. Further, although the CS modes are neutrally stable in the inviscid limit, they are known to drive transient algebraic growth. Indeed, 
Roy and Subramanian \cite{Roy_Subramanian_2014} showed that a superposition of CS modes can give rise to transient growth of perturbations, accounting for both the Orr and lift-up effects within an inviscid framework. In an analogous manner, CS mode superpositions are expected to play a role in transient growth even in the inertialess viscoelastic context \cite{jovanovic_kumar_2010,jovanovic_kumar_2011}, despite the individual modes being stable.  
In a somewhat different context, recently, Beneitez \textit{et al.} \cite{BeneitezPRF2023} showed that the inclusion of stress diffusion leads to a new type of instability (termed the `polymer diffusive instability', abbreviated as PDI) in plane Couette and Poiseuille flows. The presence of diffusion destroys the CS and a set of discrete `diffusive' modes emerges out of the (erstwhile) CS, including the unstable PDI mode. Given the complex structure of the CS in shearing flows of  FENE-P fluids, it may be anticipated that inclusion of a small amount of stress diffusion can lead to a very complicated discrete spectrum with components associated with each of the (multiple) CS predicted here; the PDI mode(s) will also likely emerge from one or more of these CS. Understanding the CS will therefore help one better understand the nontrivial relation between the discrete spectrum for $D \neq 0$ and the CS modes for $D = 0$. It is worth adding that this relation may be nontrivial, in the sense of not being one-to-one, as has been demonstrated in the context of the Orr-Sommerfeld and Rayleigh spectra for Newtonian fluids \cite{Roy_Subramanian_2014}.
 
The rest of this paper is organized as follows. In Sec.\,\ref{sec:probformulation}, we describe the flow configurations considered in this work, as well as the governing equations and the constitutive model used. We proceed to first discuss the relatively simple case of rectilinear flows in Sec.\,\ref{sec:rectilinear}. This is divided into two subsections, with Sec.\,\ref{sec:rectilinear_PC} dealing with plane Couette flow and Sec.\,\ref{sec:rectilinear_PP} dealing with pressure-driven channel flow (characterized by the plane Poiseuille profile for Newtonian and Oldroyd-B fluids). We consider curvilinear shearing flows in Sec.\,\ref{sec:curvilinear},  and
first explain the procedure to identify the CS in these flows subjected to arbitrary three-dimensional disturbances.
We then discuss the results for both the Dean (Sec.\,\ref{subsec:Dean}) and Taylor-Couette (Sec.\,\ref{sec:TaylorCouette})  configurations. Finally, in  Sec.\,\ref{sec:concl}, we provide the salient conclusions of the present effort.

\section{Problem Formulation}
\label{sec:probformulation}

The FENE-P model is one of the most widely used constitutive equations for dilute polymeric solutions \cite{Bird1980,birdvol2}. In this model, the (dimensionless) polymeric stress tensor $\boldsymbol{\tau}$ is related to the conformation tensor $\mathbf{C}$ by
\begin{equation}
    \boldsymbol{\tau} = \frac{f\mathbf{C} - \mathbf{I}}{\Wi}\, ,
    \label{eq:tau_c_f}
\end{equation}
where $f$ is the Peterlin function
\begin{equation}
    f = \frac{L^2 - 3}{L^2 - \Tr(\mathbf{C})}\, .
\end{equation}
Here $L$ is the (dimensionless) finite extensibility parameter, with $L \rightarrow \infty$ being the limit of an Oldroyd-B fluid. The equation governing the evolution of $\mathbf{C}$ is given by
\begin{equation}
   \frac{\partial \mathbf{C}}{\partial t} + \boldsymbol{v}\cdot\nabla\mathbf{C} - (\nabla \boldsymbol{v})^T \cdot \mathbf{C} - \mathbf{C}\cdot(\nabla \boldsymbol{v}) = - \boldsymbol{\tau} \, .
\label{eq:C_tau}
\end{equation}
The scheme for nondimensionalizing the conformation tensor is identical to that used in Ref.\,\cite{Khalid_etal_2025}.
The flow geometries considered in this study are (i) plane Couette and pressure-driven channel flows as examples of rectilinear shearing flows, and (ii) the Taylor-Couette and Dean flows as examples of curvilinear shearing flows. In the following sections, for plane Couette flow, the velocity has been non-dimensionalized with \( U_w \), the velocity of the upper or lower plate, whereas for pressure-driven channel flow, the characteristic velocity is taken as the centerline maximum velocity (\( U_c \)) for an Oldroyd-B fluid subjected to the same pressure gradient as that imposed on the FENE-P fluid. The characteristic velocity (\( U_c \)) for Taylor-Couette flow is taken as the velocity of the inner rotating cylinder, while that for Dean flow is taken to be
$U_c = \frac{-d^2}{2 \eta_{t0} R_1}\frac{dP}{d\theta}$ \cite{Tej2024}; the latter velocity scale was first used by Joo and Shaqfeh in their analysis of Dean flow \cite{joo_shaqfeh_1991}. Here, $\frac{dP}{d\theta}$ is the imposed pressure gradient in the azimuthal direction ($\theta$). The characteristic length scale (\( L_c \)) for the rectilinear flows is taken as half the channel width (\( H/2 \)), whereas that for the curvilinear flows (Taylor-Couette and Dean flows), is taken as the gap width between the inner and outer cylinders (\( d = R_o - R_i \)). The characteristic time scale for all the flows is \( L_c / U_c \). 
    The polymeric stress tensor is non-dimensionalized by \( \eta_{p0} U_c / L_c \), and the pressure by \( \eta_{t0} U_c / L_c \). Here, and in the expression for $U_c$ above,  \( \eta_{t0} = \eta_{p0} + \eta_s \) is the total zero-shear viscosity, given by the sum of the (zero-shear) polymeric and solvent viscosities. Note that, owing to shear thinning in the FENE-P model, it is convenient to use the zero-shear polymeric ($\eta_{p0})$ and total ($\eta_{t0}$) viscosities in the scales for stresses and the pressure field. The governing continuity and Cauchy momentum equations in dimensionless form are given as follows:
\begin{eqnarray}
    \nabla \cdot \boldsymbol{v} &=& 0\\
    Re \left(\frac{\partial \boldsymbol{v}}{\partial t} + \boldsymbol{v} \cdot \nabla \boldsymbol{v}\right) &=& - \nabla p + \beta \nabla^2 \boldsymbol{v} + (1 - \beta) \nabla \cdot \boldsymbol{\tau}\, .
\end{eqnarray}
Here, $\boldsymbol{v}$ is the velocity field, $p$ is the pressure field, and $\boldsymbol{\tau}$ is the polymeric stress tensor governed by the FENE-P model as discussed above.
We proceed along the lines of a standard temporal linear stability analysis. Infinitesimal perturbations are introduced to the velocity, pressure, and stress fields, and the equations are linearized. The perturbations are then expressed in terms of Fourier modes; the forms of the Fourier modes for rectilinear and curvilinear shearing flows are given in the pertinent sections below. The linearized stability equations for the rectilinear \cite{Khalid_etal_2025} and curvilinear flows are provided in Appendices \ref{FENEP_PCF_Appendix}  and \ref{FENEP_TC_Appendix}, respectively. These equations are solved numerically using the Chebyshev pseudo-spectral collocation method, from which the eigenspectra are obtained for the various flow configurations.

 \section{Rectilinear flows}
 \label{sec:rectilinear}

It is instructive to first discuss the nature of the CS for rectilinear shearing flows, before moving on to the more complex case of curvilinear shearing flows. Although we consider plane Couette and pressure-driven channel flows as examples here, 
a brief demonstration is given of our results  for pressure-driven channel flow being applicable to pressure-driven pipe flow as well. The perturbations to the various field variables are expressed in the form $\phi'(x,y,z,t) = \tilde{\phi}(z) \exp[i (kx - \frac{\omega}{\Wi}t) + i l y]$. Here, $k$ and $l$ are the wavenumbers in the base-flow ($x$) and vorticity ($y$) directions, and $\omega = \omega_r + i \omega_i$ is the complex frequency (nondimensionalized using $\lambda^{-1}$) with $\omega_i$ $> 0$ ($<0$) representing unstable (stable) modes; it is natural to use $\lambda^{-1}$ since, as mentioned in the Introduction, the (dimensional) decay rate of the CS modes in the Oldroyd-B limit is proportional to $-1/\lambda$. Also,
it is appropriate to use the complex frequency instead of the phase speed $c = \omega/(k\Wi)$, since we also examine the limit $k = 0$, with $\omega$ finite. 
In the Fourier mode ansatz above, $z$ is the wall-normal coordinate, and $\tilde{\phi}(z)$ denotes the eigenfunctions in this direction. For future reference, the conformation tensor ($\mathbf{C}$) in the base state (denoted by an overbar) is given by
 \begin{equation}
    \bar{\mathbf{C}} = \begin{bmatrix} \bar{C}_{xx} & \bar{C}_{xy} & \bar{C}_{xz} \\ \\
    \bar{C}_{yx} & \bar{C}_{yy} & \bar{C}_{yz} \\ \\\bar{C}_{zx} & \bar{C}_{zy} & \bar{C}_{zz}
    \end{bmatrix} = \begin{bmatrix}
        \frac{1 + \frac{2 \Wi^2}{\overline{f}^2}\left(\frac{\partial \overline{V}_x}{\partial z}\right)^2}{\overline{f}} & 0 & \frac{\Wi}{\overline{f}^2}\frac{\partial \overline{V}_x}{\partial z} \\ \\ 0 & \frac{1}{\overline{f}} & 0 \\ \\ \frac{\Wi}{\overline{f}^2}\frac{\partial \overline{V}_x}{\partial z} & 0 & \frac{1}{\overline{f}} 
    \end{bmatrix}\, ,
\end{equation}
with $\overline{V}_x$ denoting the base-state velocity profile associated with the shearing flows.
\begin{figure}
  \centering
  \subfigure[$l = 0$]
  {\includegraphics[width=0.45\textwidth]{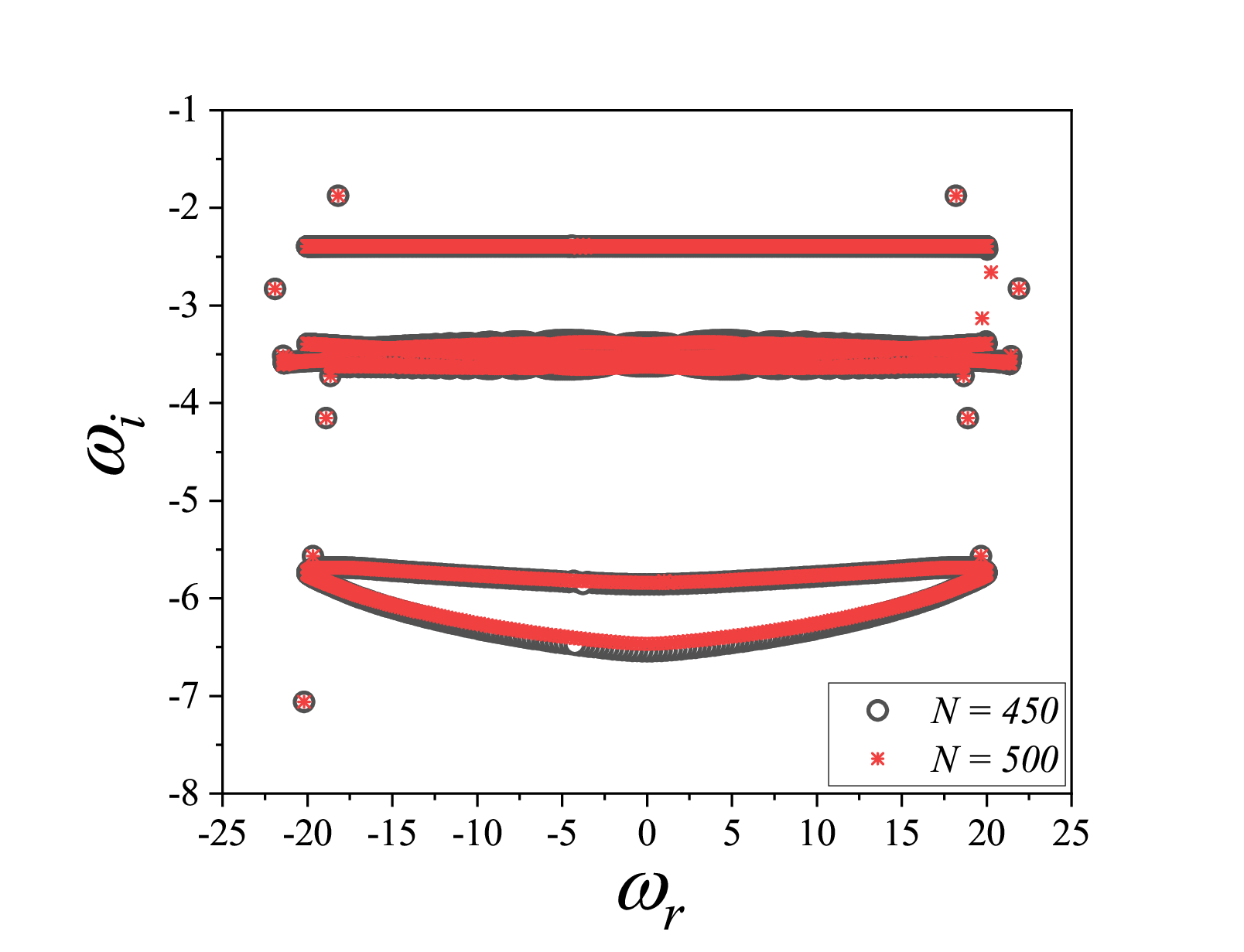}\label{fig:l0_PC}}
    \subfigure[$l = 5$]
  {\includegraphics[width=0.45\textwidth]{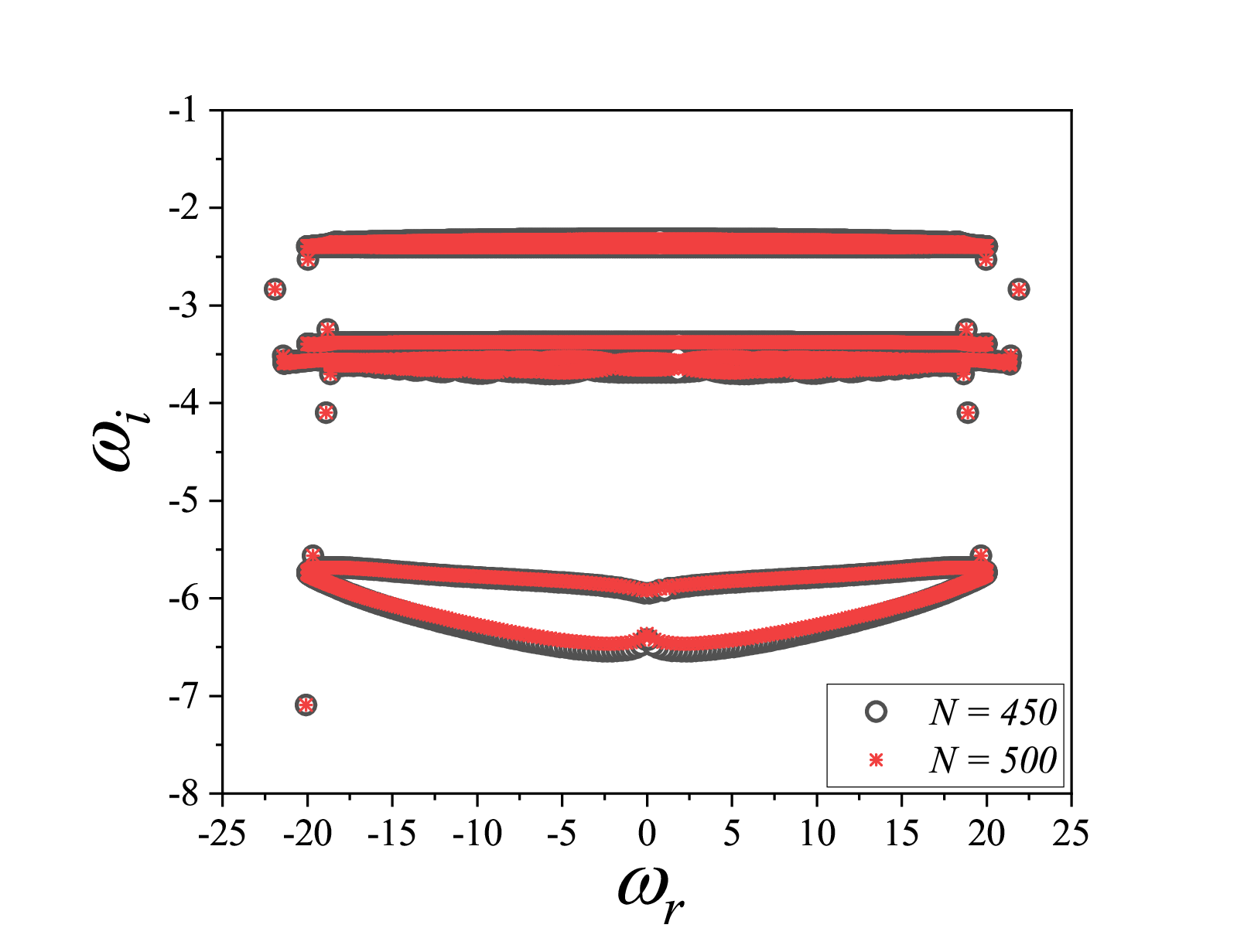}\label{fig:l5_PC}}
  \caption{Numerical  eigenspectra for plane Couette flow showing the independence of CS with respect to the spanwise wavenumber $l$; data for $k = 0.1$, $W = 200, L = 100, \beta = 0.5$.}
  \label{fig:Effect_l_PC_CS}
\end{figure}

As already evident from Fig.\,\ref{fig:Effect_Re_Dean_CS},
the location and nature of the CS is not affected by $Re$ (provided that it is finite), and so, without loss of generality, we set $Re = 0$ in the following discussion; it being understood that the results described below apply equally for $Re \neq 0$. 
Mathematically, this is because the terms proportional to $Re$ do not contribute to the highest derivative of the governing differential equation, and it is the latter that determines the CS eigenvalues.
We note, however, that there exists a distinguished limit defined by $Re, W\!i \rightarrow \infty$ such that the ratio $E = W\!i/Re$ is finite -- the so-called `elastic Rayleigh limit' \cite{azaiez_homsy_1994,Roy_Garg_Reddy_Subramanian_2022}, where the structure of the CS does depend on $Re$ (via $E$). Physically, this limit corresponds to the neglect of momentum diffusion as well as relaxation of the polymeric stress. The nature of the CS in the elastic Rayleigh limit is significantly different from what is discussed below. In this limit, one obtains three distinct CS \cite{Roy_Garg_Reddy_Subramanian_2022}, of which the first is 
the elastically-modified CS of the classical Rayleigh equation \cite{Case1960} and corresponds to the base-state interval of velocities.  In addition, there is a pair of CS  corresponding to upstream and downstream traveling elastic shear waves, which arise due to a balance of inertial and elastic effects; this pair disappears when $W\!i$ is finite.
With the above caveat,  we proceed below assuming $Re = 0$.
Further, without loss of generality, and in the interests of clarity,  we set the wavenumber in the base-state vorticity direction $l$ to zero;  the numerical spectra reveal that the structure of the CS is independent of $l$ (see Fig.\,\ref{fig:Effect_l_PC_CS}). This amounts to considering  only two-dimensional perturbations in the $x$--$z$ plane, and ignoring velocity and (normal and shear) stress components pertaining to the $y$ (base-state vorticity)-direction.

In order to locate the CS, it is necessary to convert the continuity, momentum, and constitutive equations into a single fourth-order ordinary differential equation for $\widetilde{v}_z$.
 This is achieved by first eliminating  pressure from the two momentum equations, and then eliminating $\tilde{v}_x$ using the continuity equation, resulting in:
  \begin{equation}
 \begin{aligned}
     \frac{\beta i}{k 
     } \left(\frac{d^2}{d z^2} - k^2  \right)^2 \widetilde{v}_z  + (1 - \beta)\left[ \left( \frac{d^2}{d z^2} + k^2 \right) \widetilde{\tau}_{xz} 
     + i k \frac{d }{d z} \left(\widetilde{\tau}_{xx} -  \widetilde{\tau}_{zz}\right) \right] = 0\, .
\label{eq:single4thorder_rectilinear}
 \end{aligned}
 \end{equation}
When the coefficient of the highest order derivative of the above ODE is set to zero, one obtains the locations of the CS.
Note, however, that this coefficient, as it appears in Eq.\,\ref{eq:single4thorder_rectilinear}, is not the complete version; there are other contributions that arise upon expressing the perturbation stresses  in Eq.\,\ref{eq:single4thorder_rectilinear} in terms of the velocity perturbations using the linearized constitutive relations.  The components of the linearized form of Eq.\,\ref{eq:C_tau} in terms of the perturbation stress ($\widetilde{\boldsymbol{\tau}}$) and conformation tensor ($\widetilde{\boldsymbol{C}}$) fields are given in Appendix~\ref{FENEP_PCF_Appendix}.
As explained below, the complete expression for the aforementioned coefficient is a polynomial in the eigenvalue $\omega$, and the degree of the polynomial determines the number of distinct CS; for the FENE-P fluid, we find that there are, in principle, six distinct CS.

  To obtain a closed system of equations involving the conformation tensor alone, one first uses the linearized form of Eq.\,\ref{eq:tau_c_f} to express $\widetilde{\boldsymbol{\tau}}$ in terms of $\widetilde{\boldsymbol{C}}$ in the form
 \begin{equation}
     \widetilde{\boldsymbol{\tau}} = \frac{\overline{f}^2~\overline{\mathbf{C}} ~tr(\widetilde{\mathbf{C}})}{(L^2 - 3)\Wi} + \frac{\overline{f}~\widetilde{\mathbf{C}}}{\Wi} \, .
     \label{eq:C_tau_final}
 \end{equation}
As before, base-state quantities are denoted with an overbar, with $\bar{f}$ now being the Peterlin function in the base state.
The components of Eq.\,\ref{eq:C_tau_final} may written explicitly as follows:
 \begin{equation}
     \widetilde{\tau}_{xx} = \left(\frac{\overline{f}^2~\overline{C}_{xx}}{(L^2 - 3)W} + \frac{\overline{f}}{\Wi} \right)\widetilde{C}_{xx} + \frac{\overline{f}^2~\overline{C}_{xx}}{(L^2 - 3)\Wi}\widetilde{C}_{yy} + \frac{\overline{f}^2~\overline{C}_{xx}}{(L^2 - 3)\Wi}\widetilde{C}_{zz}\, ,
     \label{tau_xx_rect}
 \end{equation}
 \begin{equation}
      \widetilde{\tau}_{xz} = \frac{\overline{f}^2~\overline{C}_{xz}}{(L^2 - 3)\Wi} \widetilde{C}_{xx} + \frac{\overline{f}^2~\overline{C}_{xz}}{(L^2 - 3)Wi}\widetilde{C}_{yy} + \frac{\overline{f}^2~\overline{C}_{xz}}{(L^2 - 3)Wi}\widetilde{C}_{zz} + \frac{\overline{f}}{Wi}\widetilde{C}_{xz}\, ,
      \label{tau_xz_rect}
 \end{equation}
  \begin{equation}
     \widetilde{\tau}_{yy} = \frac{\overline{f}^2~\overline{C}_{yy}}{(L^2 - 3)\Wi}\widetilde{C}_{xx} + \left(\frac{\overline{f}^2~\overline{C}_{yy}}{(L^2 - 3)W} + \frac{\overline{f}}{\Wi} \right)\widetilde{C}_{yy}   + \frac{\overline{f}^2~\overline{C}_{yy}}{(L^2 - 3)\Wi}\widetilde{C}_{zz}\, ,
     \label{tau_yy_rect}
 \end{equation}
 \begin{equation}
     \widetilde{\tau}_{zz} =\frac{\overline{f}^2~\overline{C}_{zz}}{(L^2 - 3)\Wi}\widetilde{C}_{xx} + \frac{\overline{f}^2~\overline{C}_{zz}}{(L^2 - 3)\Wi}\widetilde{C}_{yy} + \left(\frac{\overline{f}^2~\overline{C}_{zz}}{(L^2 - 3)\Wi} + \frac{\overline{f}}{\Wi} \right)\widetilde{C}_{zz} \, .
     \label{tau_zz_rect}
 \end{equation}

Next, we substitute the above expressions for the stress components in the linearized constitutive equations (see Eqs.\,\ref{Cxx_rect} - \ref{Czz_rect} of Appendix\,\ref{FENEP_PCF_Appendix}) to obtain:
 \begin{equation}
 \begin{aligned}
     \left(i(k\overline{V}_x - \omega) + \frac{\overline{f}^2~\overline{C}_{xx}}{(L^2 - 3)\Wi} + \frac{\overline{f}}{\Wi}\right)\widetilde{C}_{xx} - 2\frac{d \overline{V}_x}{d z}\widetilde{C}_{xz} + \frac{\overline{f}^2~\overline{C}_{xx}}{(L^2 - 3)Wi}\widetilde{C}_{yy} + &\\\frac{\overline{f}^2~\overline{C}_{xx}}{(L^2 - 3)Wi}\widetilde{C}_{zz} - \left(2ik\overline{C}_{xx} + 2\overline{C}_{xz}\frac{d}{d z}\right)\widetilde{v}_x + \frac{d \overline{C}_{xx}}{d z}\widetilde{v}_z = 0 \, ,
 \end{aligned}
 \label{eq:tildeCxx}
 \end{equation}
 \begin{equation}
 \begin{aligned}
     \left(i(k\overline{V}_x - \omega) + \frac{\overline{f}}{\Wi}\right)\widetilde{C}_{xz} + \frac{\overline{f}^2~\overline{C}_{xz}}{(L^2 - 3)\Wi} \widetilde{C}_{xx} + \frac{\overline{f}^2~\overline{C}_{xz}}{(L^2 - 3)\Wi}\widetilde{C}_{yy} + &\\ \left(\frac{\overline{f}^2~\overline{C}_{xz}}{(L^2 - 3)\Wi} - \frac{d \overline{V}_x}{d z}\right)\widetilde{C}_{zz} -  \overline{C}_{zz}\frac{d \widetilde{v}_x}{d z} + \left(\frac{d \overline{C}_{xz}}{d z} -ik\overline{C}_{xx}\right)\widetilde{v}_z = 0 \, ,
 \end{aligned}
 \label{eq:tildeCxz}
 \end{equation}
  \begin{equation}
 \begin{aligned}
     \left(i(k\overline{V}_x - \omega) + \frac{\overline{f}^2~\overline{C}_{yy}}{(L^2 - 3)\Wi} + \frac{\overline{f}}{\Wi}\right)\widetilde{C}_{yy} + \frac{\overline{f}^2~\overline{C}_{yy}}{(L^2 - 3)\Wi}\widetilde{C}_{xx} + \frac{\overline{f}^2~\overline{C}_{yy}}{(L^2 - 3)Wi}\widetilde{C}_{zz}  + \frac{d \overline{C}_{yy}}{d z}\widetilde{v}_z = 0 \, ,
 \end{aligned}
 \label{eq:tildeCyy}
 \end{equation}
 \begin{equation}
 \begin{aligned}
     \left(i(k\overline{V}_x - \omega) + \frac{\overline{f}^2~\overline{C}_{zz}}{(L^2 - 3)\Wi} + \frac{\overline{f}}{\Wi}\right) \widetilde{C}_{zz} + \frac{\overline{f}^2~\overline{C}_{zz}}{(L^2 - 3)\Wi}\widetilde{C}_{xx} + &\\ \frac{\overline{f}^2~\overline{C}_{zz}}{(L^2 - 3)\Wi}\widetilde{C}_{yy} + \left(\frac{d \overline{C}_{zz}}{d z} - 2\overline{C}_{xz}ik - 2\overline{C}_{zz}\frac{d}{d z}\right)\widetilde{v}_z = 0 \, .
 \end{aligned}
 \label{eq:tildeCzz}
 \end{equation}

The four equations above form a linear system  $ \boldsymbol{M} \boldsymbol{a} = \boldsymbol{b}$ for the vector $\boldsymbol{a}  = [\tilde{C}_{xx}, \tilde{C}_{xz}, \tilde{C}_{zz}, \tilde{C}_{yy}]$,  being `forced' by $\boldsymbol{b}$ that contains $\widetilde{v}_z$, $\widetilde{v}_x$ and their derivatives.
Here, the matrix $\boldsymbol{M}$ and $\boldsymbol{b}$ contain base-state quantities such as $\bar{f}$, and the components of $\bar{\textbf{C}}$.
This linear system is solved for $\boldsymbol{a}$, which is then substituted in Eqs.\,\ref{tau_xx_rect}--\ref{tau_zz_rect}, to obtain the perturbation stresses in terms of $\widetilde{v}_z$ and its derivatives. Next, on substituting these expressions into Eq.\,\ref{eq:single4thorder_rectilinear}, the resulting coefficient of the fourth derivative of $\widetilde{v}_z$ turns out to be a sixth-order polynomial in $\omega$, whose (in principle, distinct) roots give rise to six CS. Unlike the Oldroyd-B case, where it is easier to obtain closed-form analytical expressions for the CS eigenvalues in terms of $W\!i$ and $\beta$, for the FENE-P fluid, one obtains the roots numerically only after specifying the  values of $W\!i$, $\beta$, $k$, and $L$.  
As $L \rightarrow \infty$ (the Oldroyd-B limit), five of the six roots  degenerate to  $\omega_i = -1$, the elastic CS, while the remaining root reduces to the `solvent CS' (with $\omega_i = -1/\beta$).  For nonzero $k$, the CS are horizontal line segments in the $\omega_r-\omega_i$ plane, as expected.

The above (exact) procedure for determining the coefficient of the highest order derivative of $\widetilde{v}_z$ is implemented for rectilinear (plane Couette and pressure-driven channel) flows. However, this methodology
is algebraically cumbersome for curvilinear geometries, on account of the more complex expressions for the gradient operator in cylindrical coordinates. To this end, we propose and validate  here a simpler alternative. This procedure is first demonstrated for rectilinear flows where we have exact solutions for all the six CS locations,  and is
subsequently employed for curvilinear flows. 
The simplified procedure involves determination of the CS eigenvalues in the limit of streamwise invariant disturbances ($k = 0$), where the system of equations becomes considerably simpler. Since we have already set $l = 0$, this amounts to considering disturbances that are both streamwise ($x$) and spanwise ($y$) uniform, and are therefore unidirectional shearing flows that vary only in the wall-normal ($z$) direction. Indeed, for $k = 0$, the continuity equation (Eq.\,\ref{eq:conty_rectilinear} of the Appendix) yields $\frac{d \widetilde{v}_z}{d z} = 0$, which yields $\widetilde{v}_z = 0$ since $\widetilde{v}_z = 0$ at $z = \pm 1$  (the no-penetration boundary condition at the walls). In this limit, the $z$-momentum equation therefore gets trivially decoupled, and it suffices to examine the highest order derivative of the $x-$momentum equation (i.e., $\frac{d^2 \widetilde{v}_x}{d z^2}$) to arrive at the expressions for the CS. Four of the six roots turn out to be purely imaginary, while the remaining two roots are complex, of the form $\pm a - i b$ ($a,b > 0)$.
Out of the four purely imaginary roots, three are virtually identical for $L > 50$, this range being appropriate for high-molecular weight polymers used in experiments \cite{Bidhan2018,choueiri2021experimental,Steinberg2021,YiBaoZhang_etal}. Thus, for all practical purposes, three of the CS are identical, being given by $\omega_i = -\bar{f}$; we therefore label them as CS1a, 1b, and 1c. 
The fourth purely imaginary root is the continuation of the solvent CS in the Oldroyd-B limit, and we label this as `CS2' henceforth.  The pair of complex CS is labelled as CS3a and 3b.

Next, by comparing the results for $k = 0$ with the exact results for $k \neq 0$, we find that all the six
 CS for nonzero $k$ can be readily obtained from their $k = 0$ analogues  by an identical horizontal shift  of $\omega_r = k\Wi\overline{V}_x$.  For the special case of plane Couette flow, the points that represent the CS at $k = 0$ are stretched out  into horizontal line segments (for nonzero $k$) of length $k \Wi \Delta \bar{V}$, where, for the scalings chosen, $\Delta \bar{V} = 2$ is the range of velocity variation in the base state.  
 For inhomogeneous shearing flows, 
 owing to the variation of $\overline{V}_x$ with the wall-normal coordinate $z$, the horizontal shift is not uniform, however, leading to curves in the complex plane
The above procedure proves helpful for the determination the CS for curvilinear flows subsequently in Sec.\,\ref{sec:curvilinear}.

\begin{figure}
\centering
    \subfigure[$k = 0$]
{\includegraphics[width=0.35\textwidth]{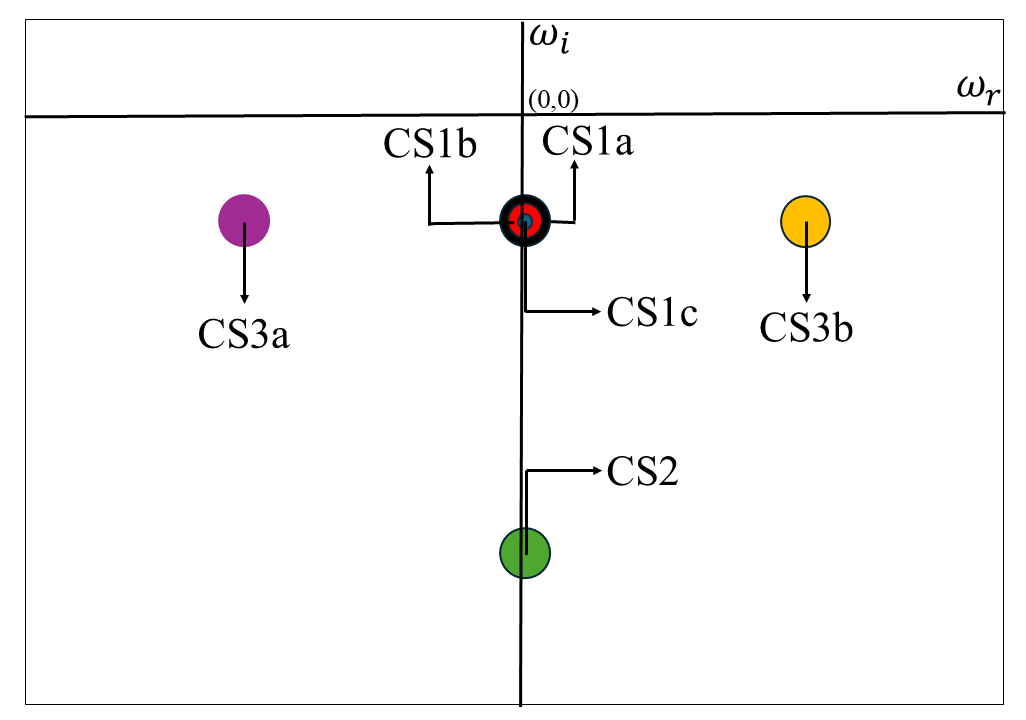}\label{fig:PC_K_0_schematics}}
\quad \quad
\subfigure[$k \neq 0$]
{\includegraphics[width=0.35\textwidth]{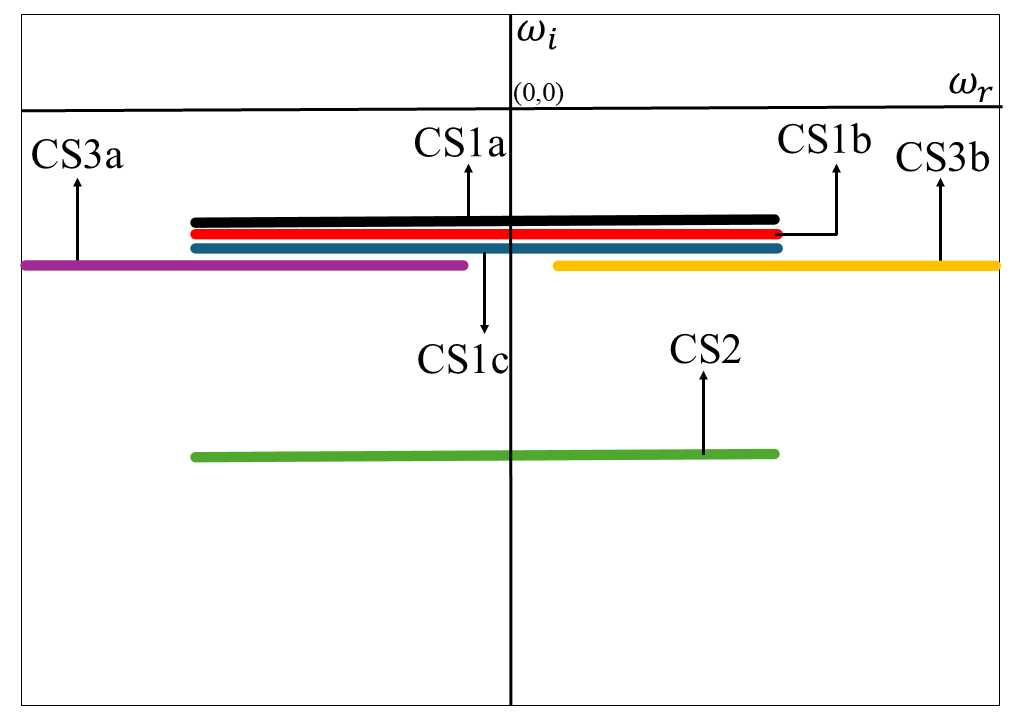}\label{fig:CPC_K_nonzero_schematics}}
  \caption{Schematic location of the various CS for plane Couette flow for $\beta \rightarrow 1$. For $L \gtrsim 50$, CS1a, 1b, and 1c are nearly identical. Their vertical loci are shown to distinct in this schematic for the purposes of illustration.}
  \label{fig:PC_CS_schematics}
\end{figure}

\begin{figure}
\centering
    \subfigure[$\Wi/L = 2$]
{\includegraphics[width=0.45\textwidth]{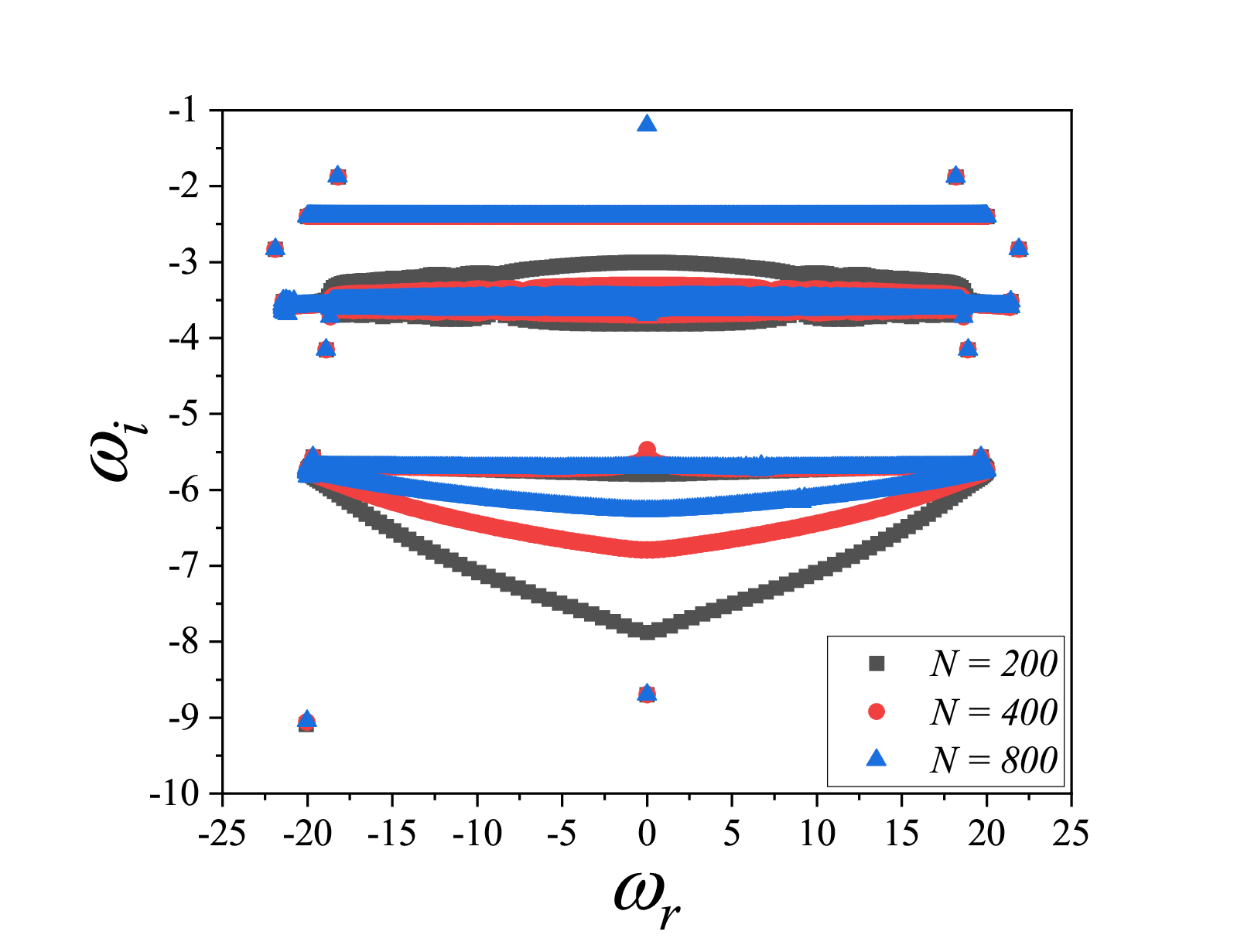}\label{fig:L_100_PCF_diff_N}}
    \subfigure[$\Wi/L = 0.02$]
{\includegraphics[width=0.45\textwidth]{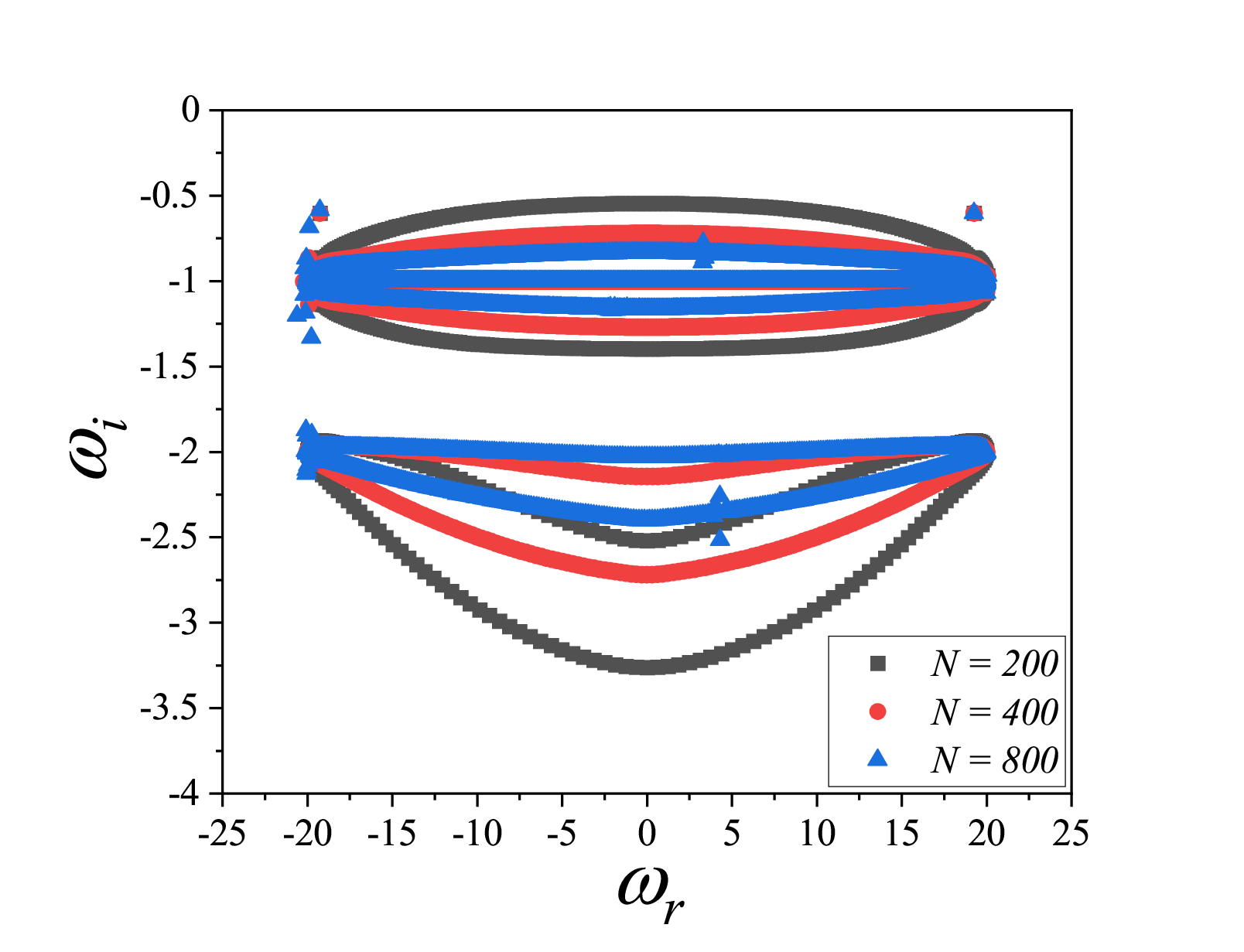}\label{fig:L_10000_PCF_diff_N}}
\caption{Numerical eigenspectra for plane Couette flow at different $N$'s illustrating the slow convergence of the various CS balloons. Data for $\Wi = 200, \beta = 0.5, k = 0.1, Re = 0$.}
\label{fig:PCF_diff_N}
\end{figure}

\begin{figure}
\centering
    \subfigure[$k = 0$ (Analytical)]{
\includegraphics[width=0.35\textwidth]{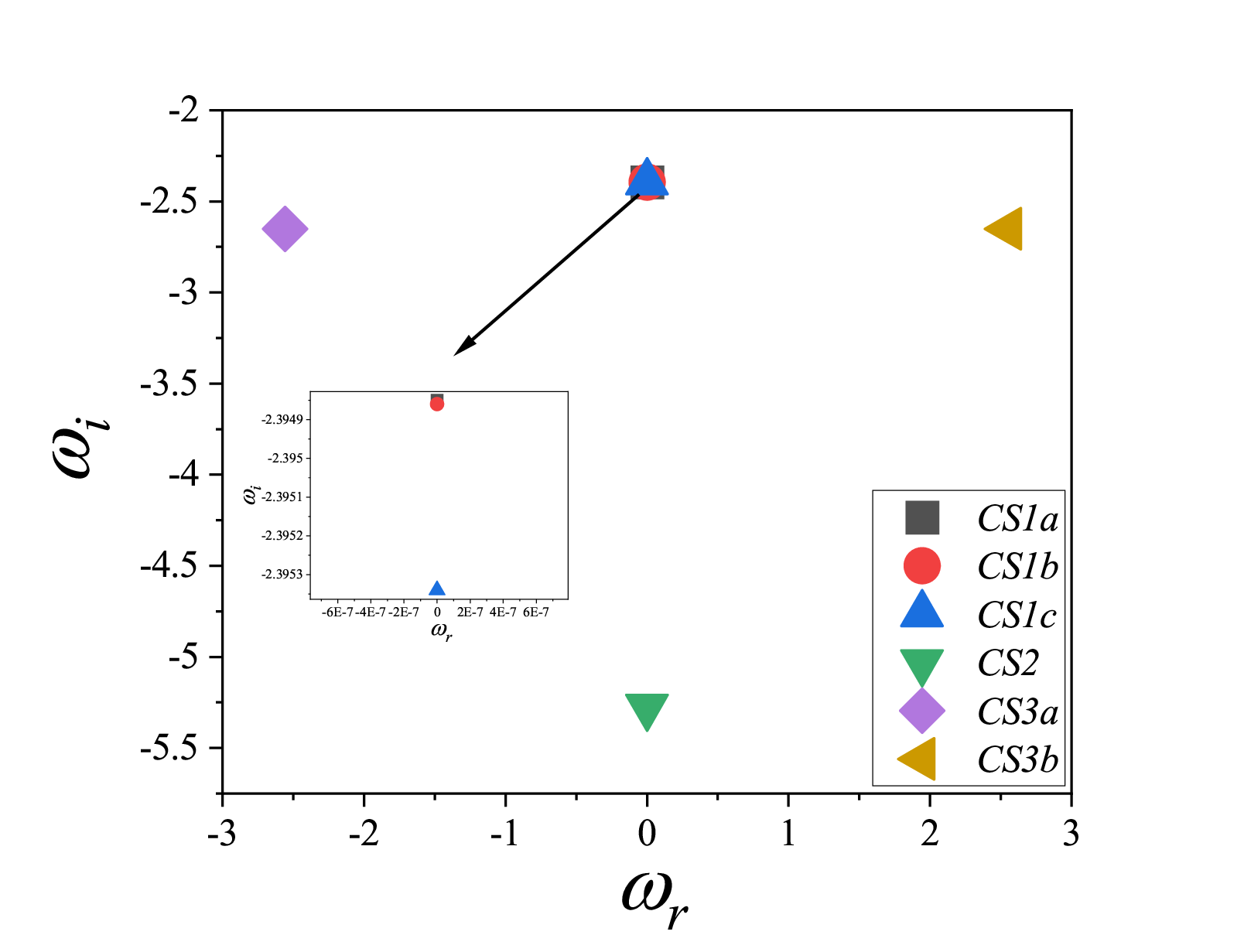}\label{fig:k0_PC_ana}
}
\subfigure[$k = 0$ (Numerical)]
{\includegraphics[width=0.35\textwidth]{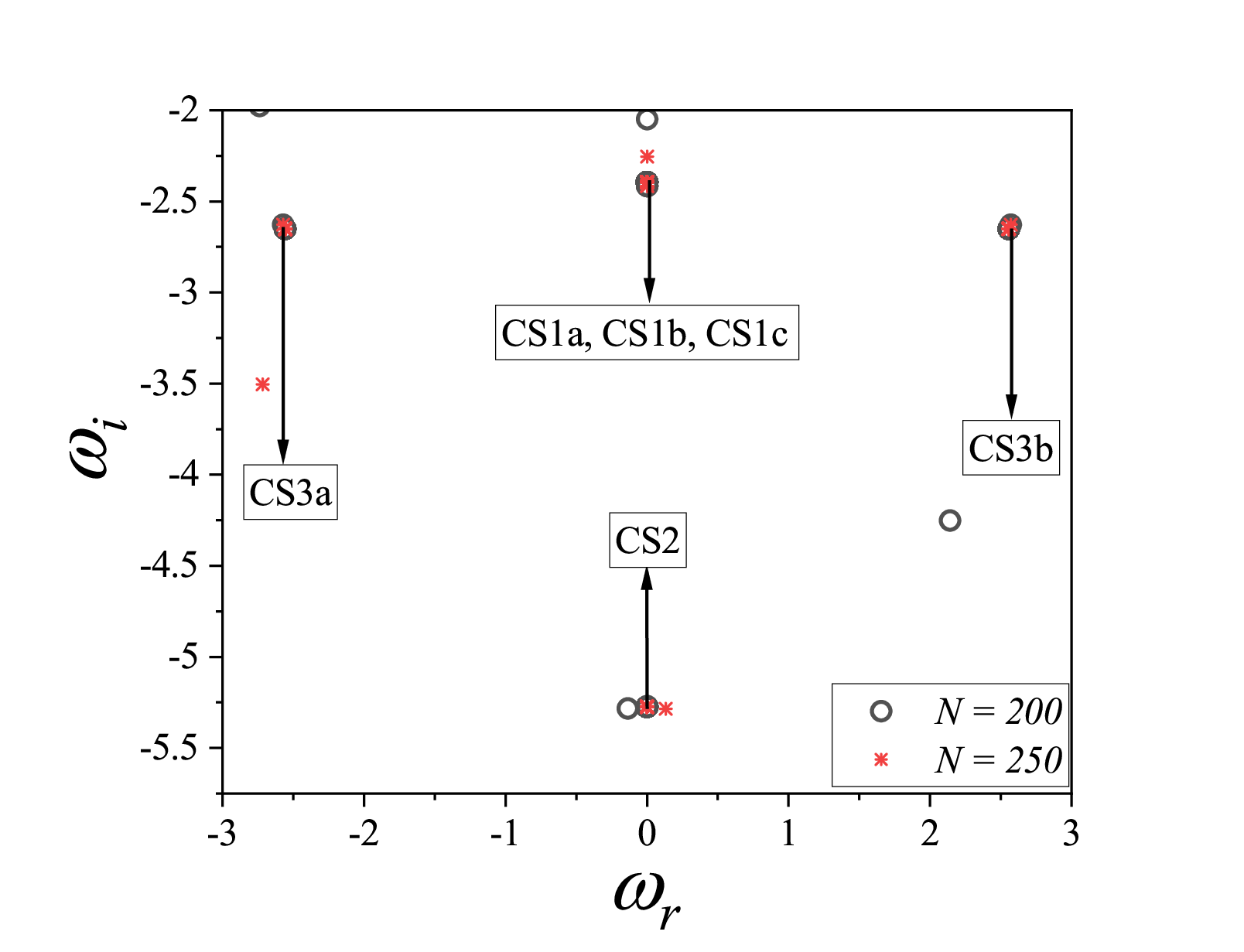}\label{fig:k0_PC_num}}
\subfigure[$k = 0.01$ (Analytical)]
{\includegraphics[width=0.35\textwidth]{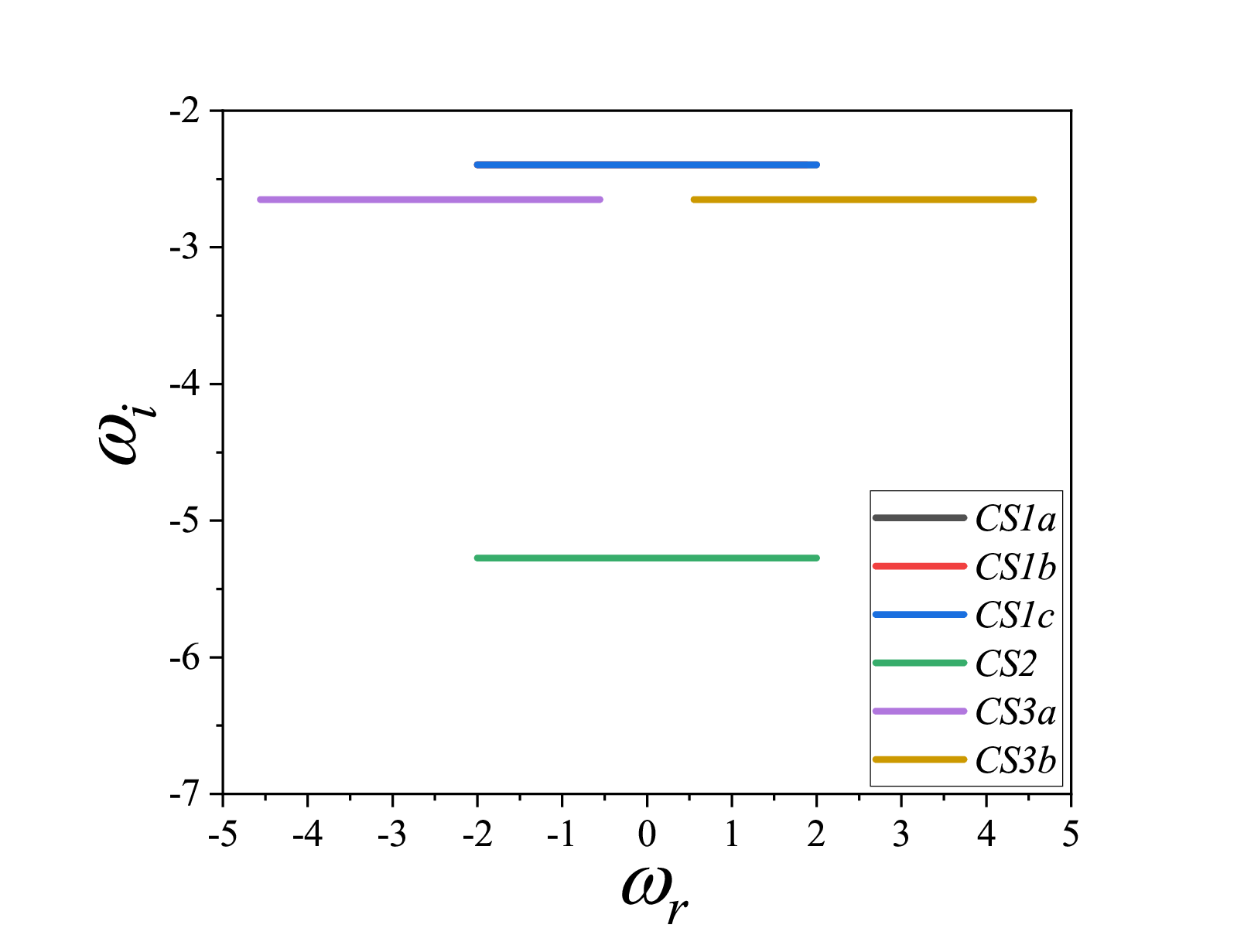}\label{fig:k0.01_PC_ana}}
\subfigure[$k = 0.01 $ (Numerical)]
{\includegraphics[width=0.35\textwidth]{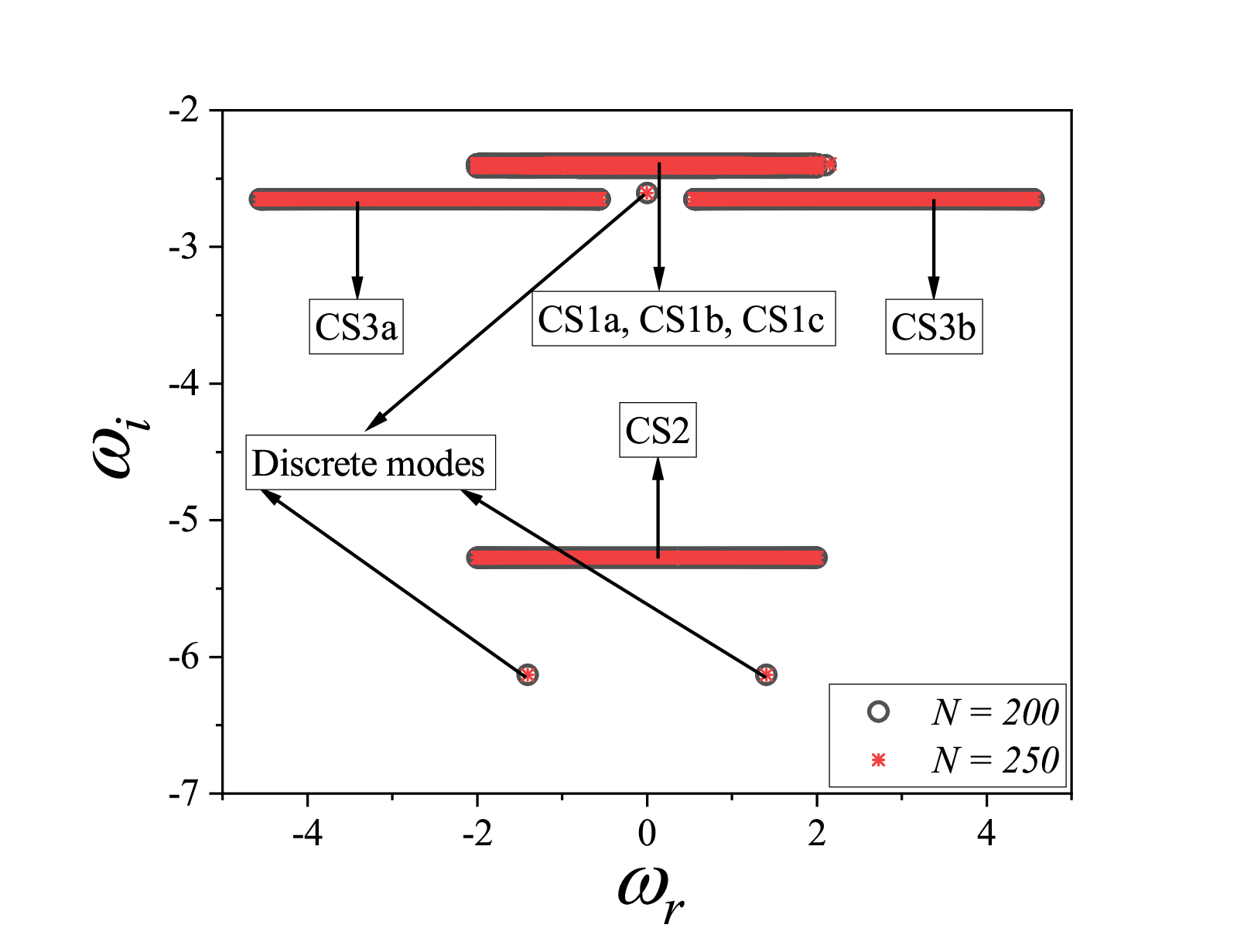}\label{fig:k0.01_PC_num}}
\subfigure[$k = 0.1$ (Analytical)]
{\includegraphics[width=0.35\textwidth]{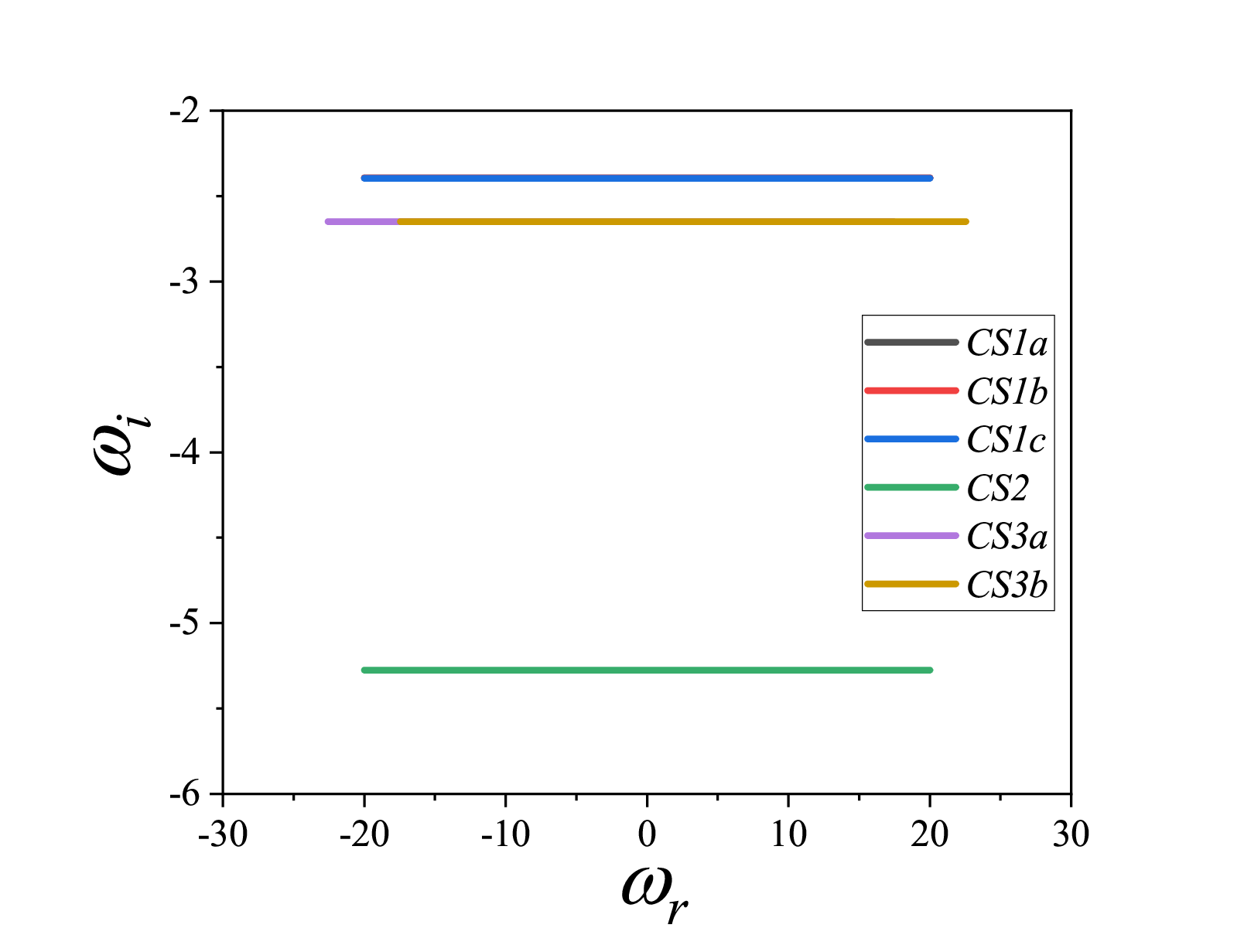}\label{fig:k0.1_PC_ana}}
\subfigure[$k = 0.1 $ (Numerical)]
{\includegraphics[width=0.35\textwidth]{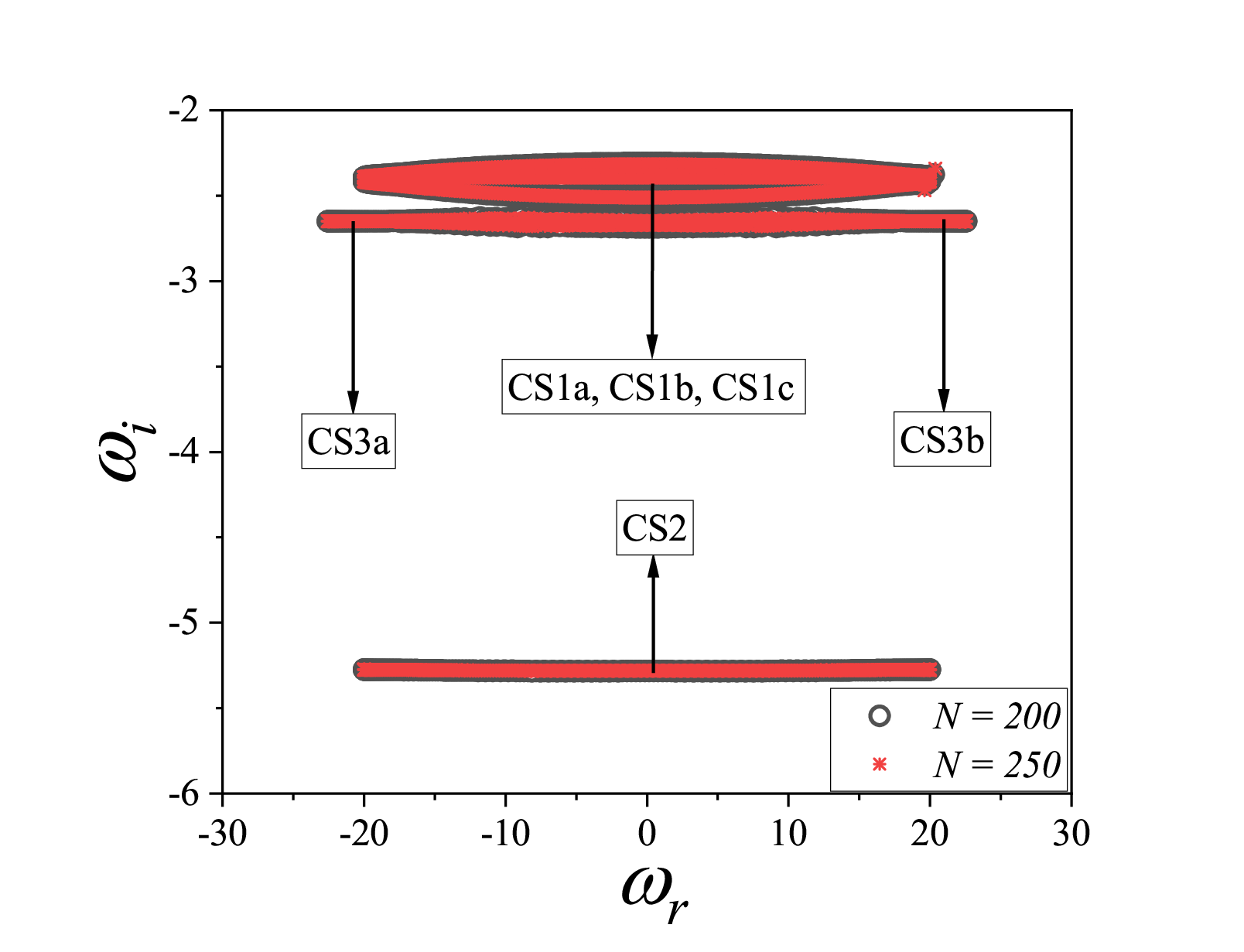}\label{fig:k0.1_PC_num}}
  \caption{Effect of wavenumber $k$ on the analytical CS and numerical spectra for plane Couette flow. Data for $\Wi = 200, L= 100, \beta = 0.98, Re = 0, l = 0$.}
  \label{fig:PC_effect_of_k}
\end{figure}

\begin{figure}
\centering
    \subfigure[$\Wi/L = 0.002$]
{\includegraphics[width=0.4\textwidth]{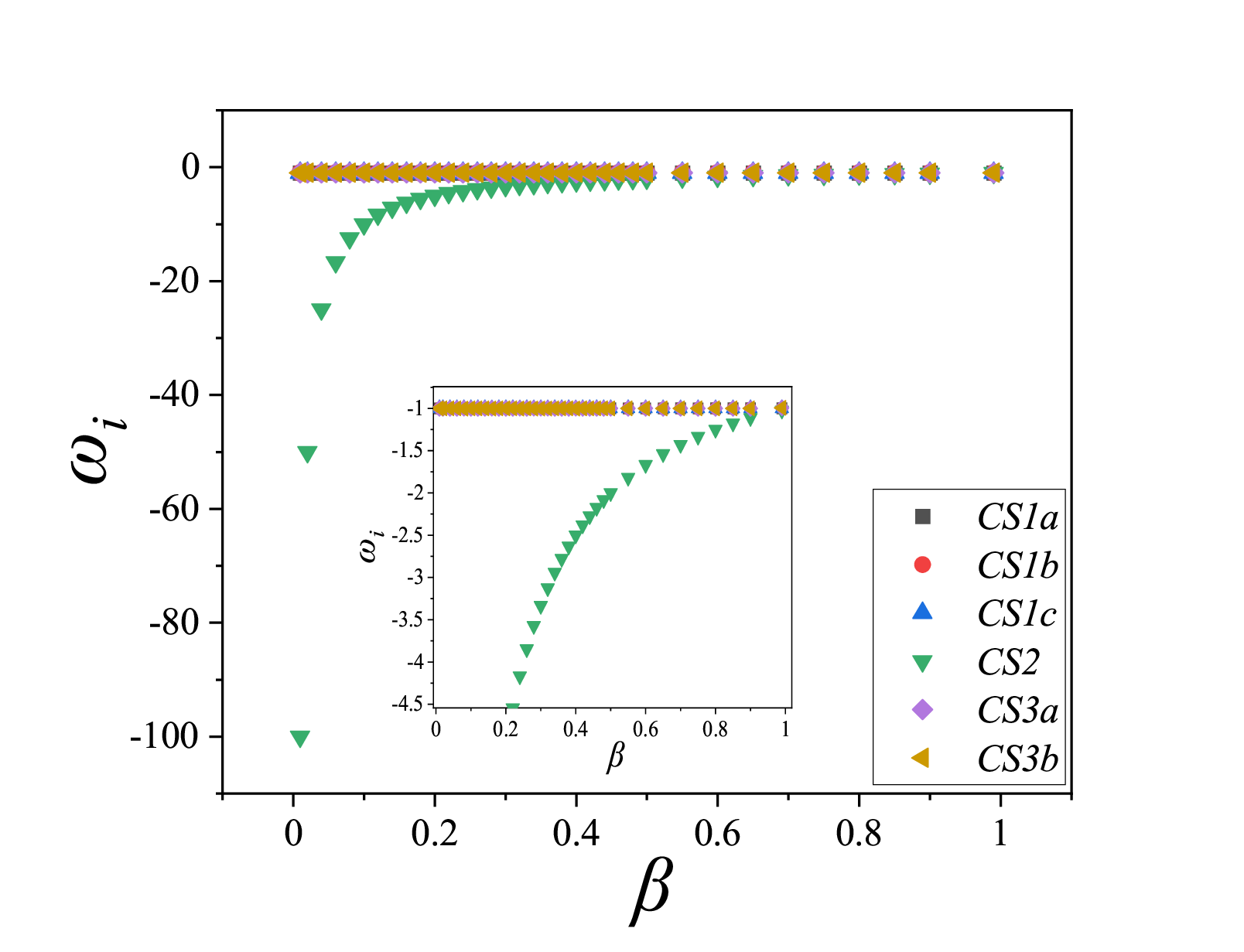}\label{fig:CS_track_beta_Wi_by_L_0.002}}
\subfigure[$\Wi/L = 0.02$]
{\includegraphics[width=0.4\textwidth]{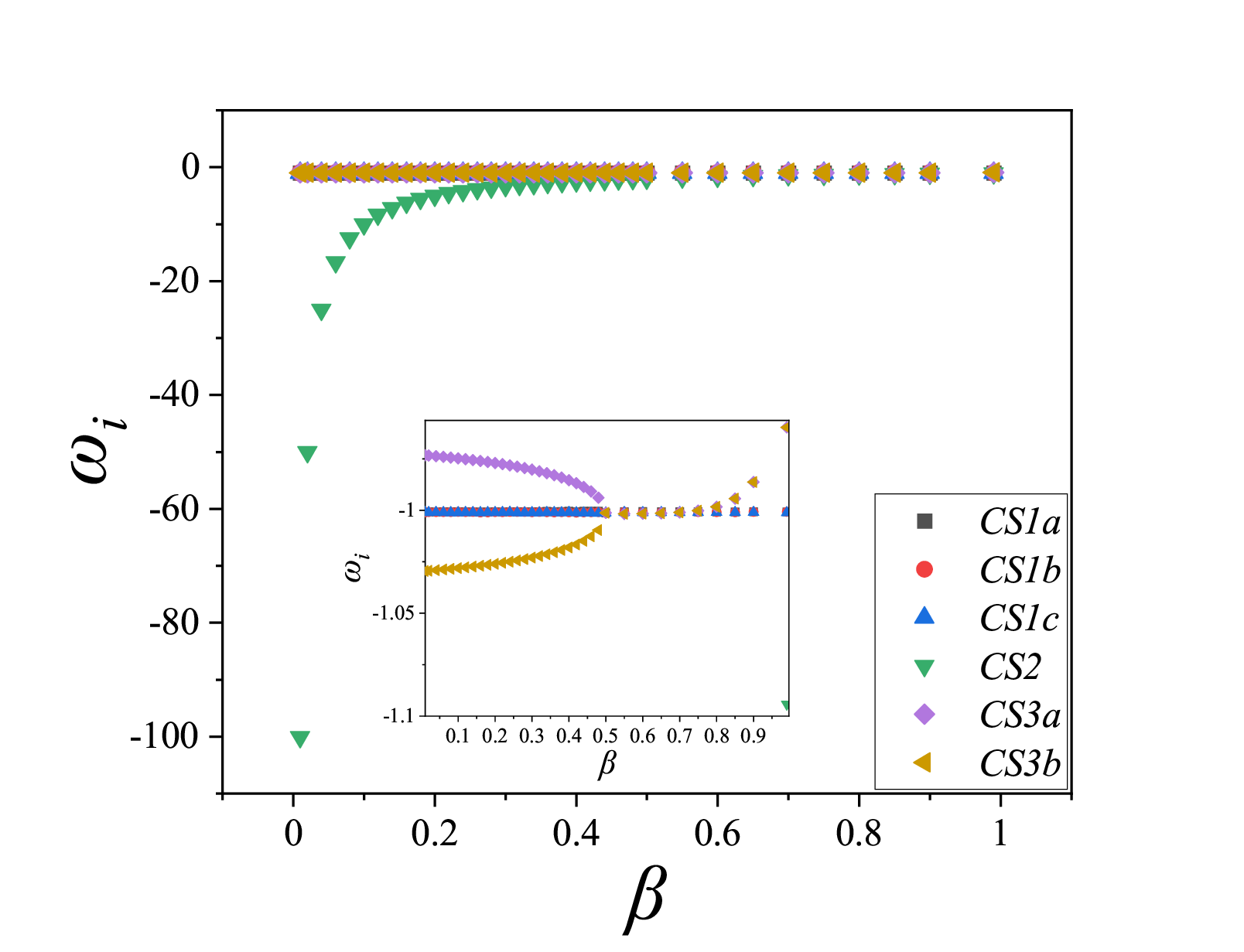}\label{fig:CS_track_beta_Wi_by_L_0.02}}
\subfigure[$\Wi/L = 0.2$]
{\includegraphics[width=0.4\textwidth]{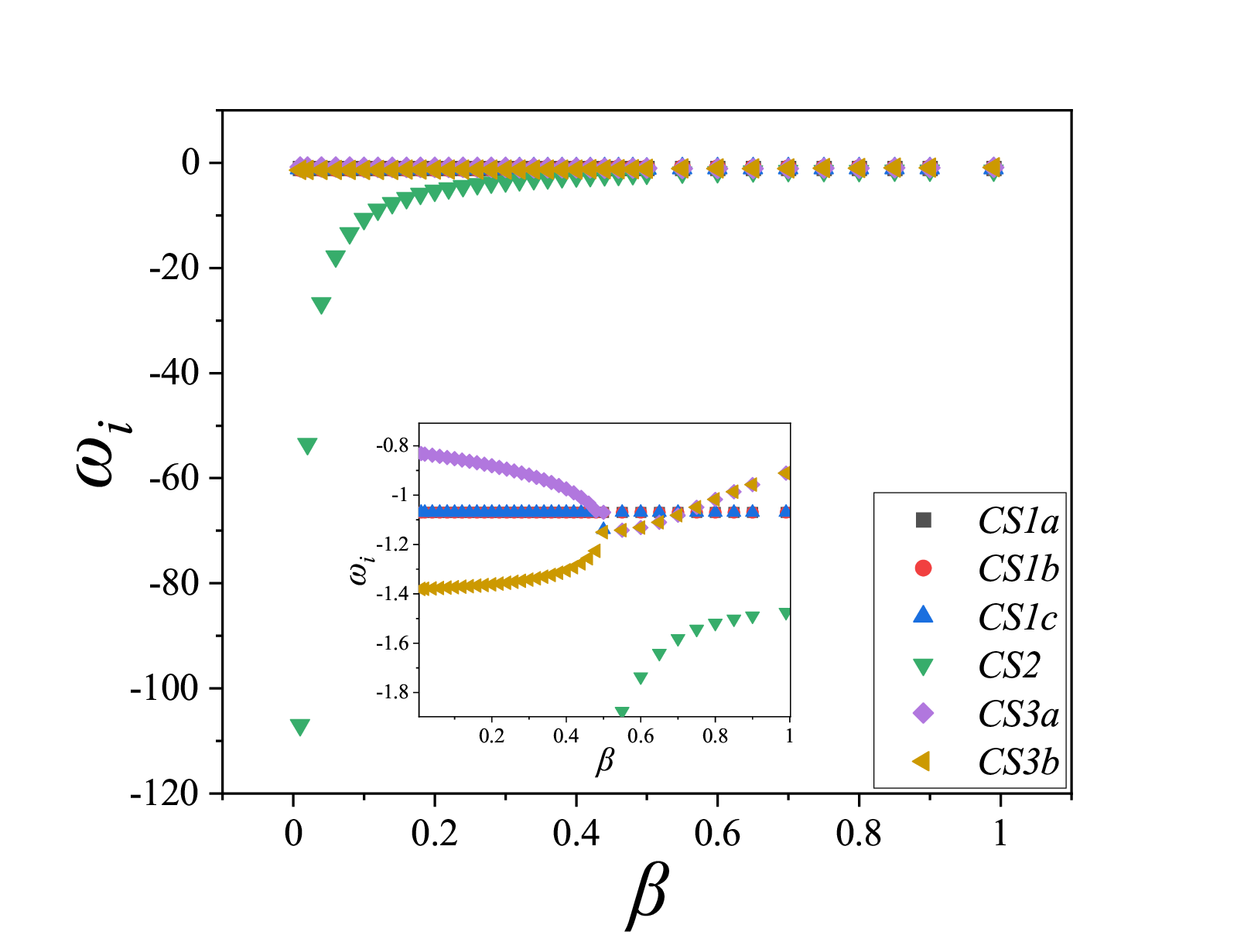}\label{fig:CS_track_beta_Wi_by_L_0.2}}
\subfigure[$\Wi/L = 2$]
{\includegraphics[width=0.4\textwidth]{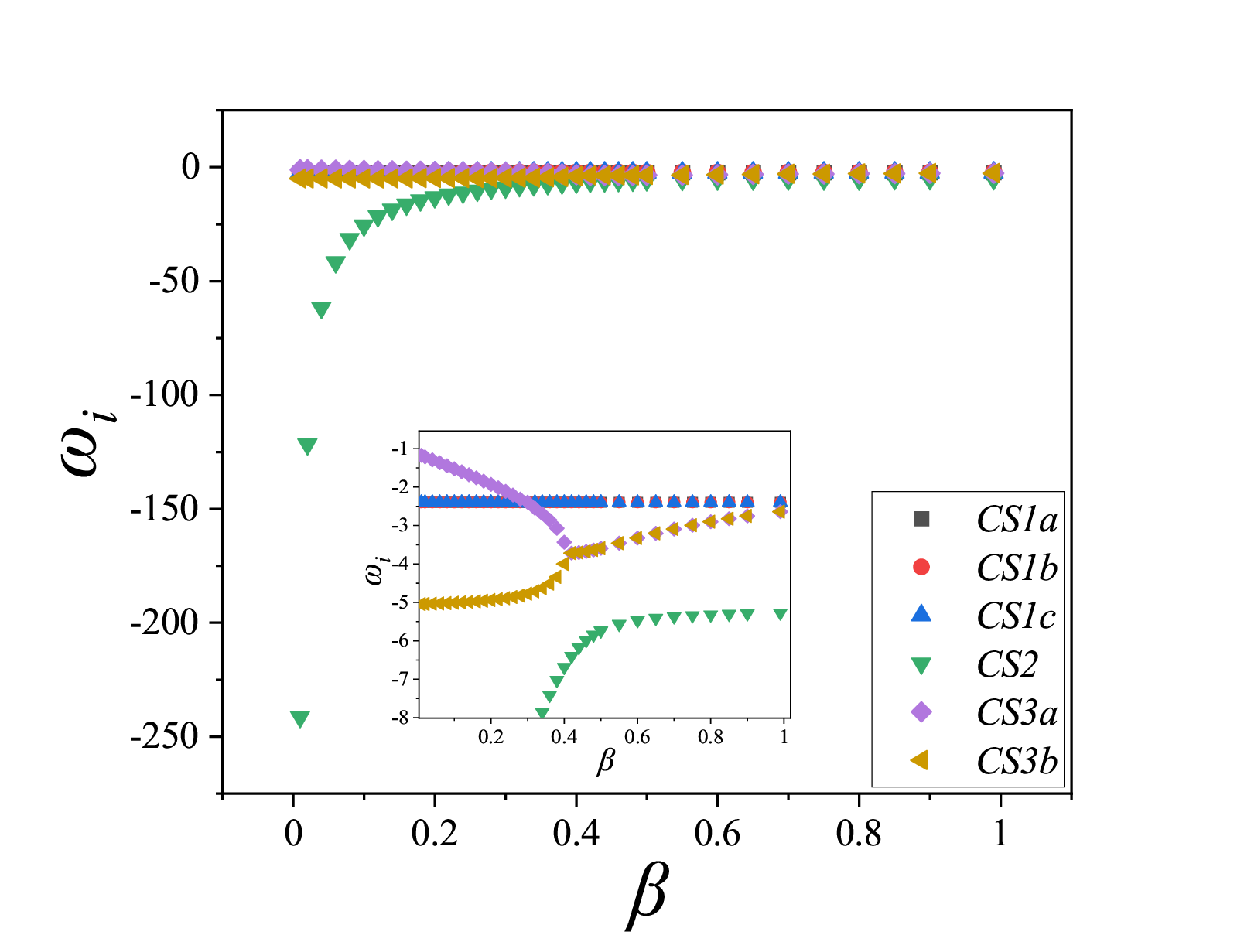}\label{fig:CS_track_beta_Wi_by_L_2}}
\subfigure[$\Wi/L = 20$]
{\includegraphics[width=0.4\textwidth]{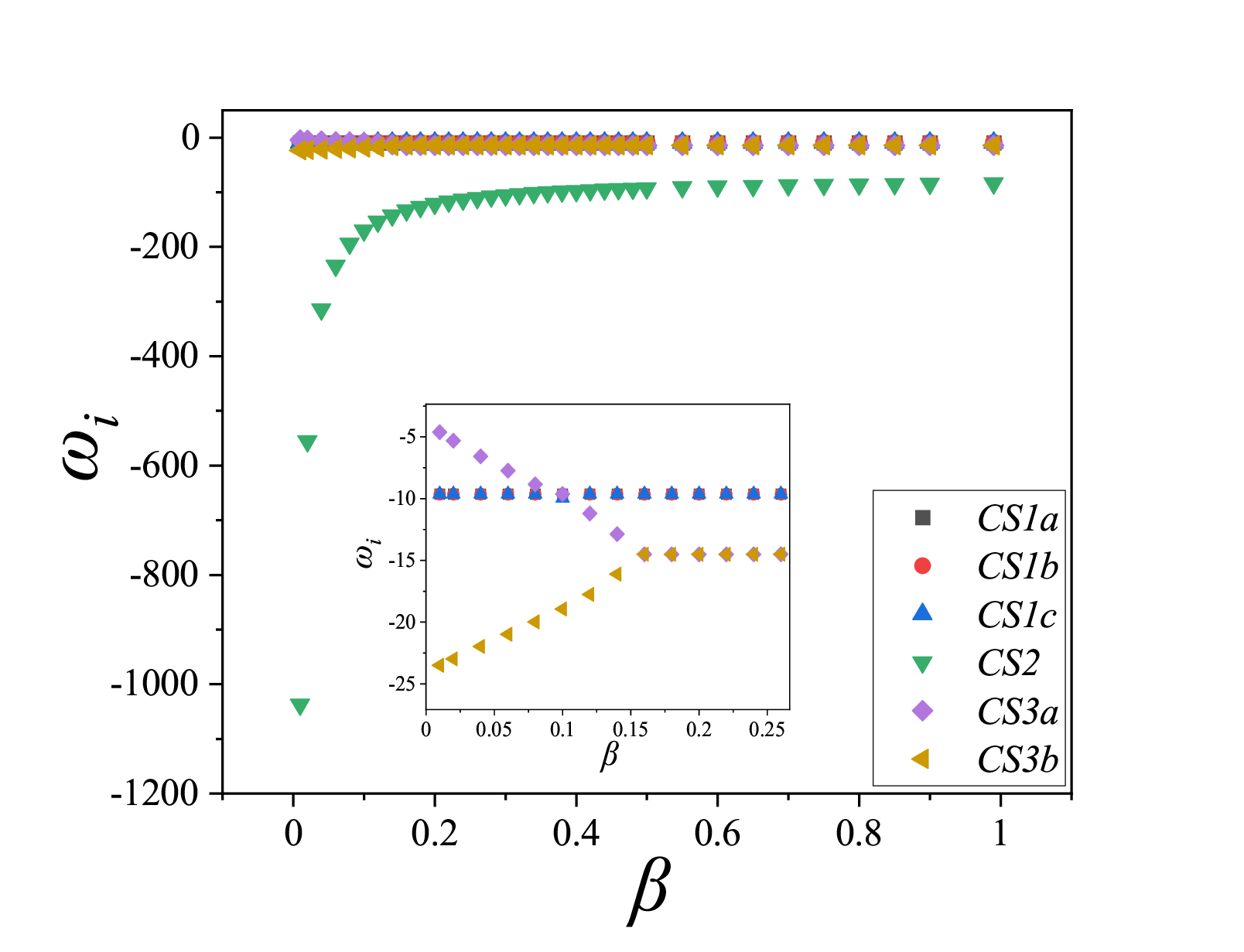}\label{fig:CS_track_beta_Wi_by_L_20}}
\subfigure[$\Wi/L = 200$]
{\includegraphics[width=0.4\textwidth]{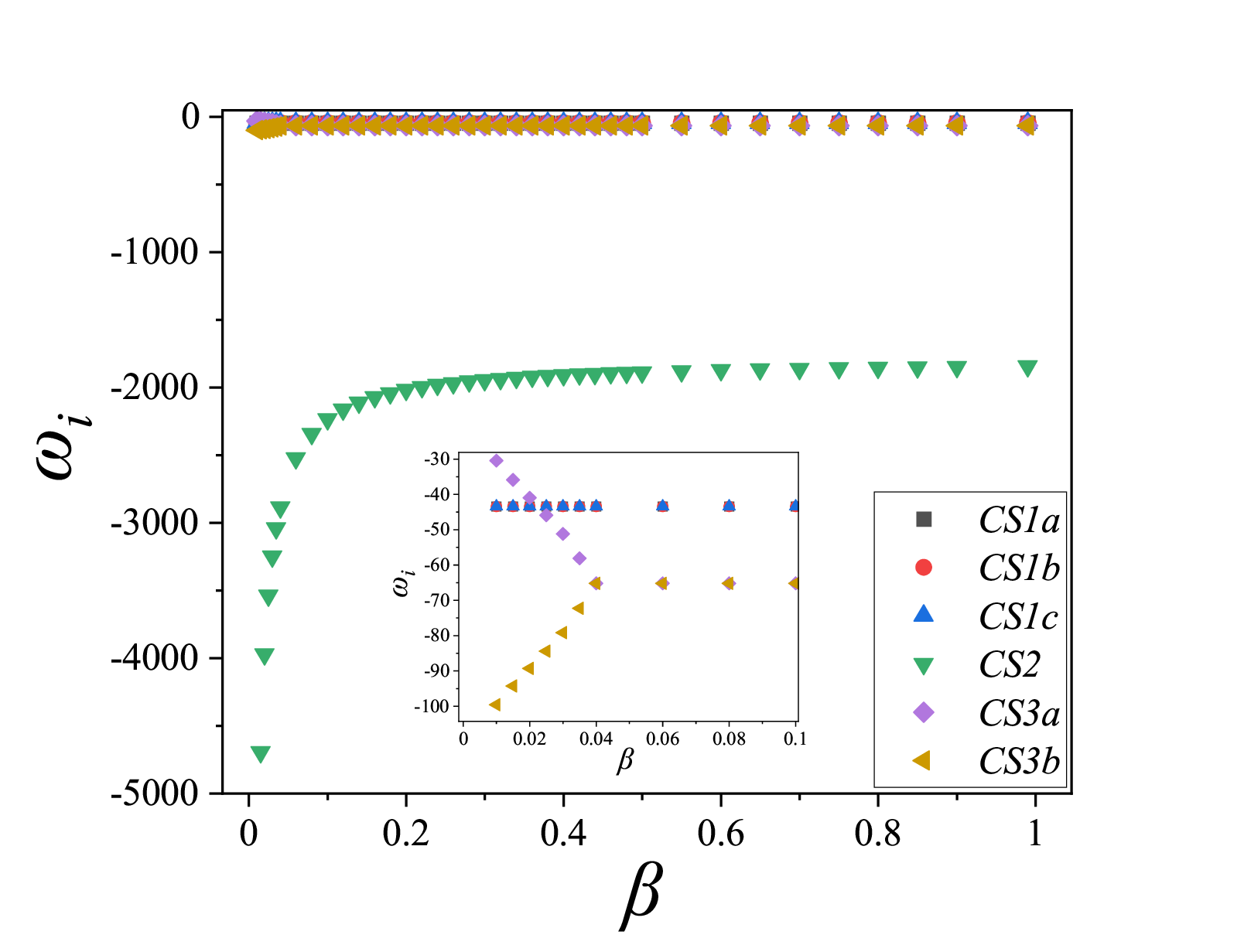}\label{fig:CS_track_beta_Wi_by_L_200}}
  \caption{Analytical predictions for the variation of the imaginary part ($\omega_i$) of the  CS with $\beta$ for plane Couette flow. Data for $k = 0$ at different $\Wi/L$ ratios.}
  \label{fig:Tracking_of_CS_beta}
\end{figure}
\begin{figure}
\centering
\subfigure[$\Wi/L = 2$]
{\includegraphics[width=0.35\textwidth]{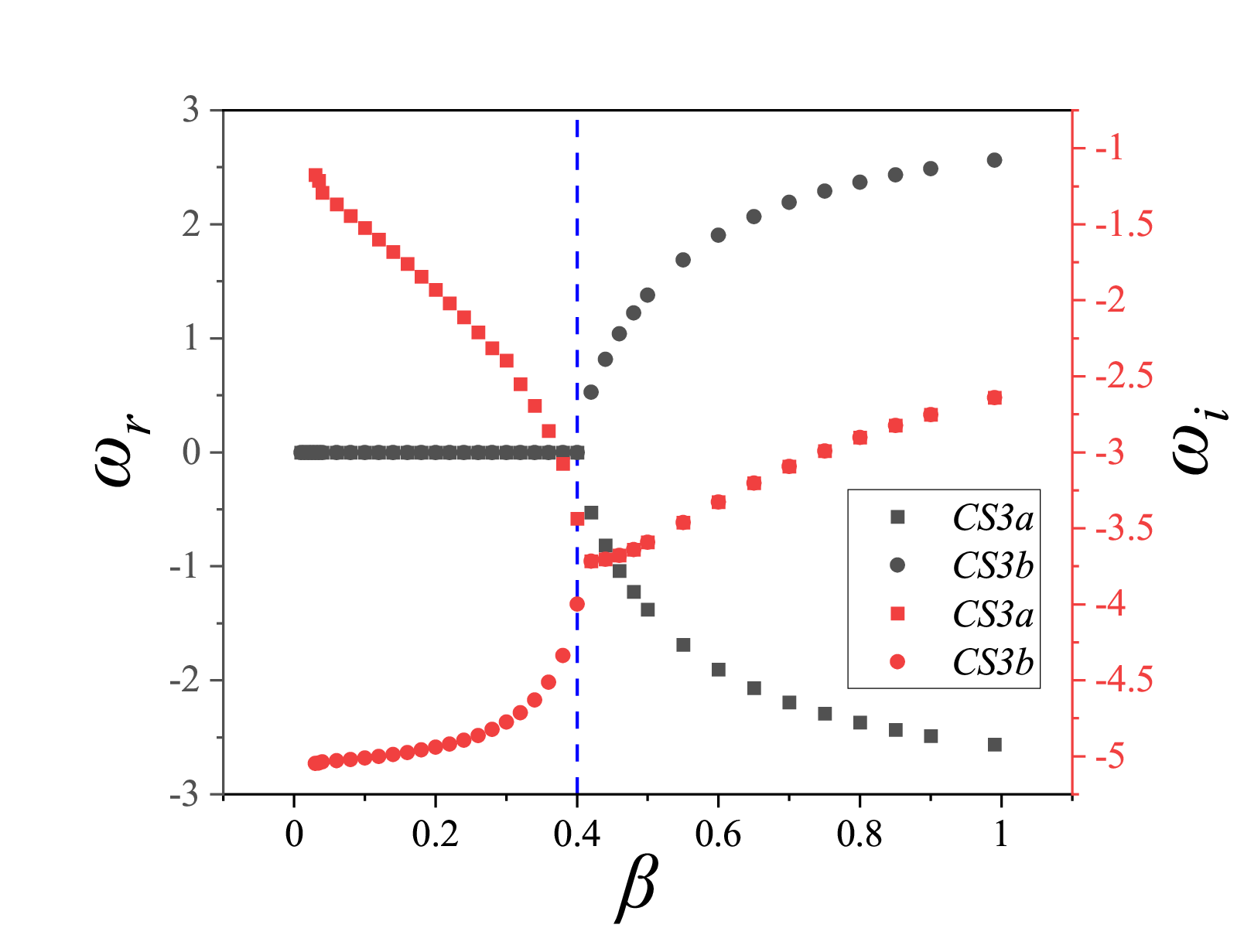}\label{fig:CS4_track_beta_Wi_by_L_2}}
\subfigure[$\Wi/L = 20$]
{\includegraphics[width=0.35\textwidth]{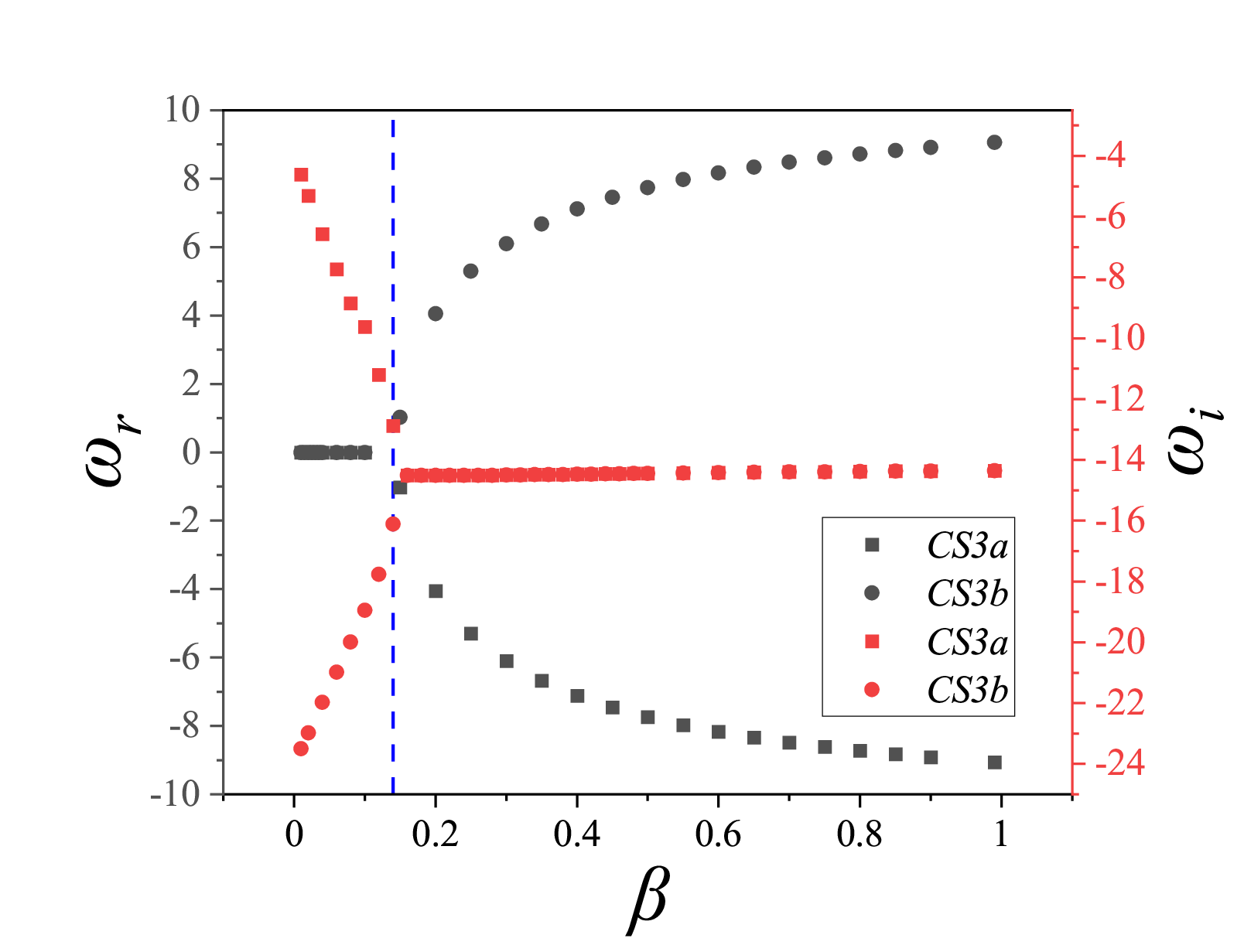}\label{fig:CS4_track_beta_Wi_by_L_20}}
   \subfigure[Varying $\beta$]
{\includegraphics[width=0.35\textwidth]{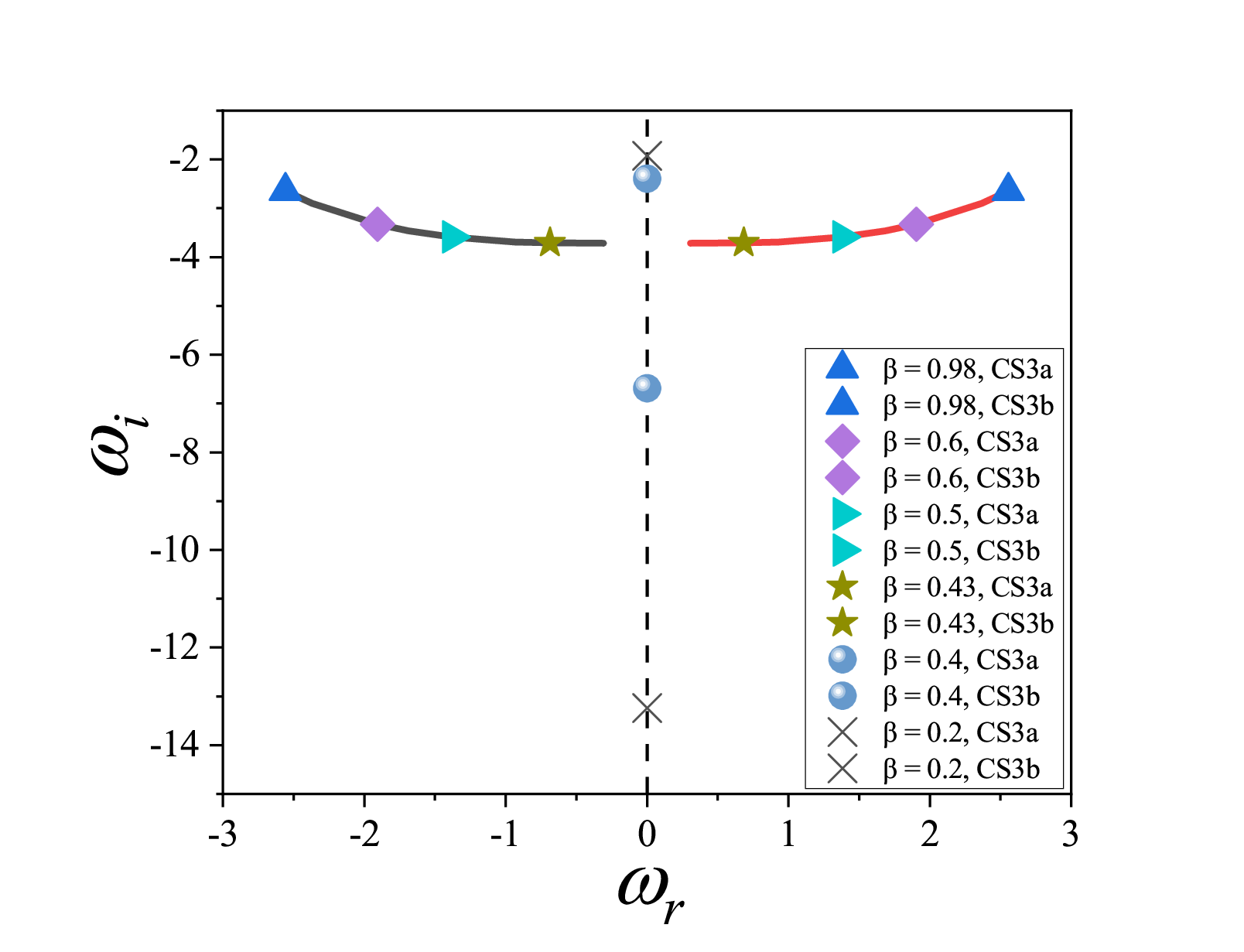}\label{fig:PC_effect_of_beta_CS4_k0}}
\subfigure[Varying $\Wi/L$]
{\includegraphics[width=0.35\textwidth]{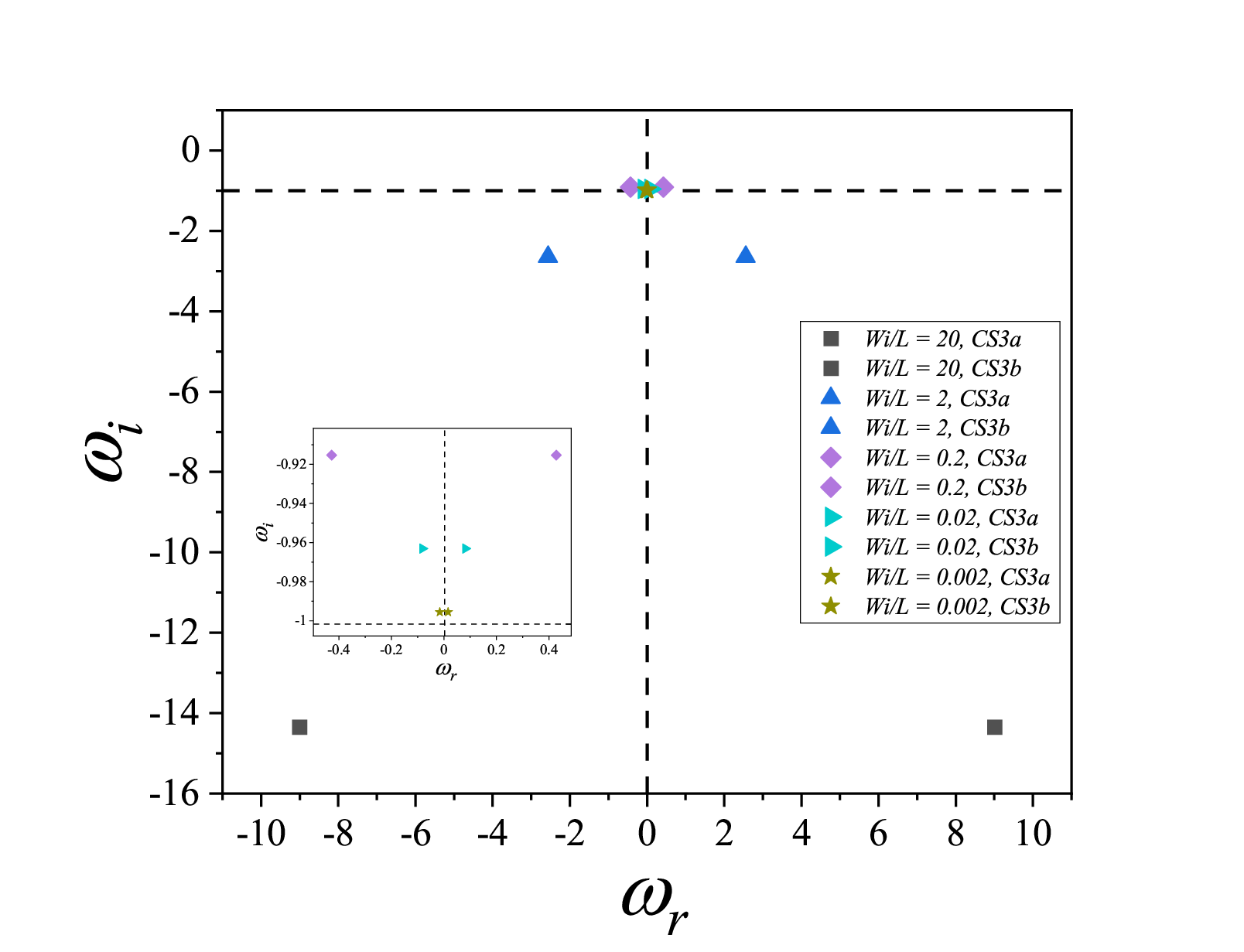}\label{fig:PC_effect_of_Wi_by_L_CS4_k0}}
  \caption{Bifurcation of CS3a,b with varying $\beta$: panels (a) and (b) show the analytical prediction of the bifurcation as $\beta$ is varied for plane Couette flow. As $\beta$ is decreased below a threshold, the modes become purely imaginary with $\omega_r = 0$, but with different $\omega_i$'s, as illustrated in Fig.\,\ref{fig:Tracking_of_CS_beta}. In panels (c) and (d), we examine, respectively, the 
  effects of $\beta$ and $\Wi/L$ on the analytically obtained CS3a, CS3b locations in the complex plane. Data for plane Couette flow for $\Wi = 200, Re = 0, k = 0, l = 0$, (c) $L = 100$; (d) $\beta = 0.98$.}
  \label{fig:Tracking_of_CS4_beta}
\end{figure}
\begin{figure}
\centering

\subfigure[$\beta = 0.1$]
{\includegraphics[width=0.35\textwidth]{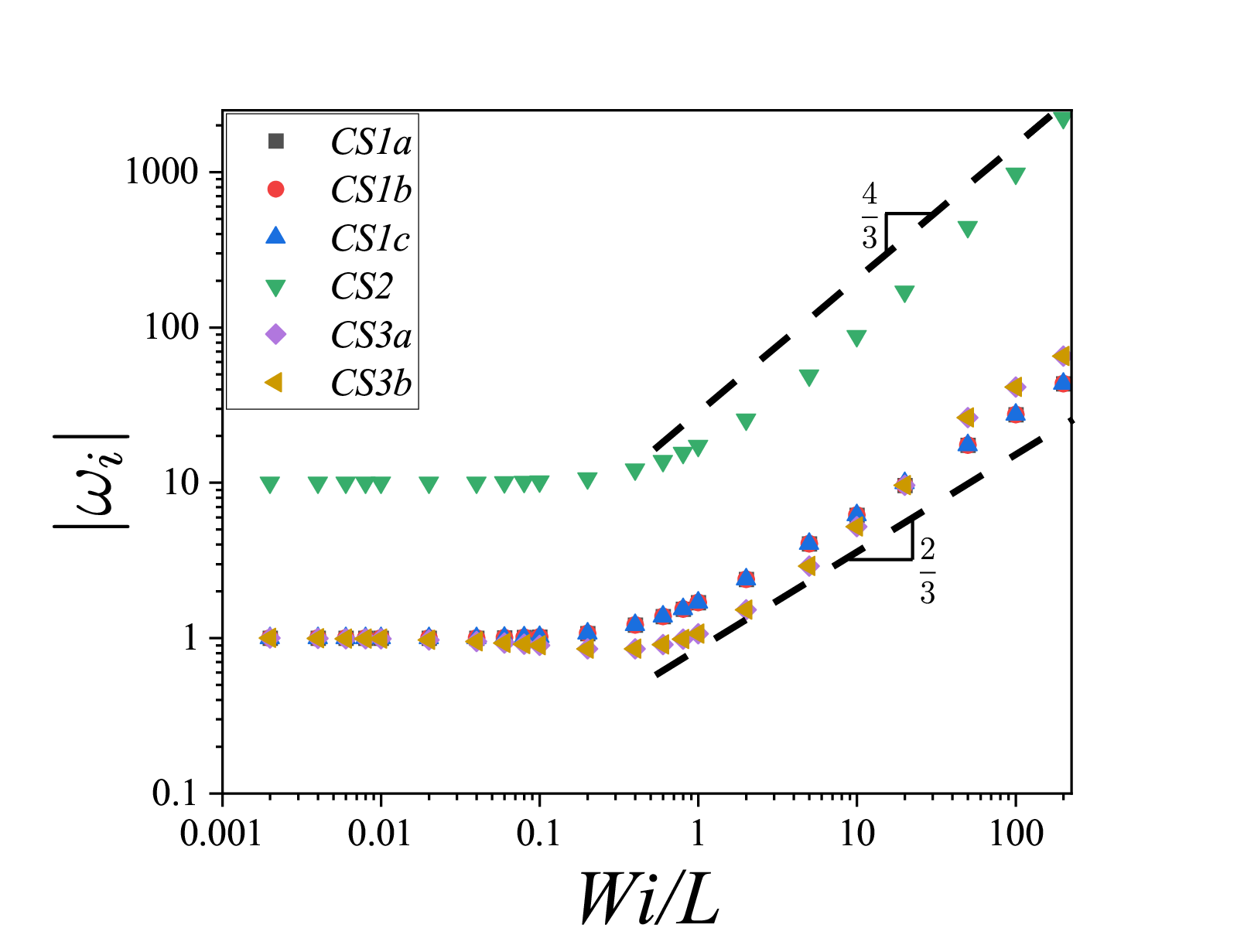}\label{fig:CS_track_Wi_by_L_beta_0.1}}
\subfigure[$\beta = 0.5$]
{\includegraphics[width=0.35\textwidth]{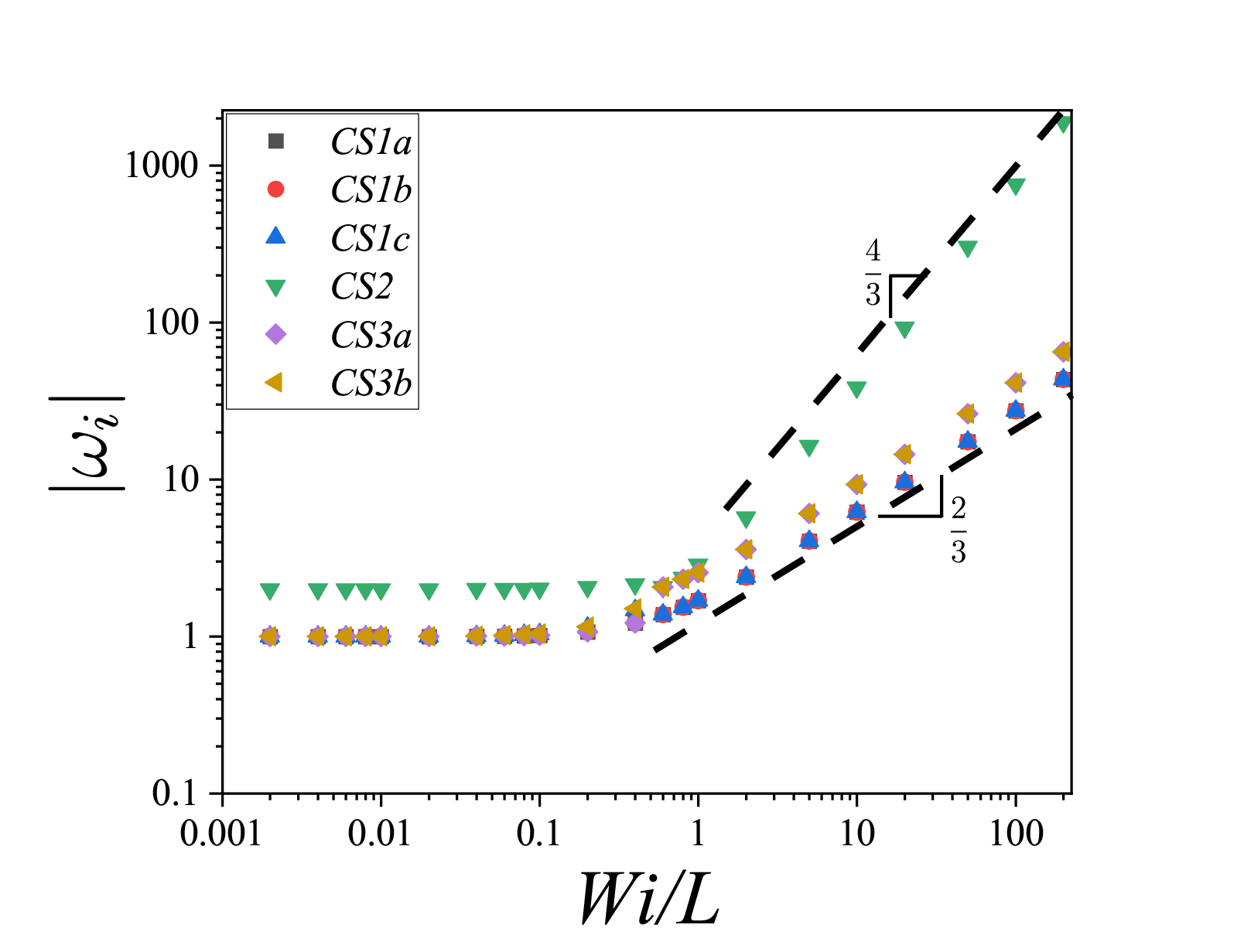}\label{fig:CS_track_Wi_by_L_beta_0.5}}
\subfigure[$\beta = 0.99$]
{\includegraphics[width=0.35\textwidth]{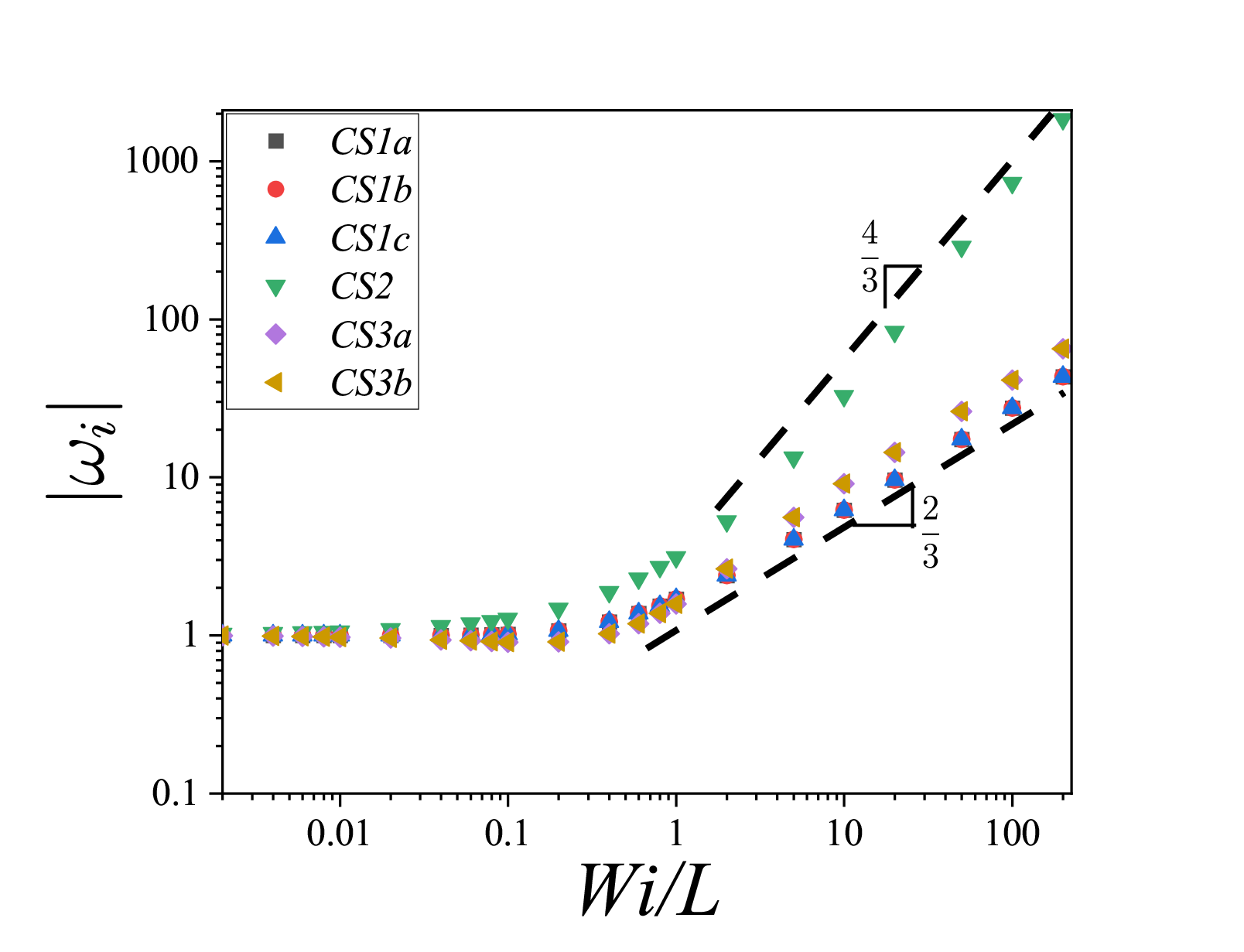}\label{fig:CS_track_Wi_by_L_beta_0.99}}
  \caption{
  Variation of the different CS with $\Wi/L$ for plane Couette flow. For $\Wi/L \ll 1$, the CS approach their respective Oldroyd-B values (viz., $\omega_i = -1$ and $-1/\beta$, for the elastic and solvent CS, respectively). For $\Wi/L \gg 1$, the magnitude of $\omega_i$ for all the CS except CS2 scale as $(\Wi/L)^{2/3}$, while that for CS2 scales as $(\Wi/L)^{4/3}$.}
  \label{fig:Tracking_of_CS_bWi_by_L}
\end{figure}

In order to better understand the structure of the CS, it is useful to examine base-state $\overline{f}$ in the limits $\Wi/L \ll 1$ and $\Wi/L \gg 1$. As shown by Yamani and McKinley \cite{Yamani_McKinley2023}, for $\Wi/L \ll 1$, shear thinning effects are negligible in the FENE-P fluid, and one recovers the Oldroyd-B limit, where $\bar{f} = 1$. Thus, for $\Wi/L \ll 1$, CS1a-c and CS3a,b become degenerate and approach the Oldroyd-B limit with $\omega_i =  -1$. 
The different CS become distinct with increasing $\Wi/L$. For $\Wi/L \gg 1$, $\overline{f} \propto (\Wi/L)^{2/3}$; in this limit, if $L \gtrsim 50$,  for CS1(a--c) $\omega_i \propto (\Wi/L)^{2/3}$ irrespective of $\beta$. The solvent CS (i.e. CS2) and CS3a,b, however, are functions of $\beta$. For $\beta \rightarrow 1$, CS2 asymptotes to  $\omega_i = -1$ for $\Wi/L \ll 1$, and $\omega_i \propto (\Wi/L)^{4/3}$ for $\Wi/L \gg 1$.  For $\beta \rightarrow 0$, the CS2 diverges as $-1/\beta$ for all $\Wi/L$. For $\beta \rightarrow 1$, CS3a and 3b also approach $\omega_i = -1$ for $\Wi/L \ll 1$, while $|\omega_i| \propto (\Wi/L)^{2/3}$ for $\Wi/L \gg 1$. The behavior of CS3a and 3b is rather complex as $\beta$ is decreased from unity, as discussed below.
These remarks are valid for both plane Couette and plane Poiseuille flows. Next, we discuss the nature of theoretical CS and their comparison with the numerical approximations for both plane Couette and pressure-driven channel flows.

\begin{figure}
\centering
    \subfigure[$\Wi/L = 2$]
{\includegraphics[width=0.35\textwidth]{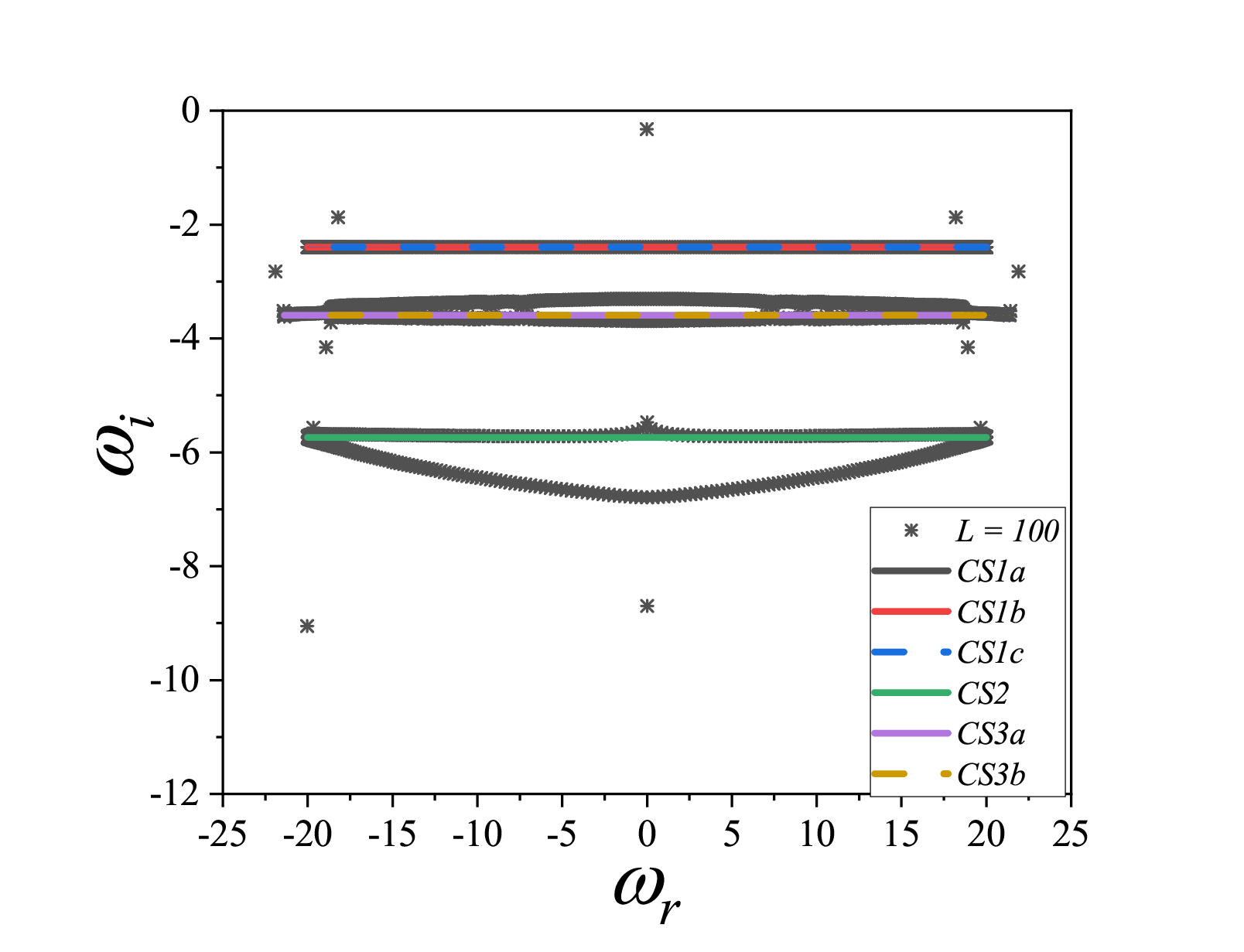}\label{fig:L_100_PC}}
\subfigure[$\Wi/L = 0.2$]
{\includegraphics[width=0.35\textwidth]{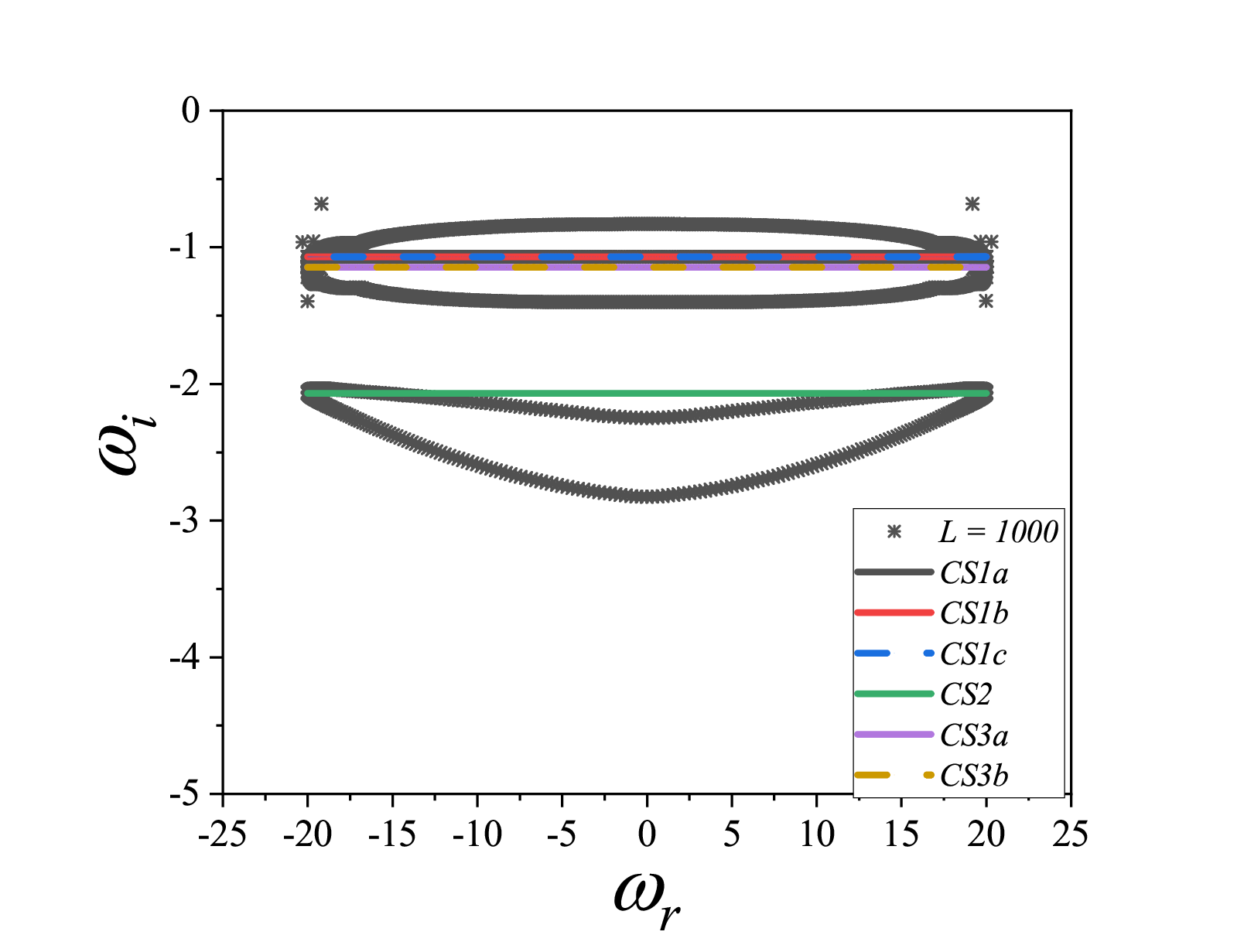}\label{fig:L_1000_PC}}
\subfigure[$\Wi/L = 0.02$]
{\includegraphics[width=0.35\textwidth]{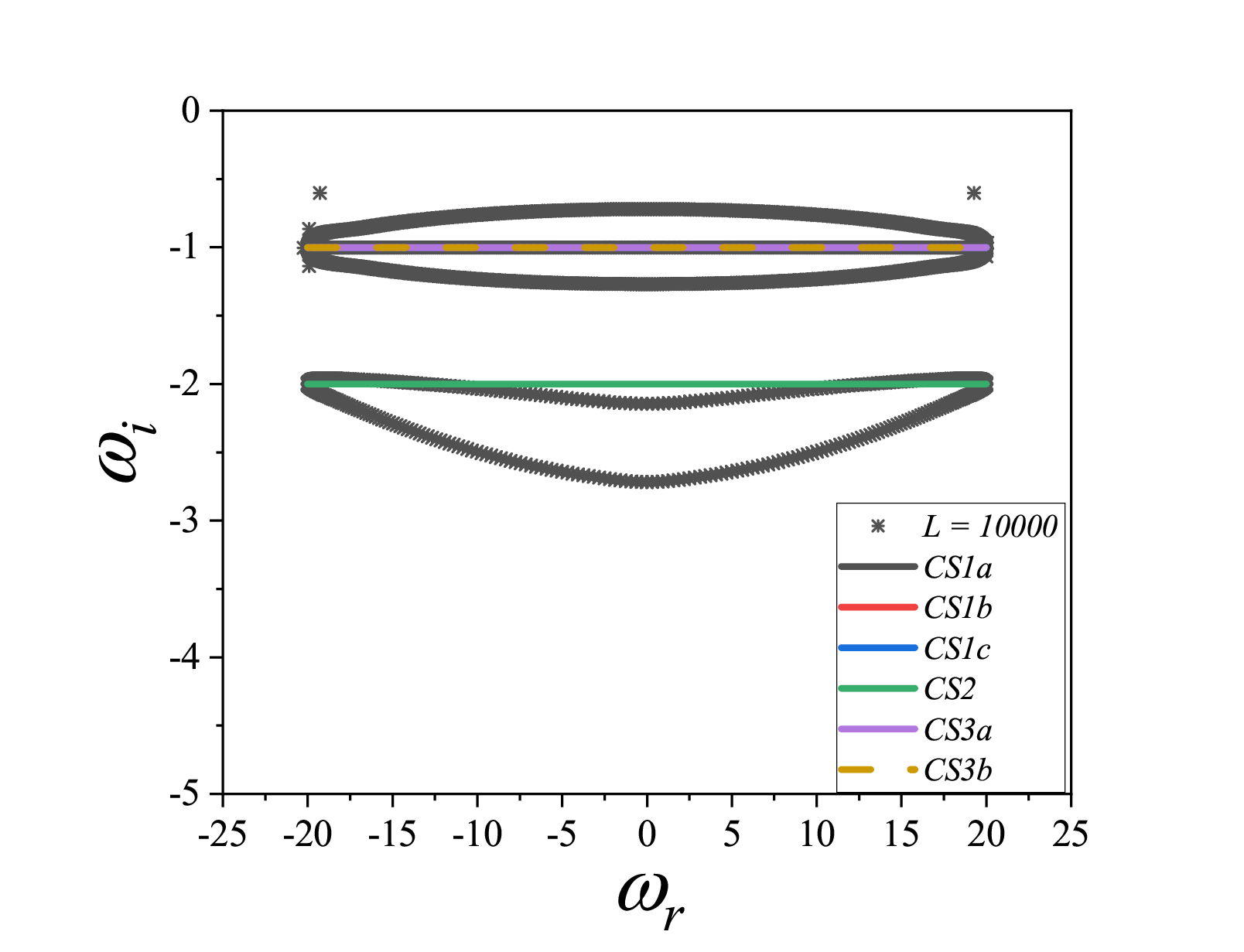}\label{fig:L_10000_PC}}
 \subfigure[Oldroyd-B]
  {\includegraphics[width=0.35\linewidth]{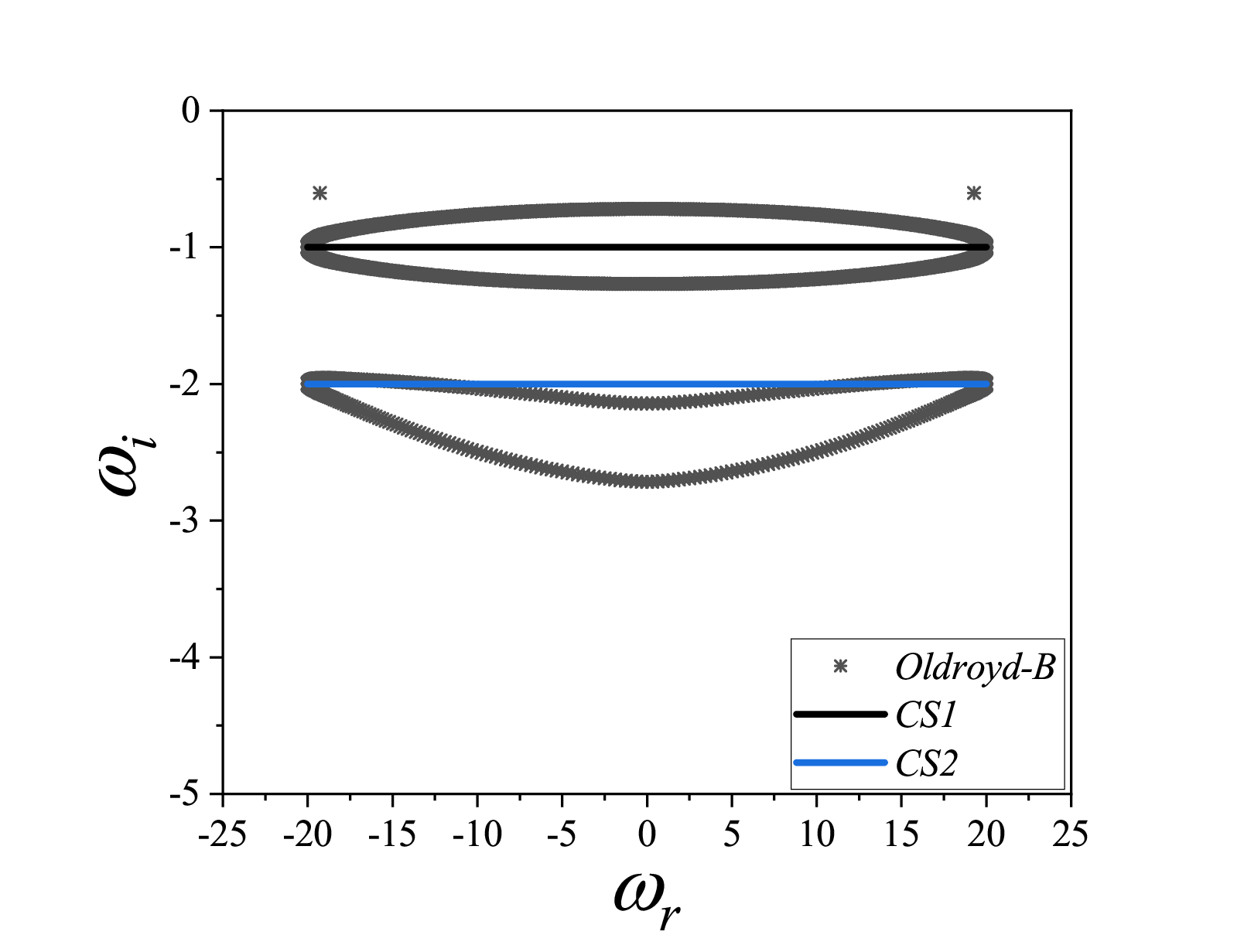}\label{fig:OldroydB_PC}}
  \caption{Numerical and analytical CS for plane Couette flow. Data for $\Wi = 200, \beta = 0.5,k = 0.1, Re = 0$. Lines represent analytical predictions and symbols denote numerical spectra obtained using $N = 400$. (Dashed lines are used to enhance the visibility of overlapping CS's)}
  \label{fig:PC_effect_of_L}
\end{figure}

\subsection{Plane Couette flow}
\label{sec:rectilinear_PC}

We first present locations of the various CS schematically in Fig.\,\ref{fig:PC_CS_schematics}; these schematics are valid  above a threshold $\beta$, and for $\Wi/L \gtrsim O(1)$, when the CS structure differs qualitatively from that of an Oldroyd-B fluid.
Figure\,\ref{fig:PC_K_0_schematics}, for $k = 0$, shows that  CS1(a--c) and CS2 are points on the imaginary axis, whereas CS3a and 3b are points symmetrically placed on either side of this axis. 
The locations of CS1(a--c) are independent of $\beta$ as mentioned above, while the $\beta$-dependence of the other three (CS2, 3a, and 3b) is discussed below. Figure\,\ref{fig:CPC_K_nonzero_schematics},
for nonzero $k$, shows that all the CS become horizontal line segments with lengths proportional to $k \triangle V$, where $\triangle V$ represents the base range of velocities. 
The presence of CS3(a,b) modes with negative phase speeds, in the inertialess regime, is somewhat counterintuitive, since such modes are usually associated with upstream propagating shear waves; the latter require inertia, as has been demonstrated in the context of CS in the elastic Rayleigh limit  \cite{Roy_Garg_Reddy_Subramanian_2022}. Even for small but finite $Re$, Gorodtsov and Leonov \cite{Gorodtsov1967} showed that there is a class of discrete modes which are (damped) upstream and downstream propagating shear waves (termed `high-frequency Gordtsov-Leonov modes' \cite{Sameer_Shankar2005,chaudhary_etal_2019}), again highlighting the role of inertia in enabling upstream propagation.
Before proceeding to compare the theoretically predicted CS with their numerical approximations, it is useful to examine the numerical spectra at different levels of discretizations (i.e., $N$, the number of collocation points), in order to get an idea of how the convergence, as manifested by the decrease in
the vertical extent of the CS balloons, improves with increasing $N$ (see Fig.\,\ref{fig:PCF_diff_N}).  
The rate of convergence is different for different CS, with the numerical approximation to CS2 exhibiting the largest vertical spread, and CS1(a--c) the smallest, for a given $N$. Focusing on CS1, the vertical extent  is more for $\Wi/L = 0.02$ (approaching the Oldroyd-B limit) compared to $\Wi/L = 2$, again, for a given $N$.  
The differing vertical extents likely represent the differing nature of the singularity (at $z = z_c$) of the eigenfunctions for a particular CS. In the rest of this paper, continuous or dashed lines are used to represent the analytical CS, and symbols represent the numerical eigenspectrum.

Figure \ref{fig:PC_effect_of_k} shows the comparison of the analytically predicted CS with the numerical spectra at different $k$'s, for $\Wi/L = 2$. Figures \ref{fig:k0_PC_ana}, \ref{fig:k0.01_PC_ana}, and \ref{fig:k0.1_PC_ana} denote the analytical spectra, while
Figs.\,\ref{fig:k0_PC_num}, \ref{fig:k0.01_PC_num}, and \ref{fig:k0.1_PC_num} depict the corresponding numerical spectra, for two different $N$'s. The numerical spectra typically contain discrete eigenvalues, in addition to the (numerical approximations of the) CS; the number of such eigenvalues being $k$-dependent. The analytical results for the CS are in excellent agreement with their numerical approximations. For $\beta \rightarrow 1$ and for small but finite $k$, CS3a and 3b are distinct line segments on either side of the imaginary axis. However, as $k$ is increased, the two segments approach each other and eventually merge into one. Thus, for $k \sim O(1)$ and higher, the overlap of  CS3a and 3b is almost complete, rendering them virtually indistinguishable, with their horizontal extents being very similar to CS1(a--c).  The vertical locations of CS3a and 3b, however, differ from those of CS1(a--c) for $\Wi/L \sim O(1)$, even for $k \sim O(1)$.

We next examine how the loci of the various CS change as $\beta$ is varied. This is best illustrated for $k = 0$, since the CS are points in the complex plane. Figure\,\ref{fig:Tracking_of_CS_beta} shows the variation of $\omega_i$ with $\beta$ for a range of $\Wi/L$ ratios. For $\Wi/L = 0.002$, close to the Oldroyd-B limit  (Fig.\,\ref{fig:CS_track_beta_Wi_by_L_0.002}), all the CS but for CS2 remain close to $\omega_i = -1$, with CS2 alone characterized by $\omega_i = -1/\beta$. As $\Wi/L$ is increased to $0.2$, CS1(a-c) 
remain independent of $\beta$, while shifting downward to more negative $\omega_i$. In contrast, CS3a and 3b  vary with $\beta$, and remain distinct from CS1(a--c) even for $\beta \rightarrow 1$. For finite $\Wi/L$, these two CS start off, for sufficiently small $\beta$, with the same $\omega_r$ ($=0$), but with different $\omega_i$.  A bifurcation at a threshold $\beta$ (that decreases with increasing $\Wi/L$), causes them to have the same $\omega_i$ but with $\omega_r$ equal and of opposite signs. For purposes of clarity, the behavior of CS3a and b alone, as a function of $\beta$, is depicted in Fig.\,\ref{fig:CS4_track_beta_Wi_by_L_2} and \ref{fig:CS4_track_beta_Wi_by_L_20}. Figures\,\ref{fig:PC_effect_of_beta_CS4_k0} and \ref{fig:PC_effect_of_Wi_by_L_CS4_k0} depict the same bifurcation on the complex $\omega$-plane, with the loci of CS3a and 3b being traced as $\beta$ and $\Wi/L$ is varied (again, for $k = 0$). For $\beta > 0.4$, as already seen, CS3a and 3b appear as points, symmetrically located about the imaginary axis. As $\beta$ is decreased further, the two begin to move downwards and towards the imaginary axis in a symmetric fashion,  coalescing at a threshold $\beta$,   corresponding
to the aforementioned bifurcation, after which the two eigenvalues are purely imaginary, moving in opposite directions along the imaginary axis (note that the labeling of the two roots becomes arbitrary post-bifurcation).  Analogous bifurcations are seen in pressure-driven channel flow as well, as discussed below in Sec.\,\ref{sec:rectilinear_PP}.
Returning to Fig.\,\ref{fig:Tracking_of_CS_beta}, interestingly, for $\Wi/L \sim O(1)$ and higher, both CS2 and CS3(a,b) remain finite and distinct from the Oldroyd-B CS, even in the limit of $\beta \rightarrow 1$, despite being absent for $\beta = 1$; this feature is indicative of the singular nature of the $\beta \rightarrow 1$ limit.

Figure\,\ref{fig:Tracking_of_CS_bWi_by_L} shows the $|\omega_i|$-loci of the different CS modes, as a function of $\Wi/L$, for $\beta = 0.1$, $0.5$, and $0.99$.  For $\Wi/L \ll 1$, CS1(a--c) and CS3(a,b) approach $\omega_i = -1$ for all $\beta$'s, while CS2 approaches $\omega_i = -1/\beta$ in the same limit. For  $\Wi/L \gg 1$, CS2 exhibits the scaling $|\omega_i| \propto (\Wi/L)^{4/3}$, while all the remaining CS exhibit $|\omega_i| \propto (\Wi/L)^{2/3}$ at all $\beta$'s. 
Next, we show the comparison of the analytically computed CS with the full numerical spectra in Fig.\,\ref{fig:PC_effect_of_L} for a fixed $\Wi = 200$, and for different $L$'s, for $\beta = 0.5$ and $k = 0.1$, and with CS3a and b having overlapped almost completely for this range of parameters.  That these two CS are distinct is only barely noticeable in Fig.\,\ref{fig:L_100_PC}, where their total horizontal extent is only slightly larger than those of CS1 and 2, for $\Wi/L = 2$. 
As $\Wi/L$ is decreased (see Figs.\,\ref{fig:L_1000_PC}-\ref{fig:OldroydB_PC}), CS3a and 3b have completely overlapped, with their total horizontal extent now  being identical to those of CS1 and 2.  The vertical locations of the three CS (viz., CS1(a--c), CS2, and CS3(a,b)) are distinct
only for $\Wi/L = 2$ in Fig.\,\ref{fig:L_100_PC}. 
 Thus, for $\Wi/L > 1$, there are three distinct line segments corresponding to CS1, 2, and 3, but with nearly the same horizontal extents, and with CS3a,b lying on the right-half plane. 
The numerical approximation of CS2 seems to show the largest vertical spread, with CS1(a--c) the least. As $\Wi/L$ is decreased to $0.2$ in Fig.\,\ref{fig:L_1000_PC}, CS3 has already moved up and is very close to CS1, while CS2 remains distinct. Figures \ref{fig:L_10000_PC} and \ref{fig:OldroydB_PC} show the spectra, respectively, for $\Wi/L = 0.02$ and $0$ (the Oldroyd-B limit). The spectra for $\Wi/L = 0.02$ has already approached the Oldroyd-B limit with only two distinct CS, as CS3(a,b) has by now merged with CS1(a--c). Thus, for $\beta$'s below the bifurcation threshold, and for $k \sim O(1)$, the CS structure for FENE-P fluids is quite similar to that for the Oldroyd-B fluid for $\Wi/L < 1$.

We have further verified  (data not shown) that the theoretical CS corresponding to CS1(a-c) and CS2, for different $(\Wi,L)$ pairs but with the same $\Wi/L$ ratio, collapse onto the same line segments. This collapse does not hold, however, for CS3(a,b). That there must be a collapse can be anticipated from the analytical expression for CS1(a--c), $\omega_i = -\bar{f}$, where the base-state value of the Peterlin function $\bar{f}$ is a function of $\Wi$ and $L$ only via the ratio $\Wi/L$, as first shown by Yamani and McKinley \cite{Yamani_McKinley2023}. Tej \textit{et al.} \cite{Tej2024} subsequently showed that the  velocity profiles for both rectilinear and curvilinear unidirectional flows  are identical  for different $(\Wi,L)$ pairs with the same $\Wi/L$ ratio. The collapse of the CS demonstrated in this study is an extension of this idea.

\begin{figure}
\centering
    \subfigure[$k = 0$]
{\includegraphics[width=0.35\textwidth]{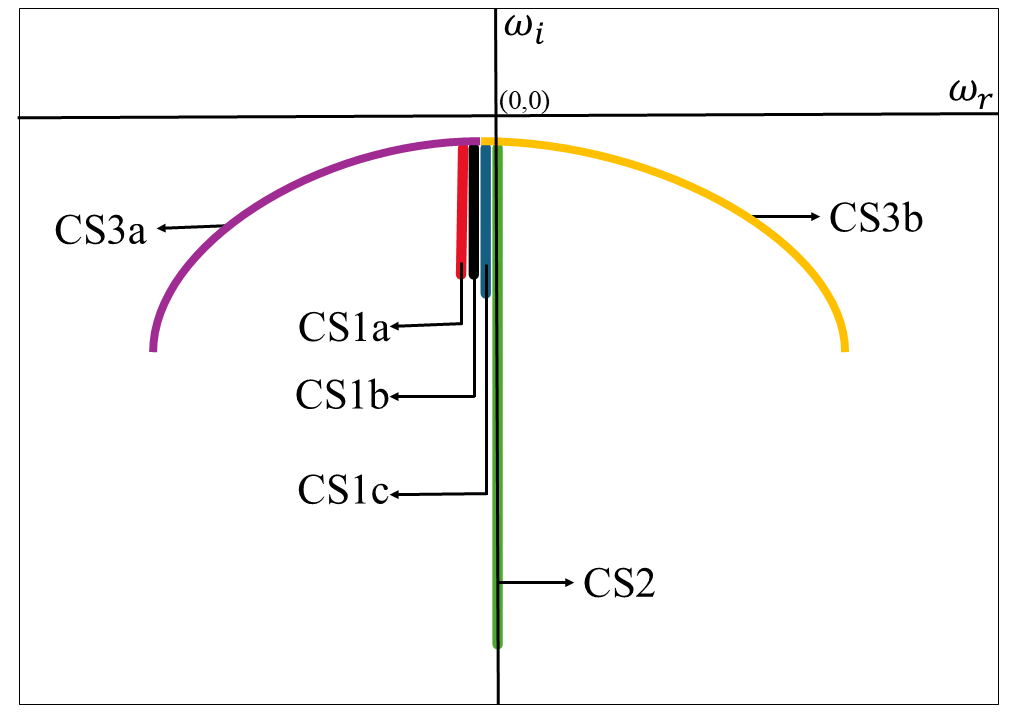}\label{fig:Dean_axis_n_0_schematic}}
\quad \quad
\subfigure[$k \neq 0$]
{\includegraphics[width=0.35\textwidth]{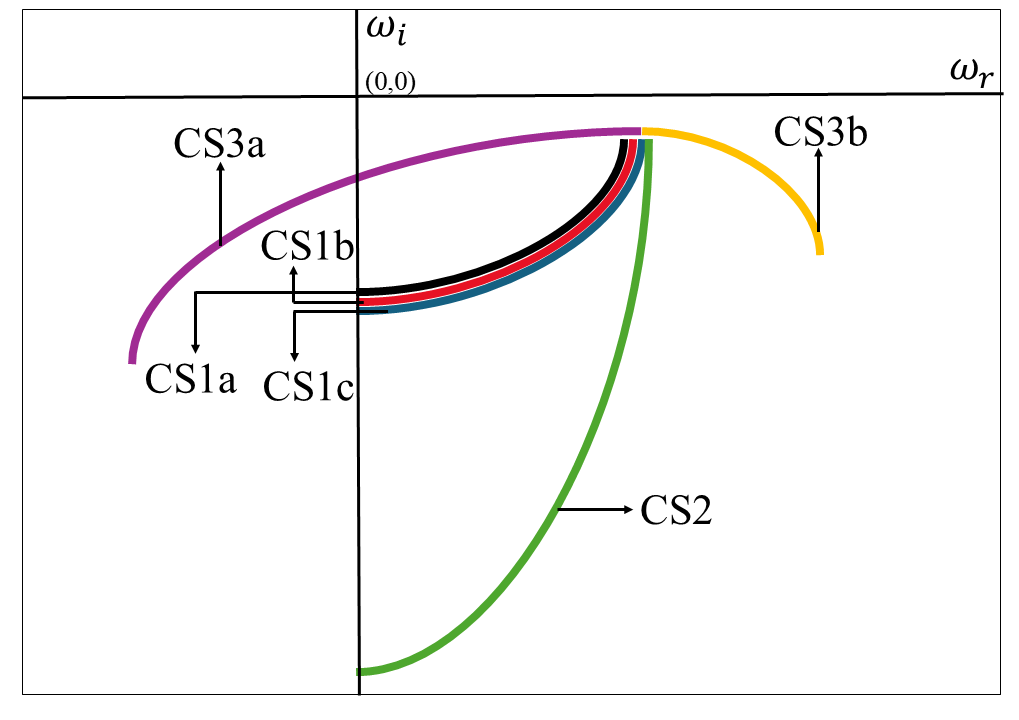}\label{fig:Dean_nonaxis_n_1_schematic}}
  \caption{Schematic location of the various CS for pressure-driven channel flow for $\beta \rightarrow 1$.  For $L \gtrsim 50$, CS1a, 1b, and 1c are nearly identical.}
  \label{fig:PPF_CS_schematics}
\end{figure}

 \subsection{Pressure-driven channel flow}
 \label{sec:rectilinear_PP}
 \begin{figure}
\centering
    \subfigure[$k = 0$ (Analytical)]
{\includegraphics[width=0.35\textwidth]{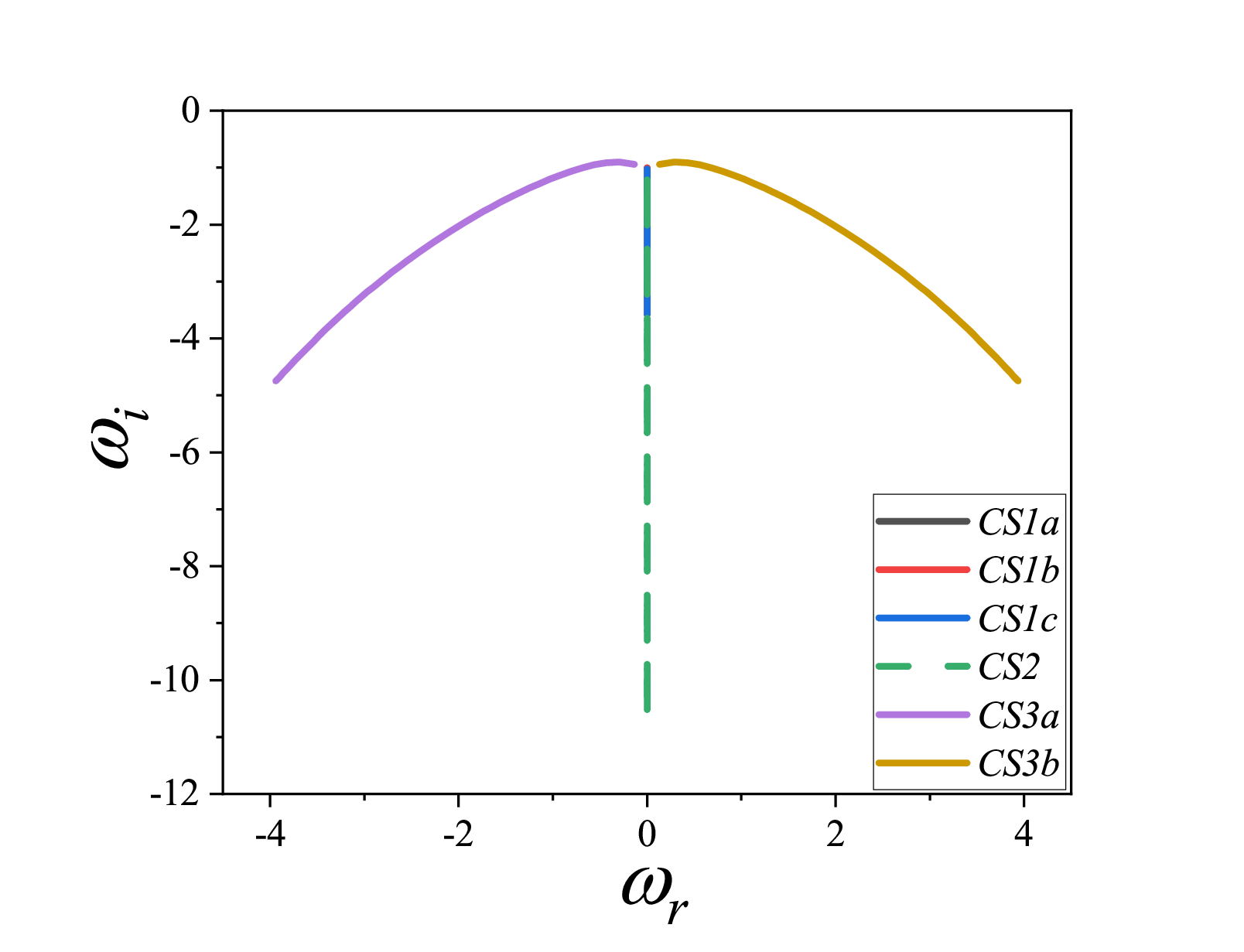}\label{fig:k0_PPF_ana}}
\subfigure[$k = 0$ (Numerical)]
{\includegraphics[width=0.35\textwidth]{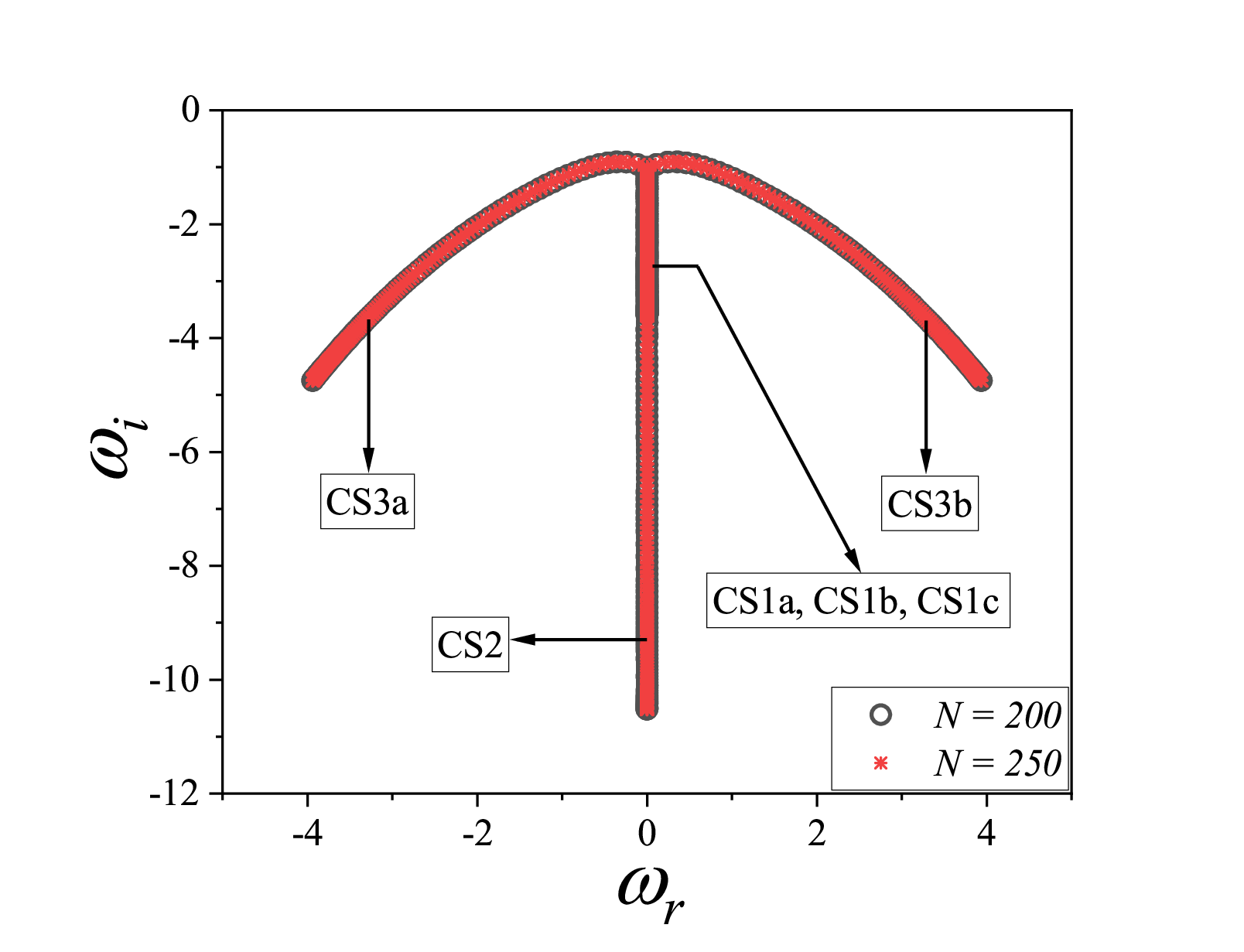}\label{fig:k0_PPF_num}}
\subfigure[$k = 0.01$ (Analytical)]
{\includegraphics[width=0.35\textwidth]{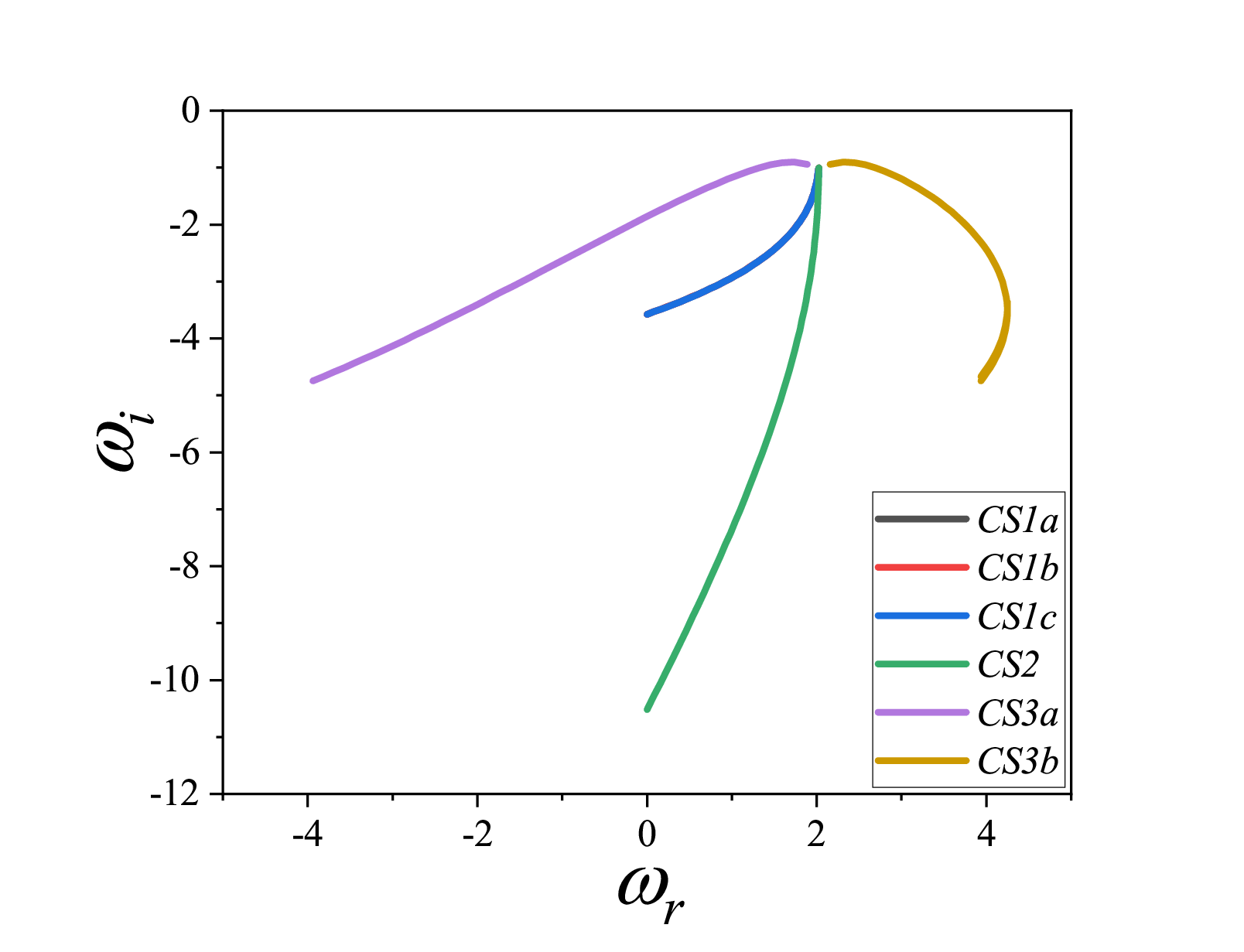}\label{fig:k0.01_PPF_ana}}
\subfigure[$k = 0.01 $ (Numerical)]
{\includegraphics[width=0.35\textwidth]{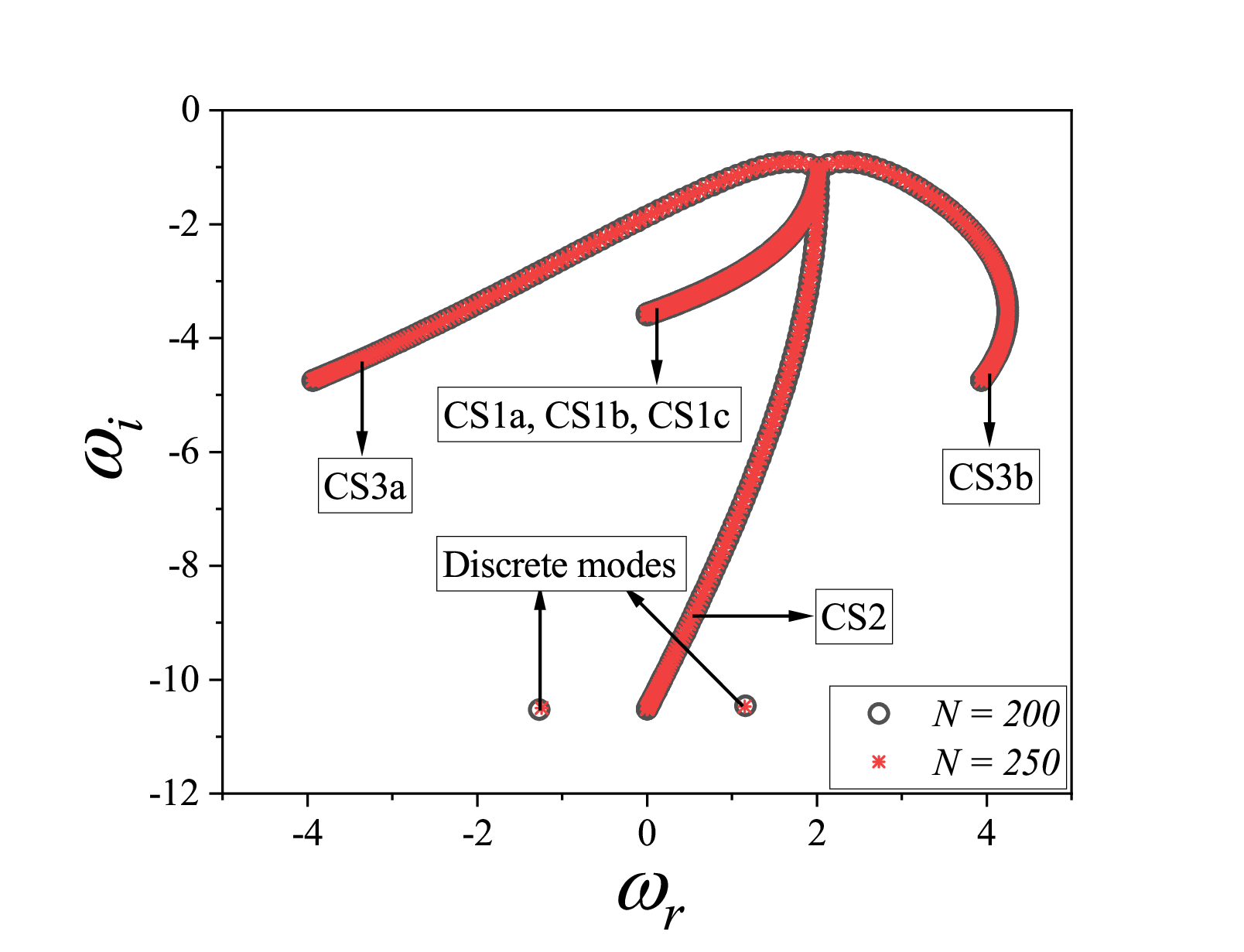}\label{fig:k0.01_PPF_num}}
\subfigure[$k = 0.1$ (Analytical)]
{\includegraphics[width=0.35\textwidth]{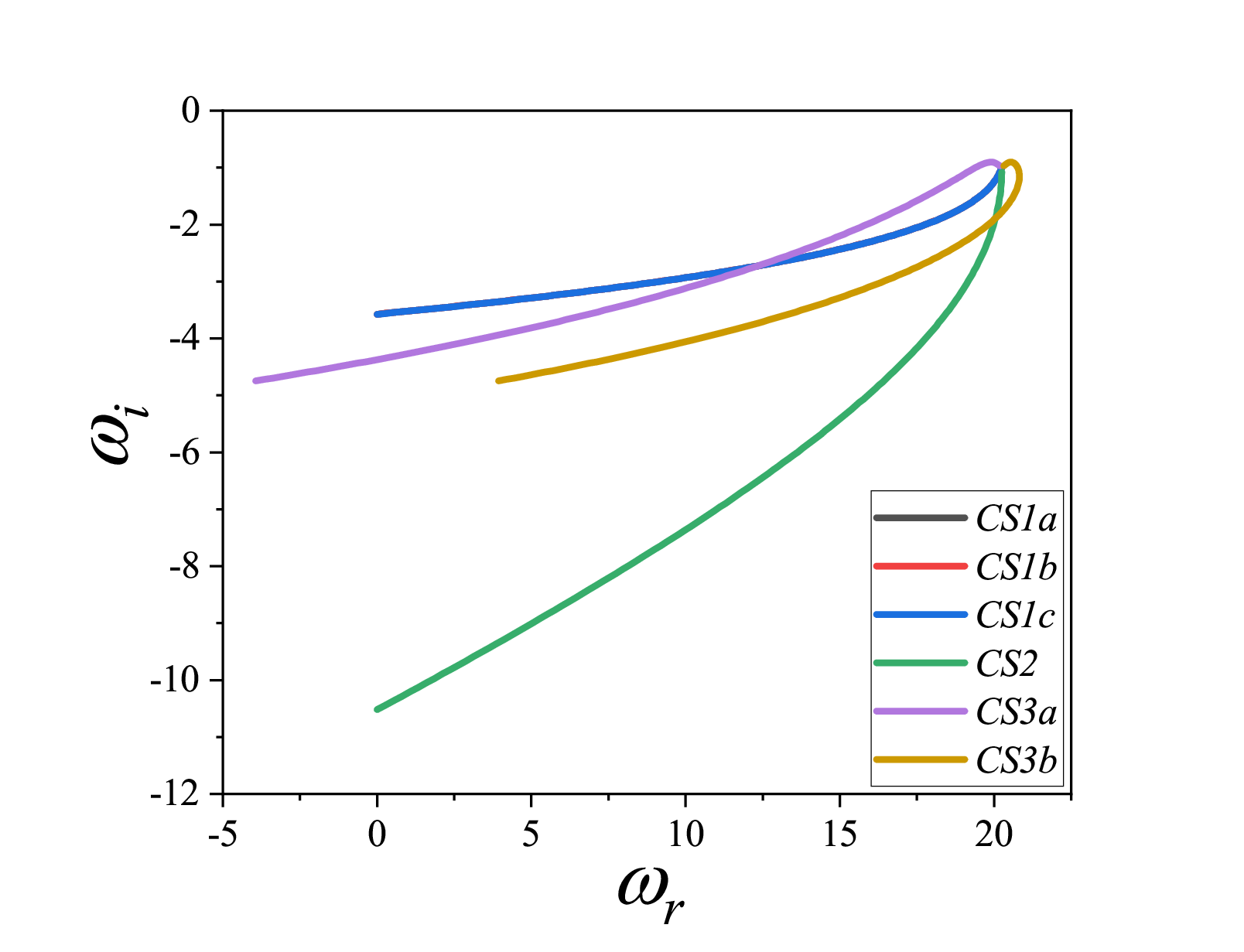}\label{fig:k0.1_PPF_ana}}
\subfigure[$k = 0.1 $ (Numerical)]
{\includegraphics[width=0.35\textwidth]{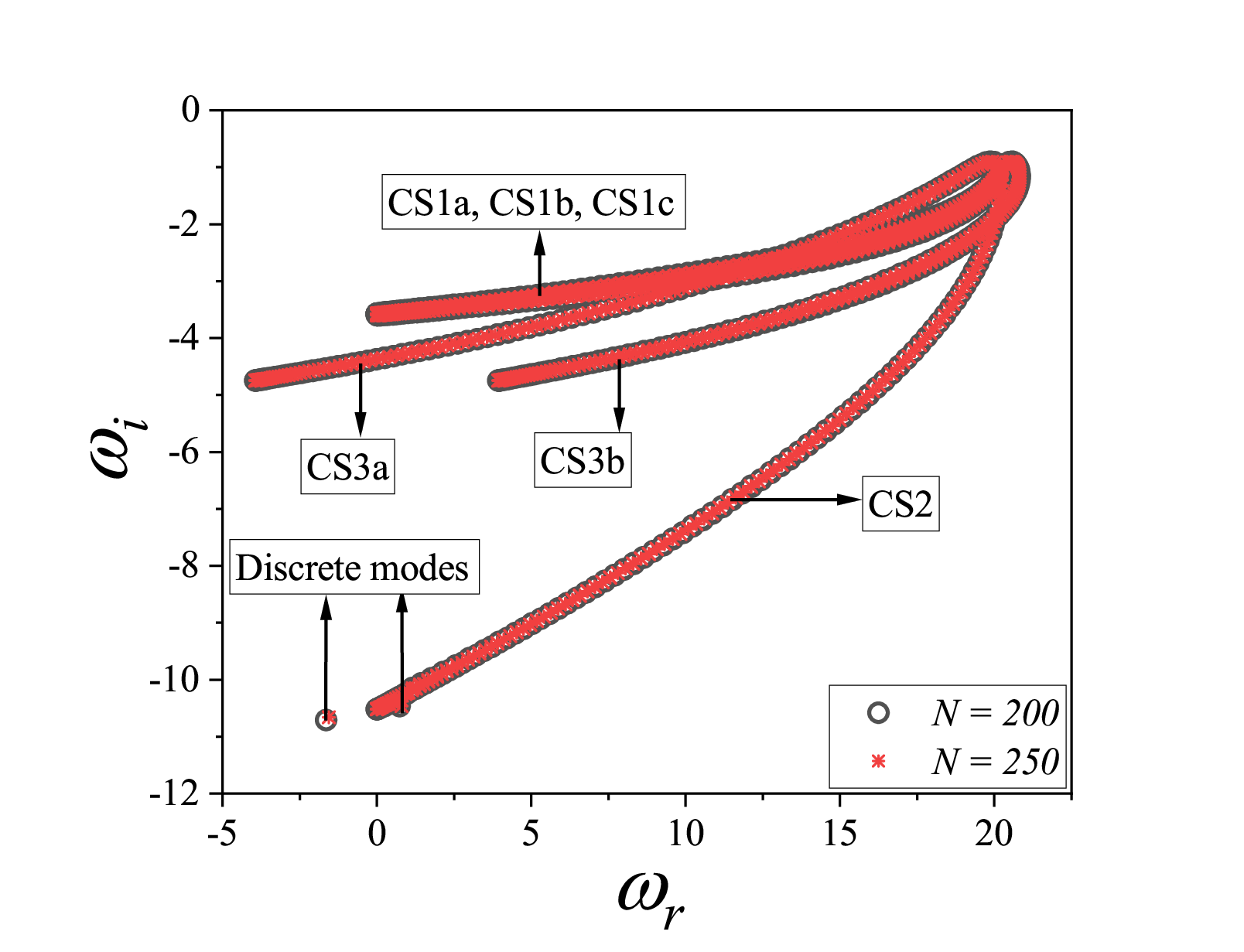}\label{fig:k0.1_PPF_num}}
  \caption{Analytical CS and numerical spectra for pressure-driven channel flow. Data for $\Wi = 200, L= 100, \beta = 0.98, Re = 0, l = 0$. (Dashed lines are used to enhance the visibility of overlapping CS's)}
  \label{fig:PPF_effect_of_k}
\end{figure}

\begin{figure}
  \centering
  \subfigure[$\Wi/L = 2$]
  {\includegraphics[width=0.35\linewidth]{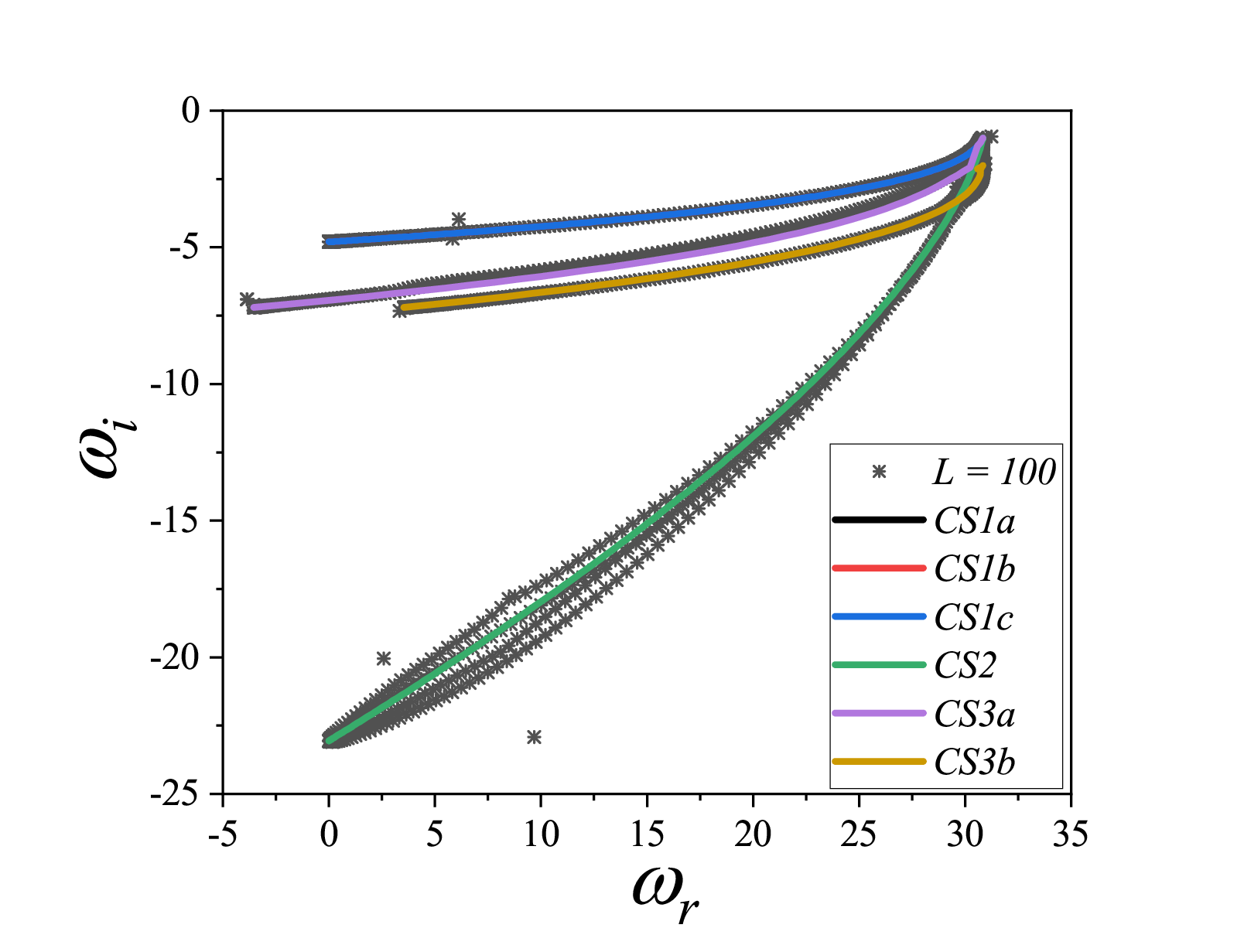}\label{fig:L_100_PP}}
    \subfigure[$\Wi/L = 0.2$]
  {\includegraphics[width=0.35\linewidth]{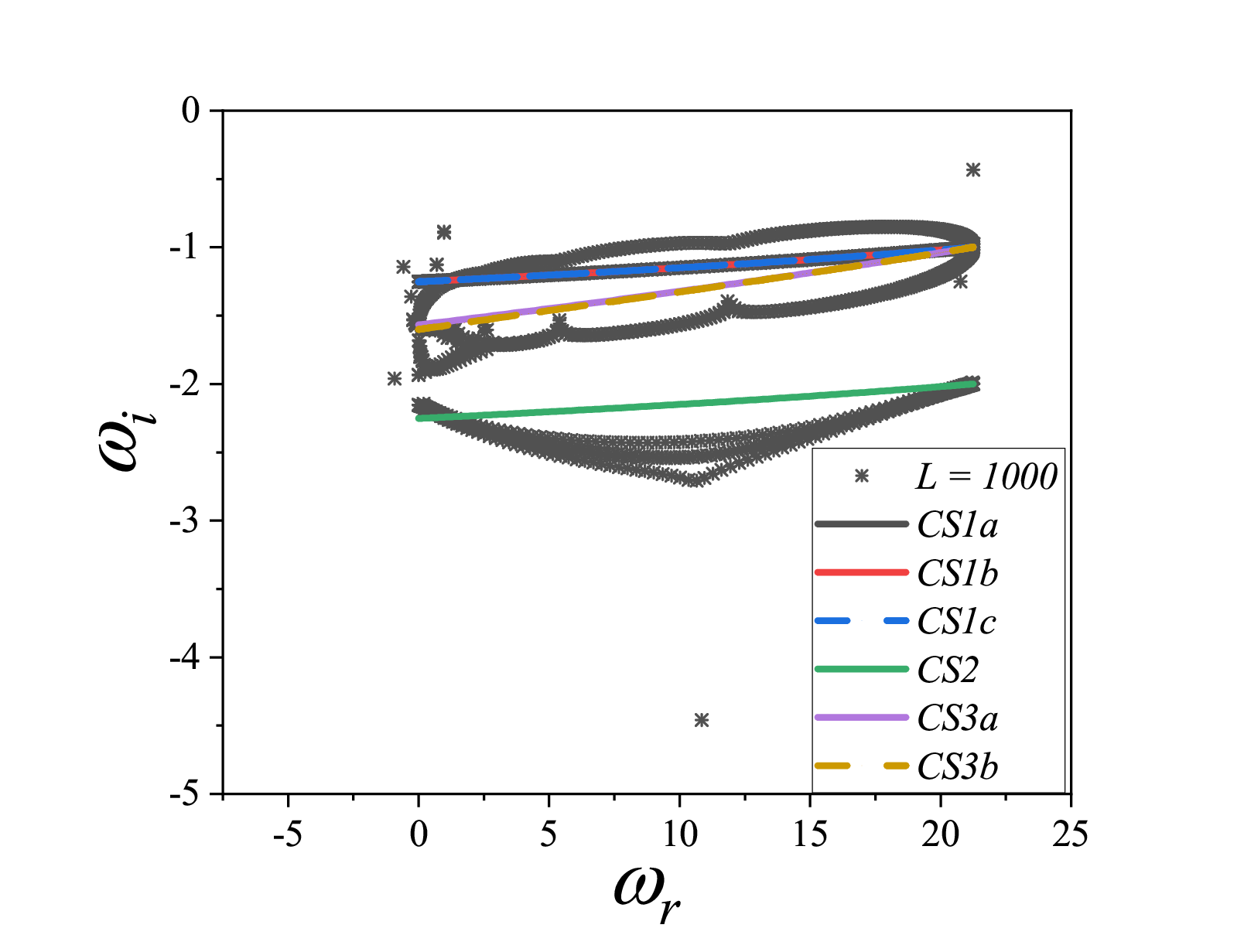}\label{fig:L_1000_PP}}
    \subfigure[$\Wi/L = 0.02$]
  {\includegraphics[width=0.35\linewidth]{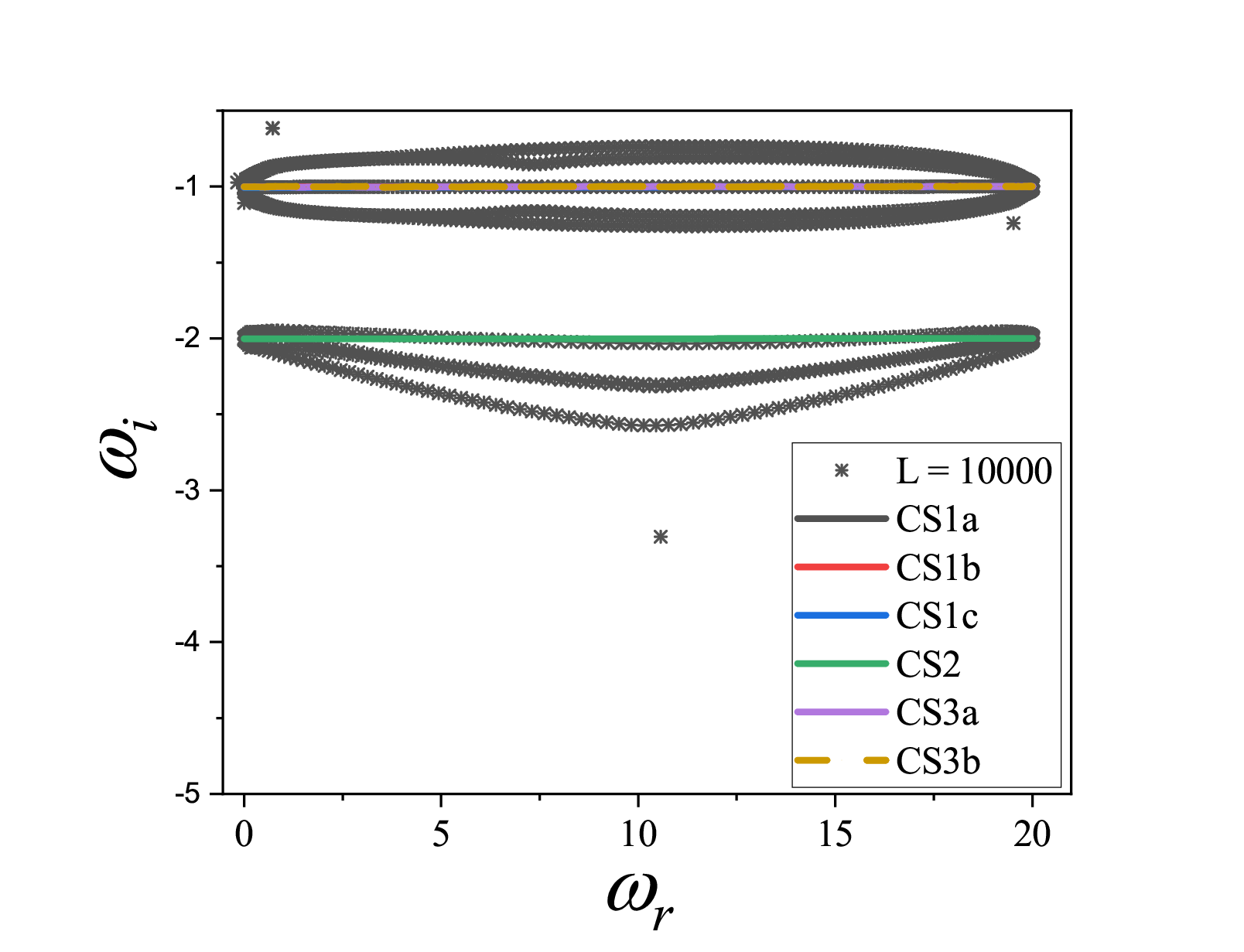}\label{fig:L_10000_PP}}
   \subfigure[Oldroyd-B]
  {\includegraphics[width=0.35\linewidth]{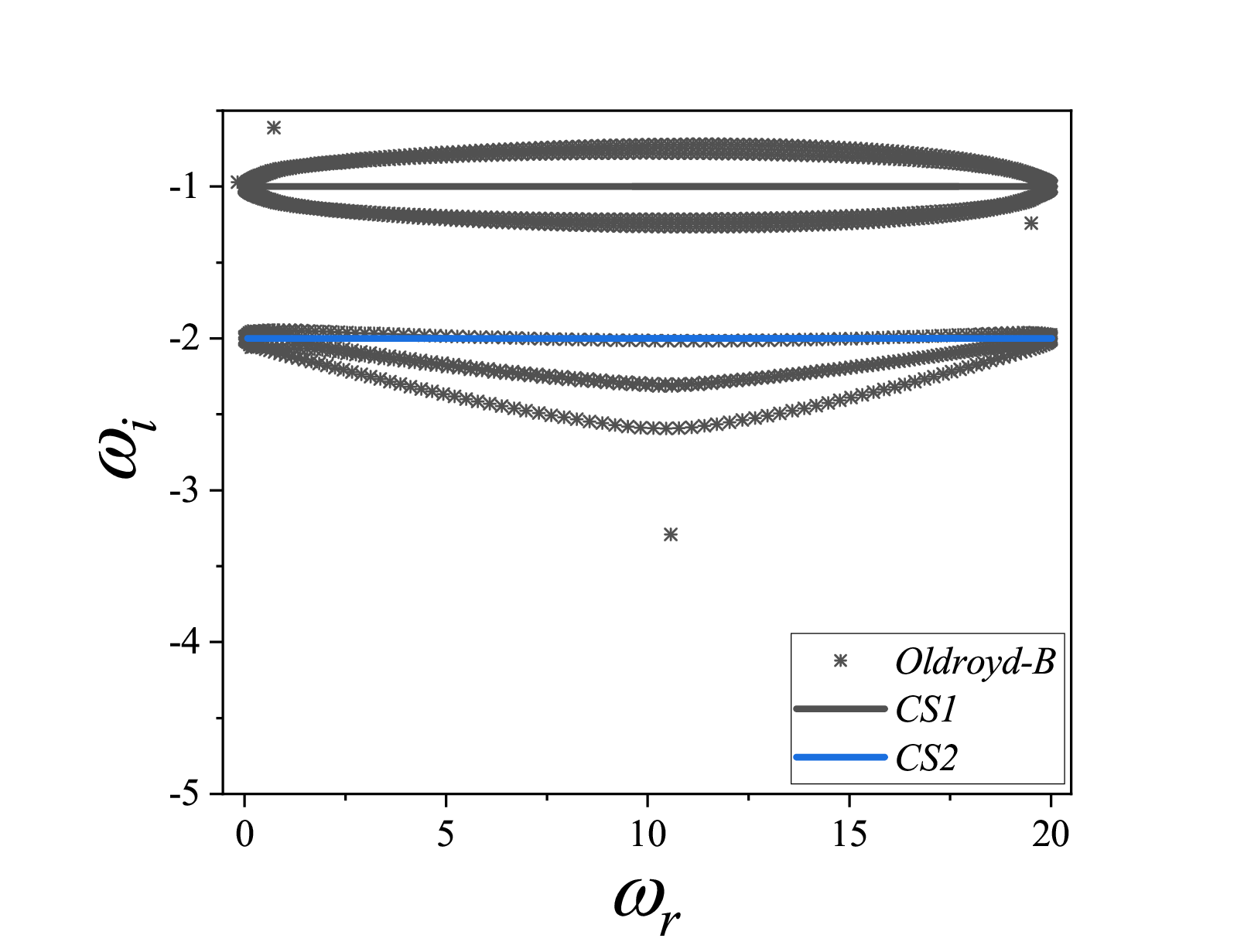}\label{fig:OldroydB_PP}}
  \caption{Numerical and analytical CS for pressure-driven channel flow. Data for $\Wi = 200, \beta = 0.5, k = 0.1, Re = 0$, and $N = 400$. (Dashed lines are used to enhance the visibility of overlapping CS's)}
  \label{fig:PP_effect_of_L}
\end{figure}
 
The base-state velocity profile for pressure-driven channel flow of a FENE-P fluid is required for the prediction of the CS, and we follow the procedure outlined by Zhang \textit{et al.} \cite{Zaki2013} to obtain the same.
We first show, in Fig.\,\ref{fig:PPF_CS_schematics}, the schematics of the CS, for both $k = 0$ and $k \neq 0$. Figure\,\ref{fig:Dean_axis_n_0_schematic} shows that, unlike plane Couette flow discussed above, the CS in plane Poiseuille flow are extended (vertical) line segments or curves even for $k = 0$. This is readily understood for CS1(a--c), for which $\omega_i \simeq -\bar{f}$ for $L \gtrsim 50$, with $\bar{f}$ now being a function of the wall-normal coordinate $z$ (owing to the varying base-state shear rates in the wall-normal direction), which turns the CS into vertical line segments on the imaginary axis. Similarly, the solvent CS (CS2) is also a vertical line segment on the imaginary axis, albeit of a larger length. Finally, both the real and imaginary parts of  CS3(a,b) are also functions of $\bar{f}$, which results in CS3(a,b) attaining a wing-like structure symmetric about the imaginary axis.  
Note that the wing structure shown in Fig.\,\ref{fig:Dean_axis_n_0_schematic} is only valid above a threshold $\beta$.  
As $\beta$ is decreased, there is a bifurcation at a threshold $\beta$, below which the two wings turn into two distinct vertical line segments on the imaginary axis. This bifurcation is analogous to the one discussed earlier in plane Couette flow (see Fig.\,\ref{fig:Tracking_of_CS4_beta}). 
For $k \neq 0$, all the CS are shifted horizontally by the factor $\omega_r = k \Wi \bar{V}_x$, as shown in Fig.\,\ref{fig:Dean_nonaxis_n_1_schematic}. Different points on  CS1(a--c) and CS2 (for $k = 0$)  get shifted by different amounts (on account of the $z$-dependence of $\bar{V}_x$), resulting in these vertical segments turning into curves for $k \neq 0$.  As a result, CS1(a--c) and CS2 turn into curves for $k \neq 0$. CS3a and 3b, which were already curves for $k = 0$, but symmetric about the $\omega_i$ axis,  now become asymmetric on account of the horizontal shift factor. 
As mentioned in the Introduction, the (dimensional) decay rate of the CS1 and 2 eigenmodes is proportional to $1/\lambda$ in the Oldroyd-B limit, with $\lambda$ being the (constant) relaxation time. 
Owing to the varying shear rate in pressure-driven channel flow, the effective relaxation time of the FENE-P fluid decreases from the channel centre (zero shear rate) to the wall (maximum shear rate), resulting in an increase in the decay rate as $\omega_r$ samples the base range of velocities from the channel center to the wall. In the spectrum, this implies that the CS eigenvalues with maximum $\omega_r$ (with phase-speeds close to the base-state maximum) would have smaller decay rates compared to those with $\omega_r$ close to zero (walls). 
Thus, the lowest points on all the CS, except CS3(a,b), must terminate at the imaginary axis, as illustrated in Fig.\,\ref{fig:Dean_nonaxis_n_1_schematic}.
Accordingly, the CS eigenfunctions associated with  the upper (lower) portions of CS1(a-c) and 2 will exhibit singularities close to the center (wall).
\begin{figure}
\centering
\subfigure[$k = 0.01$ (Analytical)]
{\includegraphics[width=0.35\textwidth]{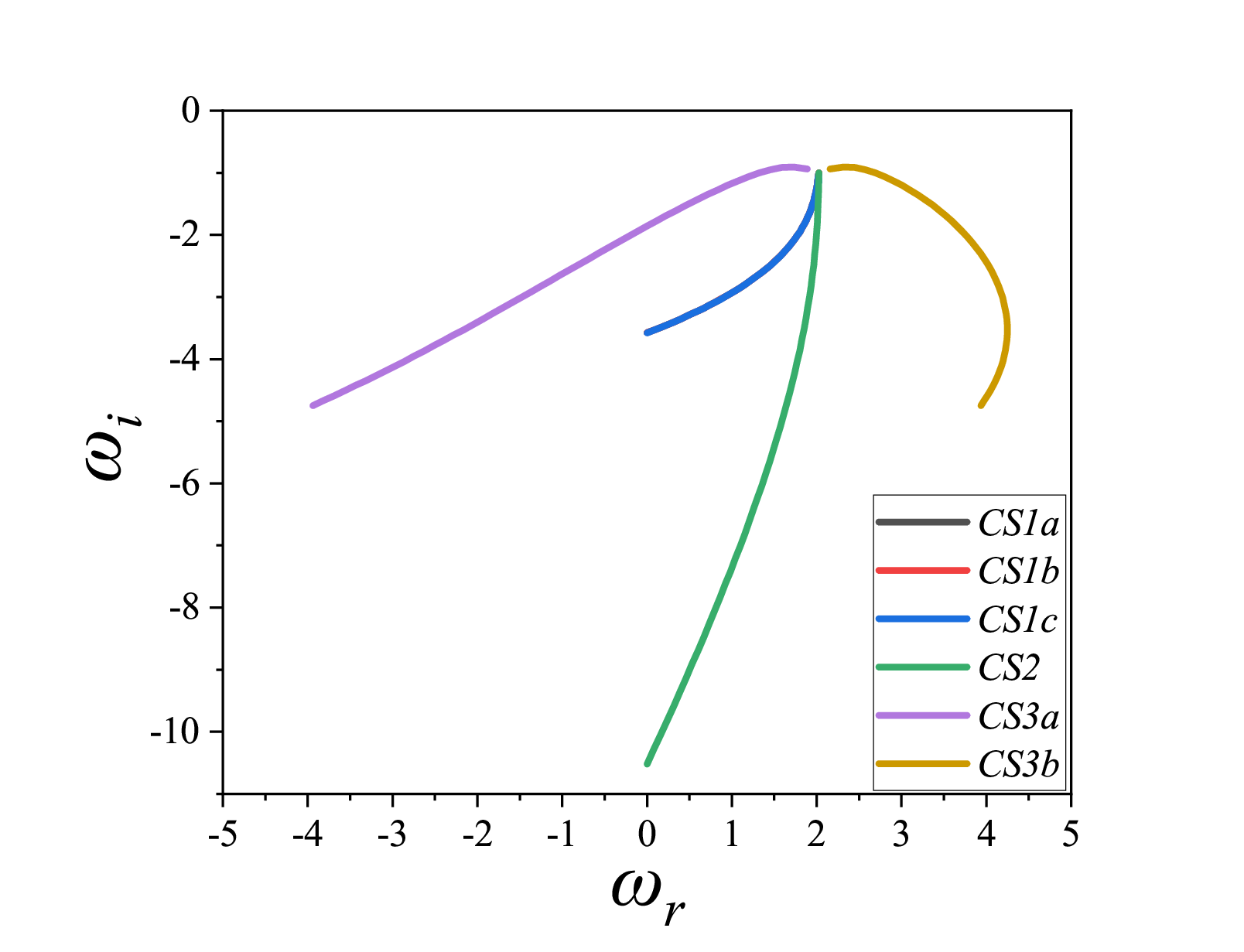}\label{fig:k0.01_Pipe_ana}}
\subfigure[$k = 0.01 $ (Numerical)]
{\includegraphics[width=0.35\textwidth]{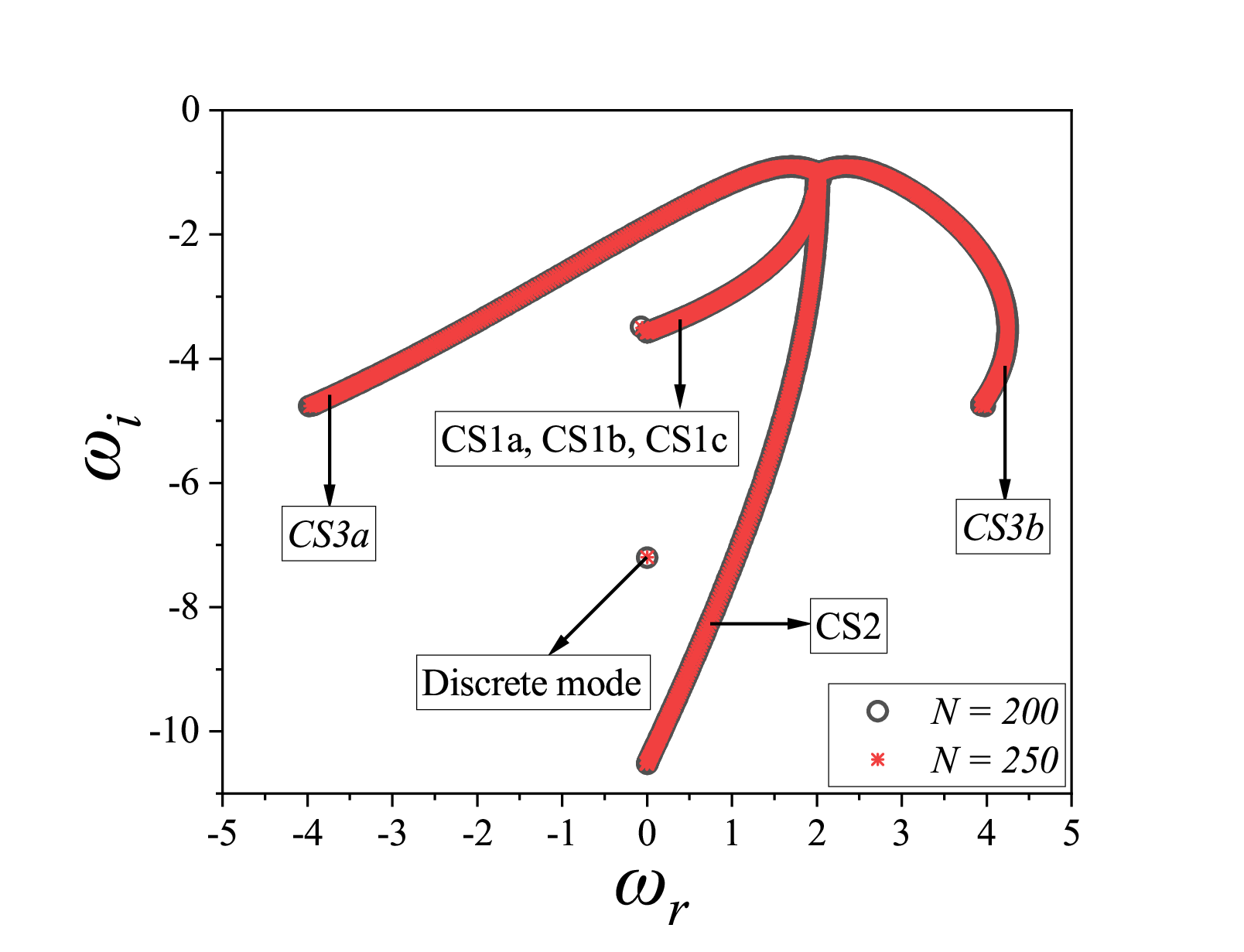}\label{fig:k0.01_Pipe_num}}
  \caption{Analytical CS and numerical spectra for pressure-driven pipe flow for $\Wi = 200, L= 100, \beta = 0.98, Re = 0$, $n = 0$.}
  \label{fig:Pipe_effect_of_k}
\end{figure}

Figure\,\ref{fig:PPF_effect_of_k} shows a comparison of the theoretical CS predictions with the numerically obtained spectra, for both $k = 0$ (Figs.\,\ref{fig:k0_PPF_ana} and \ref{fig:k0_PPF_num}) and $k \neq 0$  (Figs.\,\ref{fig:k0.01_PPF_ana}--\ref{fig:k0.1_PPF_num}), highlighting the excellent agreement between the two. 
As $k$ is increased to $O(1)$ values, the horizontal extents of CS3a and 3b become identical to that of CS1(a--c), but their vertical loci are different for $\Wi/L \sim O(1)$ and higher. In Fig.\,\ref{fig:PP_effect_of_L}, we show the comparison between the theoretically predicted CS and the full numerical spectra at different $\Wi/L$ ratios. We note, again, a very good agreement between the analytical CS and their numerical approximations in the spectrum. The vertical spread of the CS balloon is again highest for CS2 for $\Wi/L = 2$, with negligible spread in the numerical CS for CS1(a--c) and CS3(a,b). Modes CS1(a-c) and CS3(a,b) remain distinct down to $\Wi/L = 0.2$, but merge with each other for a lower $\Wi/L = 0.02$, closer to the Oldroyd-B limit.  For comparison, we also show the spectrum for the Oldroyd-B fluid for the same set of parameters in Fig.\,\ref{fig:OldroydB_PP}. Similar to plane Couette flow, CS1(a-c) and CS2 in pressure-driven channel flow also exhibit a collapse for the same $\Wi/L$ for different $(\Wi, L)$ pairs; the collapse does not hold for CS3(a,b).

Finally, we have verified that an identical procedure is valid for determining the CS in  pressure-driven pipe flow of a FENE-P fluid. We show a representative comparison of the analytical CS with the numerical spectrum for pipe flow in  Fig.\,\ref{fig:Pipe_effect_of_k}. It is worth mentioning that while the structure of the CS in Figs.\,\ref{fig:k0.01_PPF_num} (for channel) and \ref{fig:k0.01_Pipe_num} (for pipe) is identical  for the same set of parameters, the discrete modes are clearly not the same in the two geometries.

\section{Curvilinear flows}
\label{sec:curvilinear}

The base-state velocity profile $\bar{V}_{\theta}$ and the Peterlin function $\bar{f}$ are important ingredients to predict the CS for the Dean and Taylor-Couette flows discussed in this section. We use the numerical procedure discussed in the work of Tej \textit{et al.}\,\cite{Tej2024} to compute the base-state quantities required for this purpose.
For curvilinear flows, the perturbations to the various fields are expressed as 
$\phi' = \tilde{\phi}(\zeta) \exp[\frac{-i \omega t}{\Wi} ] \exp[i \alpha z] \exp[i n \theta] $. Here, $\omega$ is the complex frequency (nondimensionalized using $\lambda$)  as before, $\alpha$ is the axial wavenumber, and $n$ is the integer-valued azimuthal wavenumber. Further, instead of the dimensional radial coordinate $r^*$ varying from $R_1$ to $R_2$ ($R_1$ and $R_2$ being the inner and outer cylindrical radii),  it is convenient to measure radial distances from the inner cylinder in units of the gap width, by defining the dimensionless coordinate $\zeta = (r^* - R_1)/d$, $\zeta \in [0,1]$, which appears in the ansatz for the perturbation above; here, $d = \epsilon R_1$ is the gap width defined earlier in Sec.\,\ref{sec:probformulation}.

We first discuss the procedure to identify the CS in curvilinear flows, and the features that are common to both Dean and Taylor-Couette geometries.  Similar to rectilinear flows, where the continuous spectra are independent of the spanwise wavenumber, here too, we find from numerically computed spectra that the CS is independent of $\alpha$ (see Fig.\,\ref{fig:Effect_a_Dean_CS}). Hence, without loss of generality, we set $\alpha = 0$ for the results to be shown below.
In view of the complex nature of the spectra, similar to the case of  rectilinear flows,  it is instructive to discuss axisymmetric ($n = 0$) disturbances first and then proceed to  nonaxisymmetric ($n \neq 0$) disturbances. 
The CS for the latter disturbances are obtained via a horizontal shift, $\omega_r = \frac{n \epsilon \Wi \overline{V}_\theta}{1 + \epsilon \zeta}$, analogous to rectilinear flows. The nomenclature used to label the different CS modes is again identical to the one adopted for rectilinear flows above.

For $n = 0$ (and $\alpha = 0$), it suffices to consider only the $\theta$ momentum balance,  which governs the uni-directional azimuthal flow that arises in this limit:
\begin{eqnarray}
    \beta \left(\left(\frac{d^2}{d\zeta^2} - \frac{\epsilon^2}{(1 + \epsilon \zeta)^2} + \frac{\epsilon}{1 + \epsilon \zeta}\frac{d}{d\zeta}\right)\widetilde{V}_\theta\right) + (1 - \beta)\left(\frac{\widetilde{\tau}_{r\theta}}{d\zeta}+\frac{2\epsilon}{1 + \epsilon \zeta}\widetilde{\tau}_{r\theta}\right) = 0
    \label{eq:curvilinear_theta_momentum}
\end{eqnarray}
The coefficient of the second derivative of $\tilde{v}_{\theta}$ is identified from the above equation. As already seen, the coefficient explicitly multiplying the second derivative in Eq.\,\ref{eq:curvilinear_theta_momentum} is only a partial contribution, and to obtain the full contribution (a sixth-order polynomial in $\omega$), one again starts expressing the components of $\widetilde{\boldsymbol{C}}$ in terms the perturbation velocity components and their derivatives.
 Next, we substitute the relevant components of the stress tensor into the momentum equation in terms of the components of $\boldsymbol{C}$:
\begin{equation}
     \widetilde{\tau}_{rr} =\left(\frac{\overline{f}^2~\overline{C}_{rr}}{(L^2 - 3)Wi} + \frac{\overline{f}}{\Wi} \right)\widetilde{C}_{rr} + \frac{\overline{f}^2~\overline{C}_{rr}}{(L^2 - 3)\Wi}\widetilde{C}_{\theta \theta} + \frac{\overline{f}^2~\overline{C}_{rr}}{(L^2 - 3)\Wi}\widetilde{C}_{zz} 
 \end{equation}
  \begin{equation}
      \widetilde{\tau}_{r \theta} = \frac{\overline{f}^2~\overline{C}_{r \theta}}{(L^2 - 3)\Wi}\widetilde{C}_{rr} + \frac{\overline{f}}{\Wi}\widetilde{C}_{r \theta}+ \frac{\overline{f}^2~\overline{C}_{r \theta}}{(L^2 - 3)\Wi} \widetilde{C}_{\theta \theta} + \frac{\overline{f}^2~\overline{C}_{r \theta}}{(L^2 - 3)\Wi}\widetilde{C}_{zz}  
 \end{equation}
\begin{equation}
     \widetilde{\tau}_{\theta \theta} = \frac{\overline{f}^2~\overline{C}_{\theta \theta}}{(L^2 - 3)\Wi}\widetilde{C}_{rr} + \left(\frac{\overline{f}^2~\overline{C}_{\theta \theta}}{(L^2 - 3)W} + \frac{\overline{f}}{\Wi} \right)\widetilde{C}_{\theta \theta} + \frac{\overline{f}^2~\overline{C}_{\theta \theta}}{(L^2 - 3)\Wi}\widetilde{C}_{zz} 
 \end{equation}
 \begin{equation}
     \widetilde{\tau}_{zz} = 
     \frac{\overline{f}^2~\overline{C}_{zz}}{(L^2 - 3)\Wi}\widetilde{C}_{rr} + \frac{\overline{f}^2~\overline{C}_{zz}}{(L^2 - 3)\Wi}\widetilde{C}_{\theta \theta} 
 + \left(\frac{\overline{f}^2~\overline{C}_{zz}}{(L^2 - 3)W} + \frac{\overline{f}}{\Wi} \right)\widetilde{C}_{zz} 
 \end{equation}
 The equations governing the various components of $\boldsymbol{C}$ in cylindrical coordinates are given in Appendix\,\ref{FENEP_TC_Appendix}.
 The end result of the above steps is that the $\theta$-momentum balance (Eq.\,\ref{eq:curvilinear_theta_momentum}) is now expressed only in terms of $\tilde{v}_{\theta}$ and its derivatives, and the coefficient of the second derivative of $\tilde{v}_{\theta}$ yields the sixth-order polynomial in $\omega$.  We now discuss the results obtained for the CS specific to the Dean and Taylor-Couette flows in two separate subsections.

 \begin{figure}
  \centering
  \subfigure[$\alpha = 1$]
  {\includegraphics[width=0.45\textwidth]{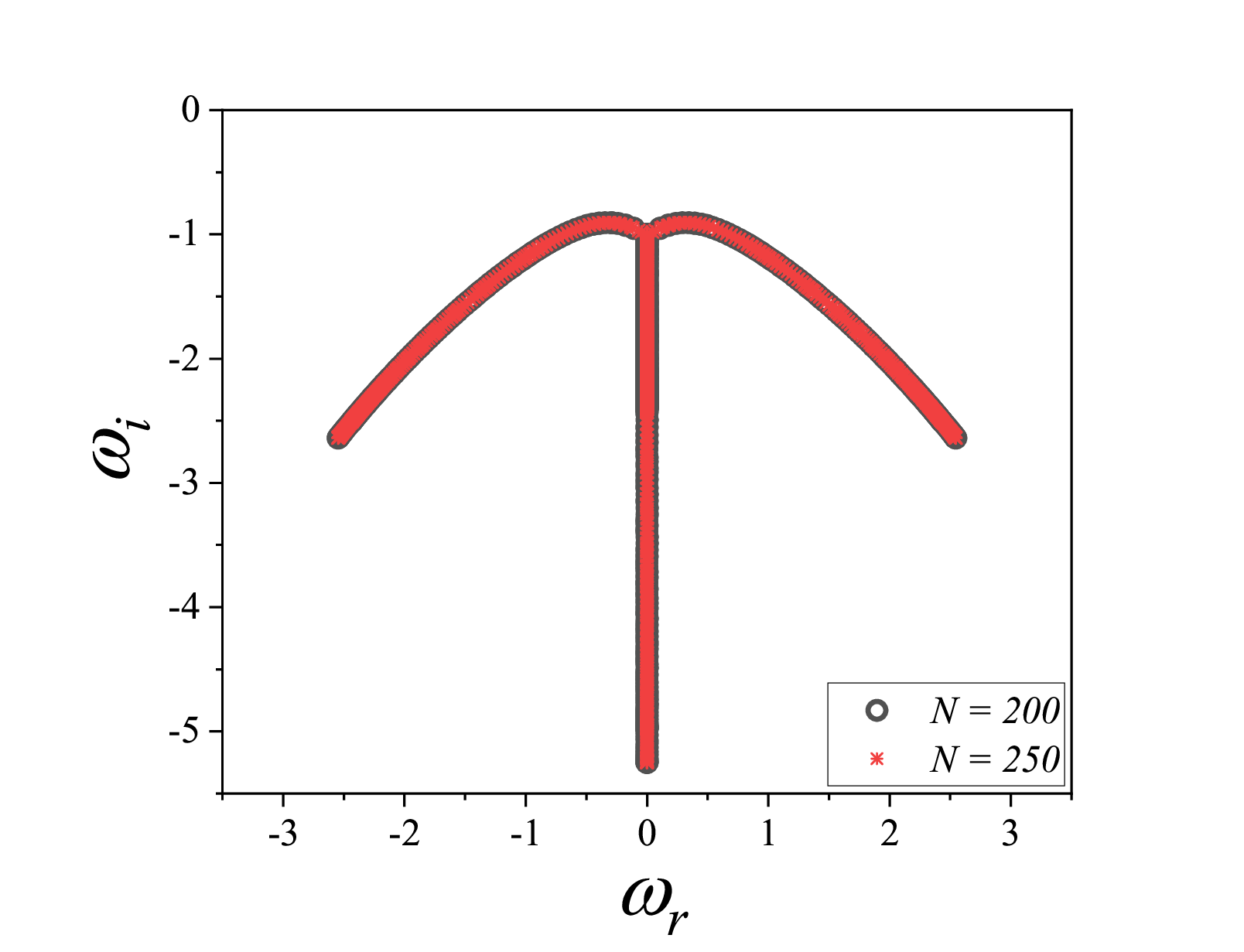}\label{fig:a0_dean}}
    \subfigure[$\alpha = 10$]
  {\includegraphics[width=0.45\textwidth]{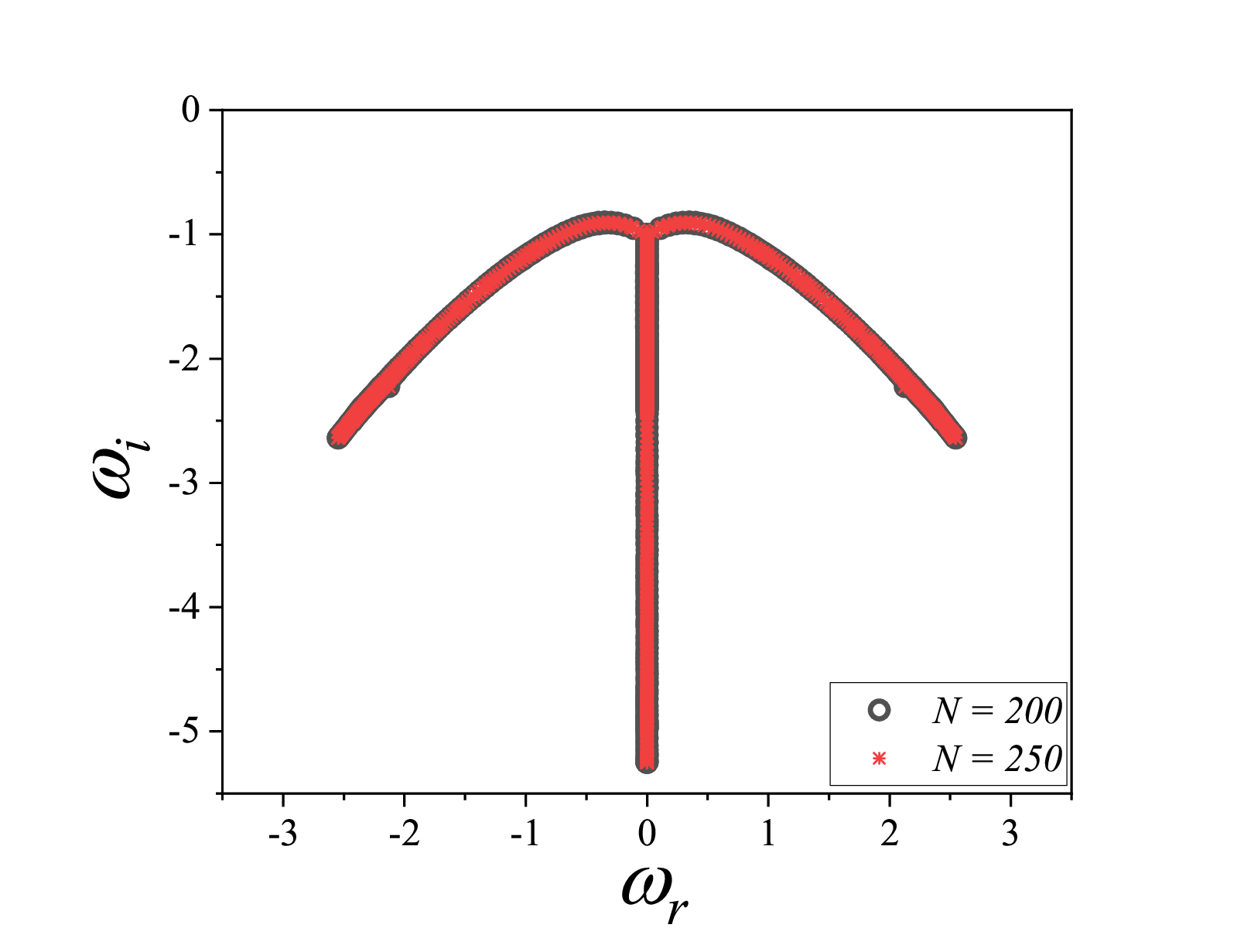}\label{fig:a10_dean}}
  \caption{Numerically computed spectra showing the independence of CS for two different $\alpha$'s; data for $n = 0$, $W = 200, L = 100, \beta = 0.98, \epsilon = 0.1$. For this parameter set, there are no discrete modes at either $\alpha$.}
  \label{fig:Effect_a_Dean_CS}
\end{figure}

\begin{figure}
\centering
    \subfigure[$n = 0$]
{\includegraphics[width=0.35\textwidth]{Dean_axis_n_0_schematic.png}\label{fig:Dean_axis_n_0_schematic_2}}
\quad \quad
\subfigure[$n \neq 0$]
{\includegraphics[width=0.32\textwidth]{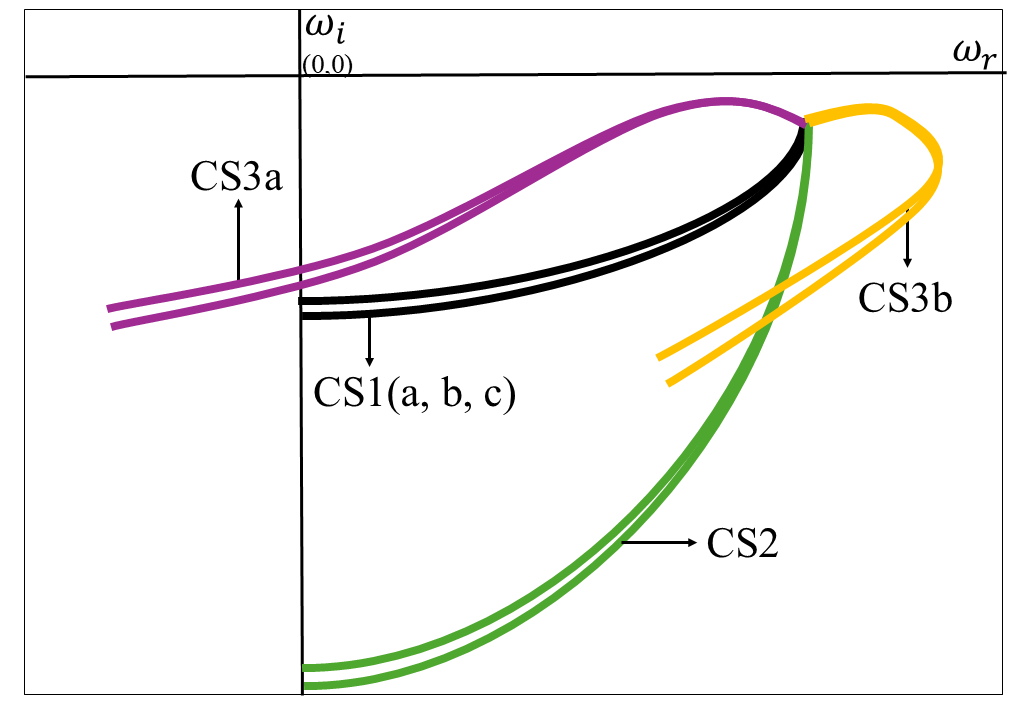}\label{fig:CS_schematics_Dean_n_1}}
  \caption{Schematics of the various CS for Dean flow.}
  \label{fig:Dean_CS_schematics_2}
\end{figure}

\begin{figure}
  \centering
  \subfigure[$L = 10^2$ (analytical)\label{fig:L_100_dean_axis_ana}]
  {\includegraphics[width=0.35\linewidth]{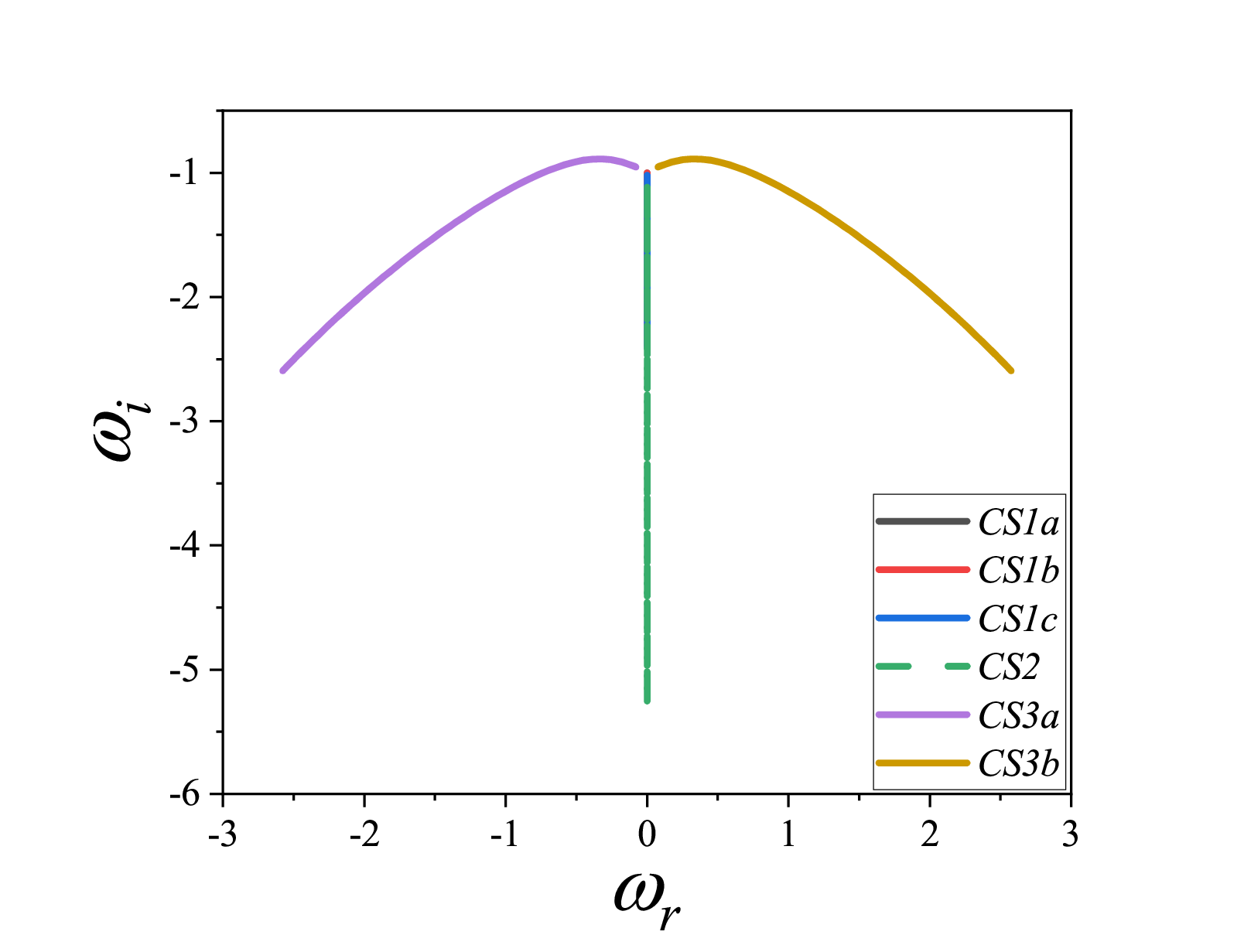}}
  \subfigure[$L = 10^2$ (numerical)\label{fig:L_100_dean_axis_num}]
  {\includegraphics[width=0.35\linewidth]{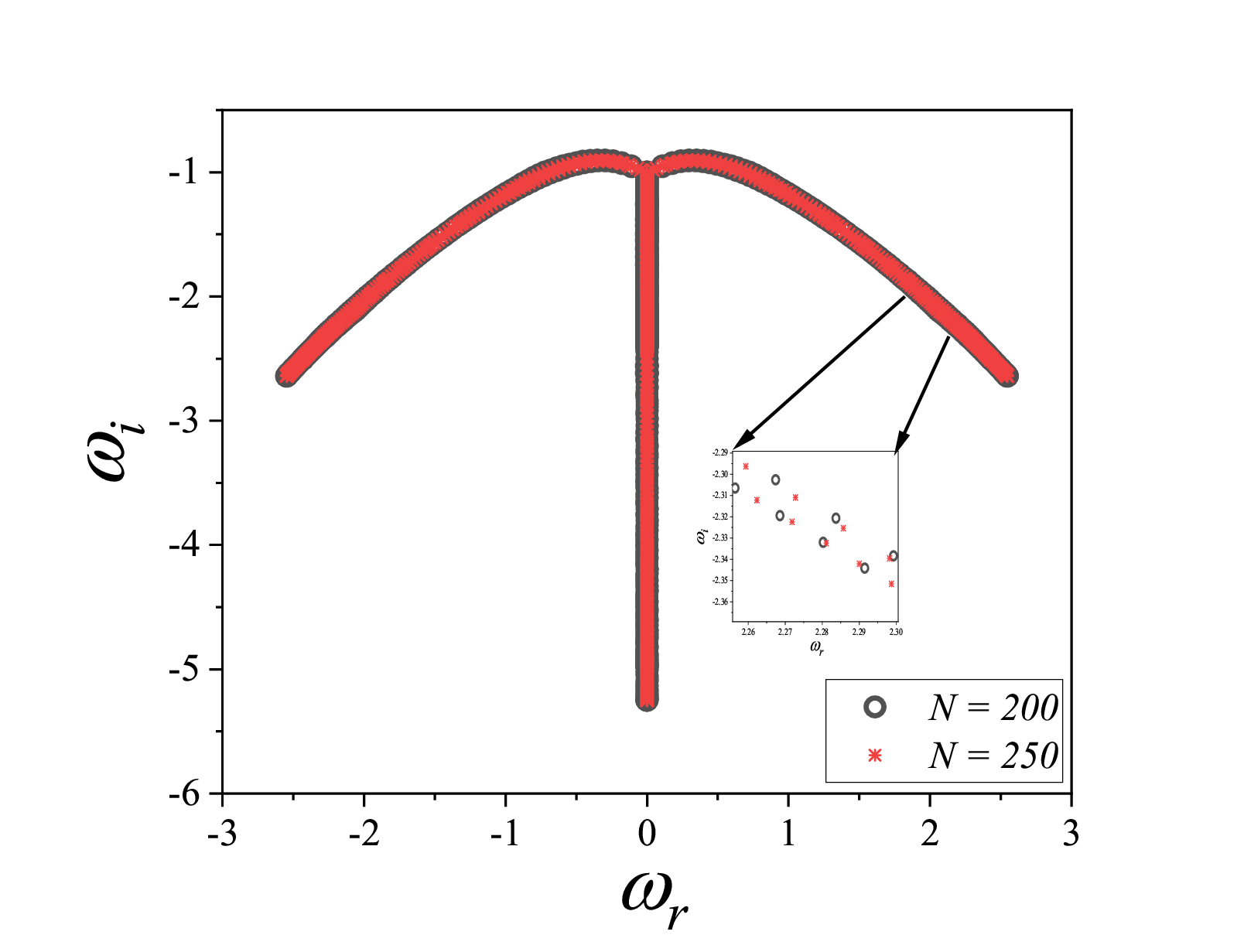}}
  \subfigure[$L = 5 \times 10^2$ (analytical)\label{fig:L_500_dean_axis_ana}]
  {\includegraphics[width=0.35\linewidth]{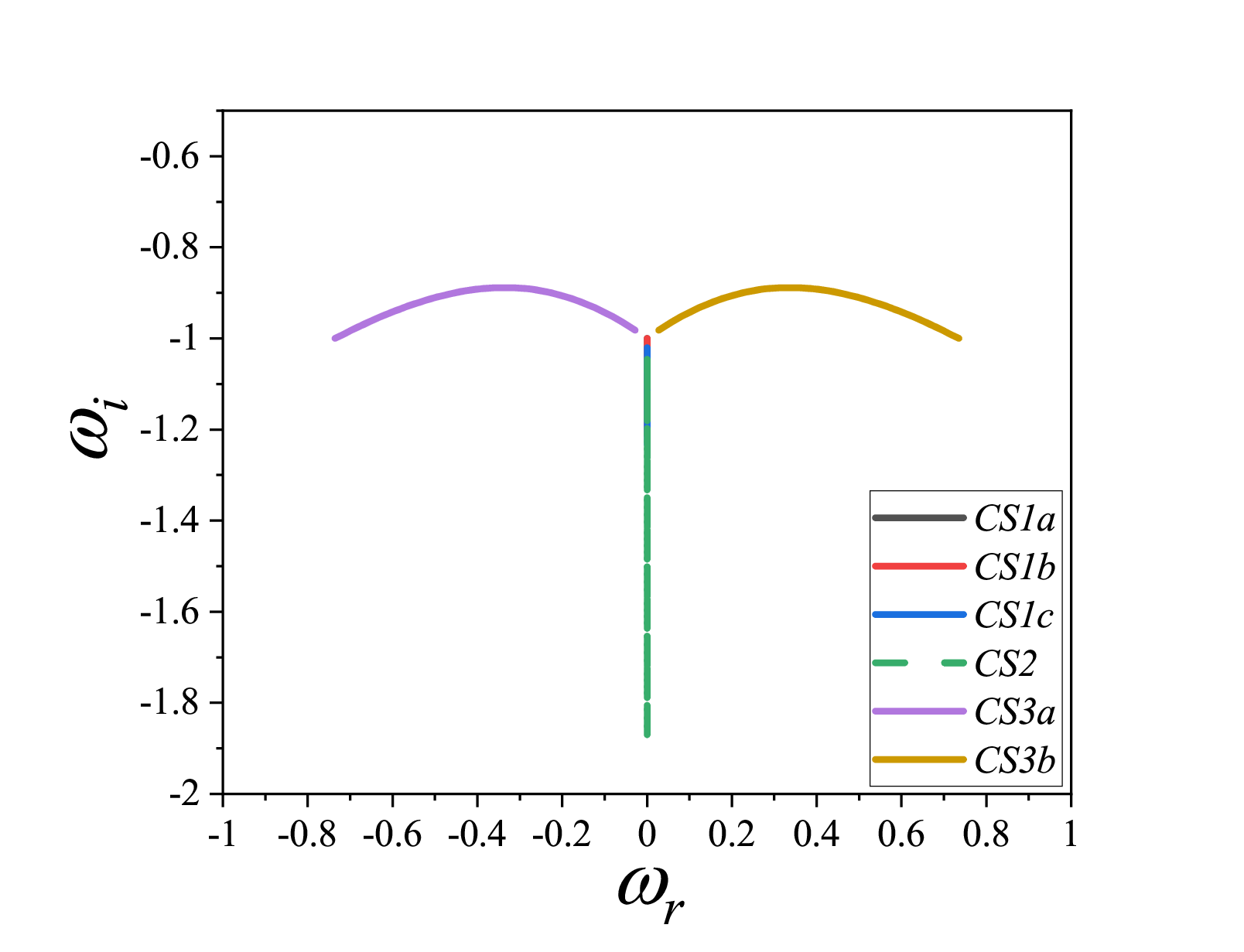}}
  \subfigure[$L = 5 \times 10^2$ (numerical)\label{fig:L_500_dean_axis_num}]
  {\includegraphics[width=0.35\linewidth]{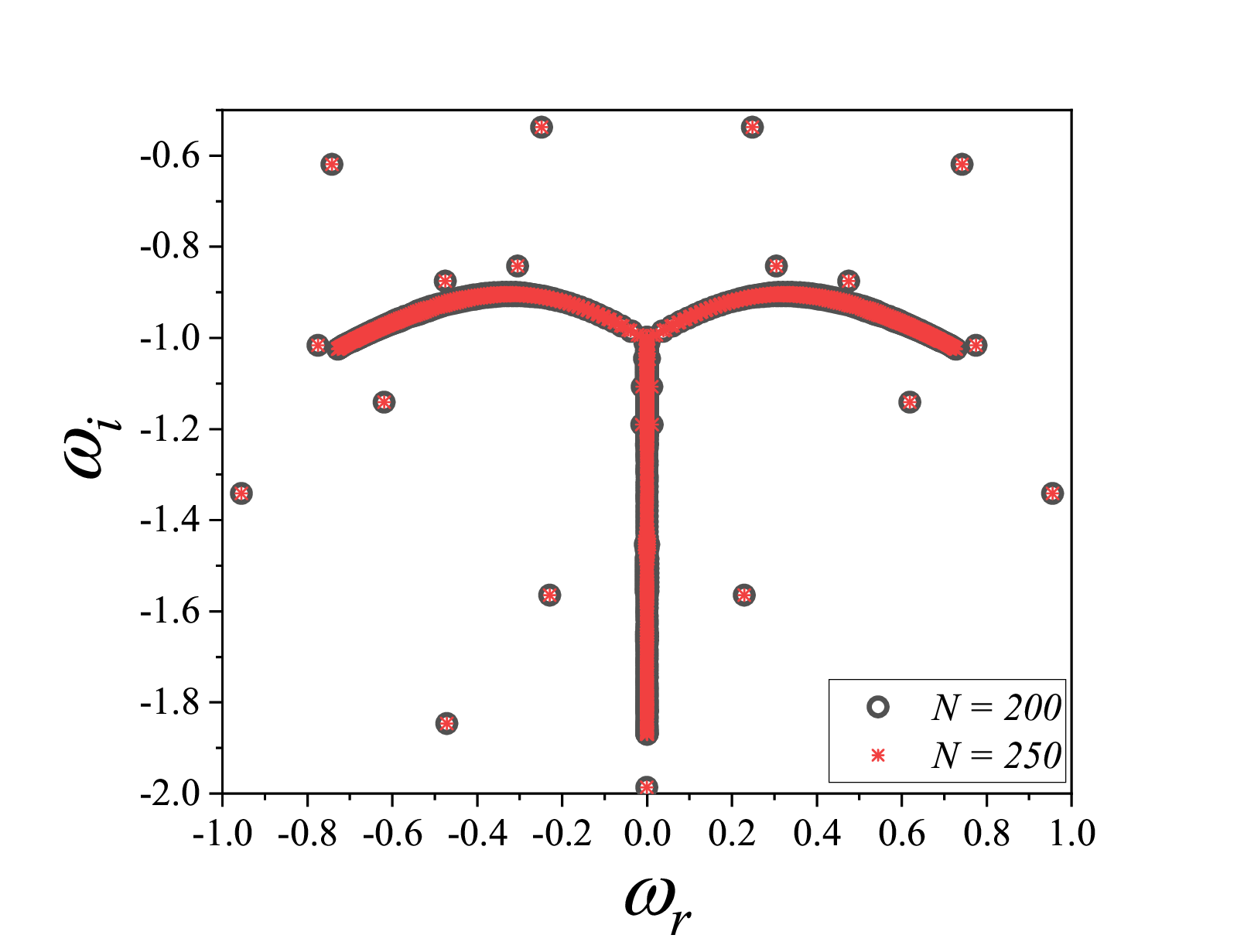}}
  \subfigure[$L = 10^3$ (analytical)\label{fig:L_1000_dean_axis_ana}]
  {\includegraphics[width=0.35\linewidth]{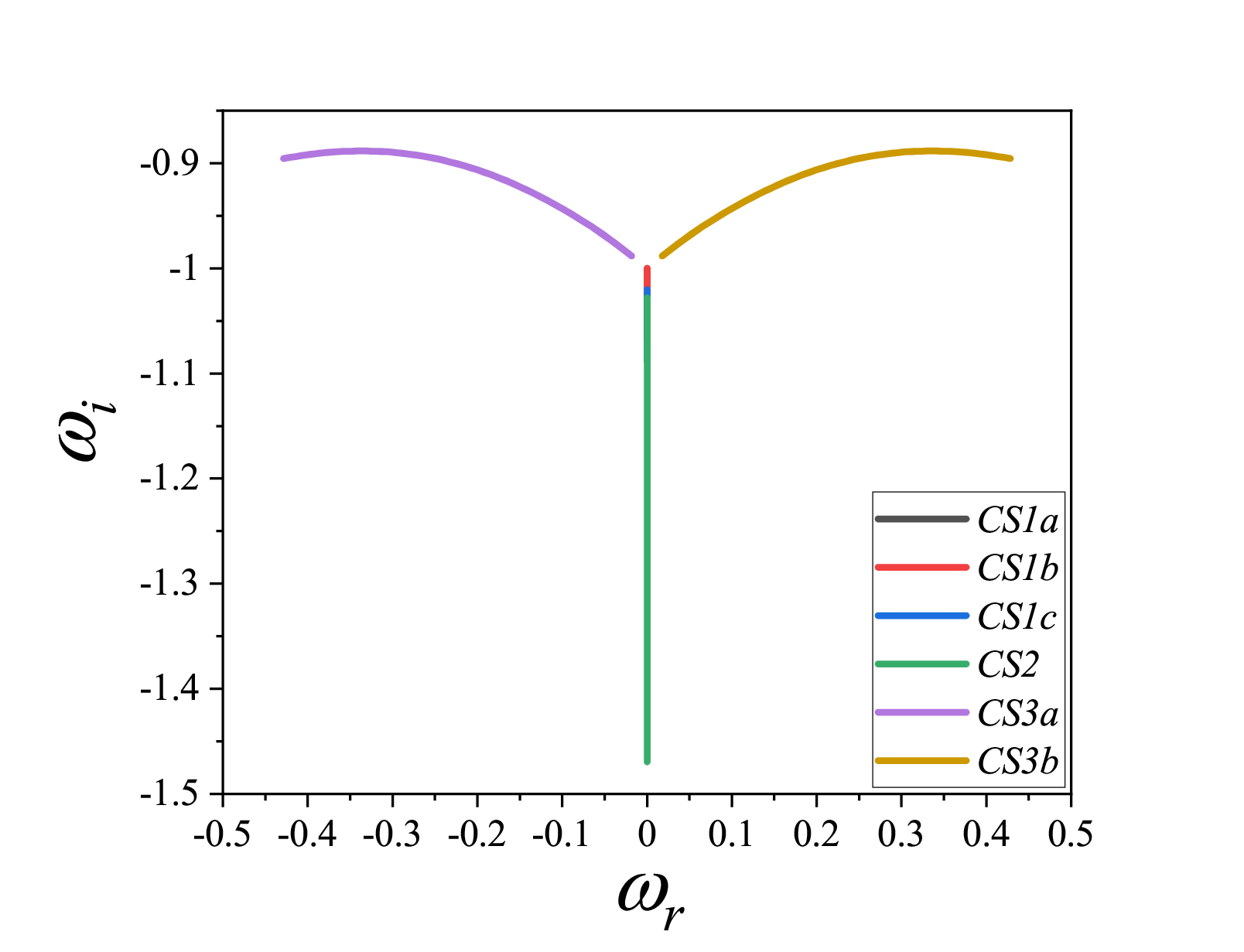}}
  \subfigure[$L = 10^3$ (numerical)\label{fig:L_1000_dean_axis_num}]
  {\includegraphics[width=0.35\linewidth]{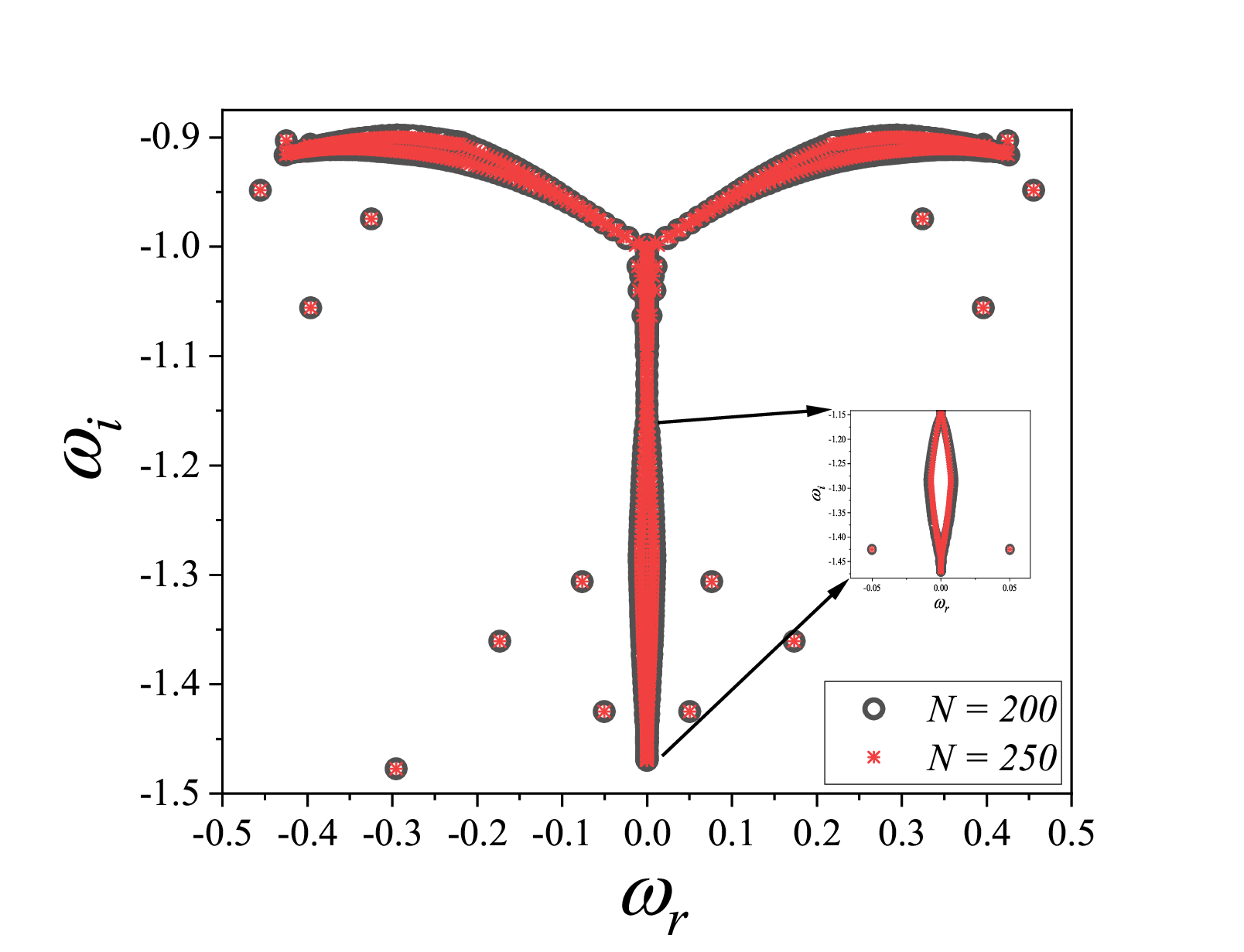}}
  \subfigure[$L = 10^4$ (analytical)\label{fig:L_10000_dean_axis_ana}]
  {\includegraphics[width=0.35\linewidth]{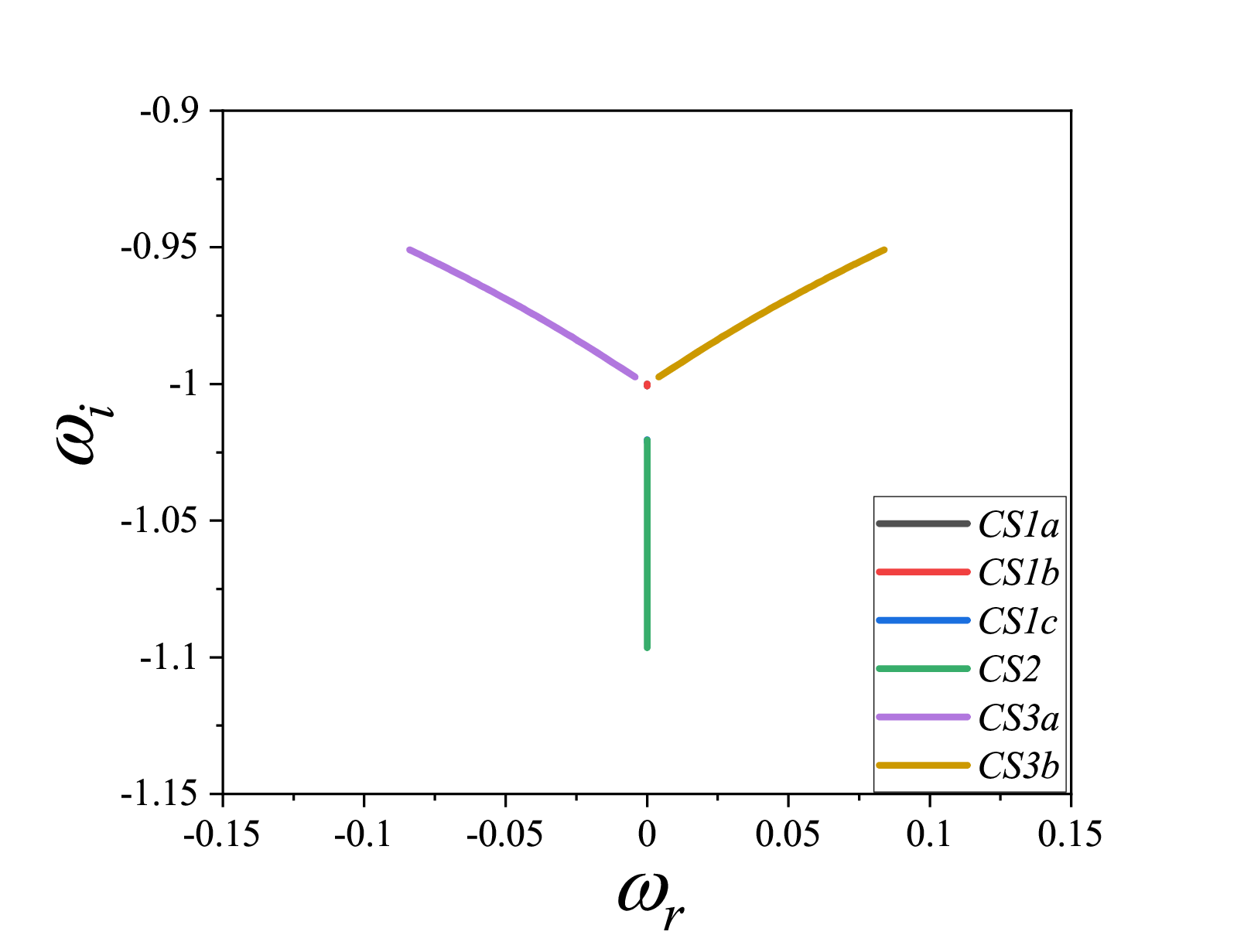}}
  \subfigure[$L = 10^4$ (numerical)\label{fig:L_10000_dean_axis_num}]
  {\includegraphics[width=0.35\linewidth]{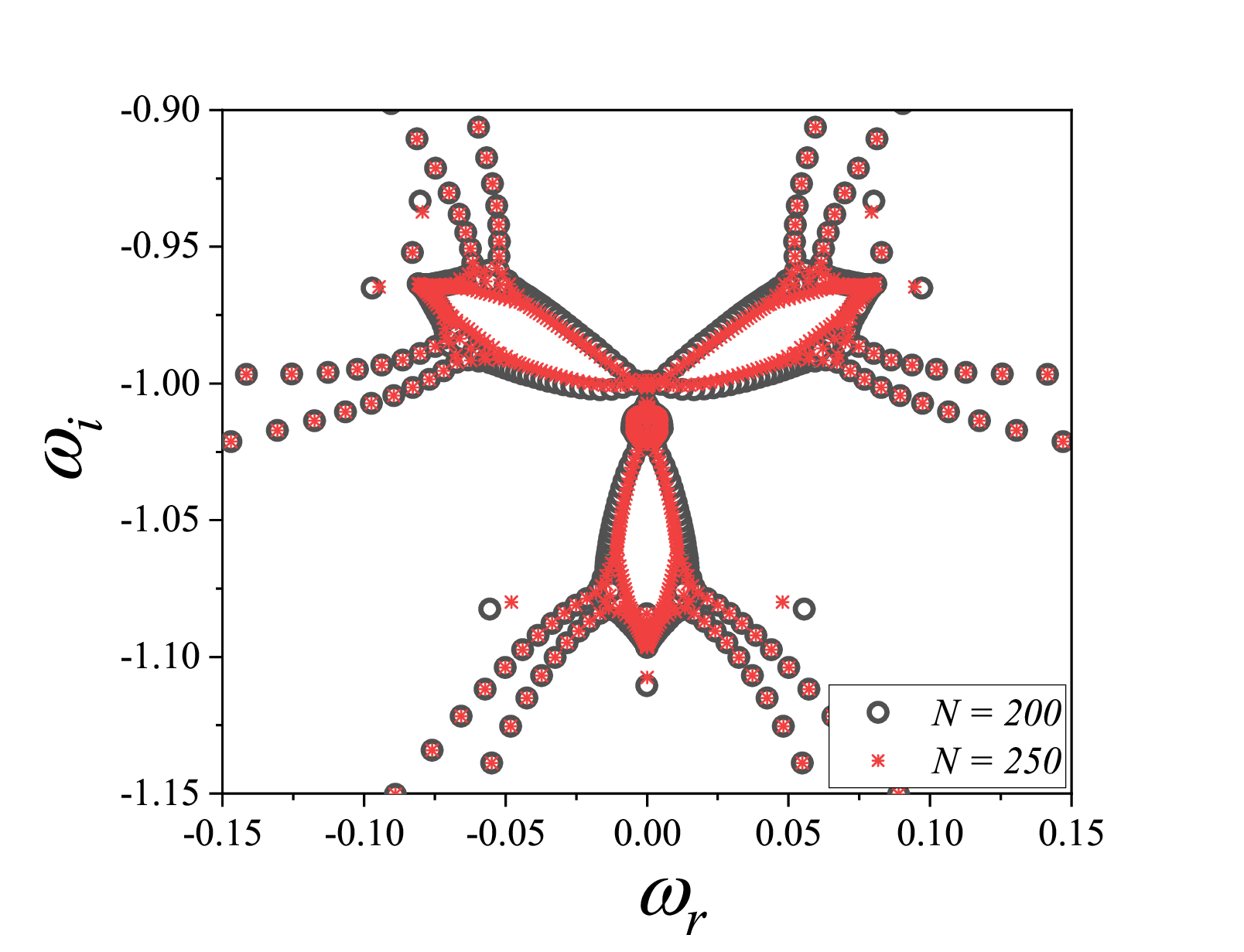}}
  \caption{Analytical CS and numerical spectra for Dean flow subjected to axisymmetric disturbances ($n = 0$). Data for $\Wi = 200, \beta = 0.98, \epsilon = 0.1, \alpha = 7$. Note the reduction in the scale from top to bottom. (Dashed lines are used to enhance the visibility of overlapping CS).}
  \label{fig:Effect_of_L_dean_axis}
\end{figure}

\subsection{Dean flow}
\label{subsec:Dean}

The structure of the CS for Dean flow, shown schematically in Fig.\,\ref{fig:Dean_CS_schematics_2}, shares some similarities to that of pressure-driven channel flow, especially in the narrow-gap limit of $\epsilon \ll 1$, where Dean flow approaches pressure-driven channel flow. Here too, there are, in principle, six CS eigenvalues. However, even for $\epsilon \gtrsim 0.1$, there are some differences in detail, especially for nonaxisymmetric disturbances, as discussed below. This difference pertains to the presence of two branches for each of the CS in Dean flow (Fig.\,\ref{fig:Dean_CS_schematics_2}), in contrast to plane Poiseuille flow (Fig.\,\ref{fig:PPF_CS_schematics}), where each CS is characterized by a single curve. The reason underlying this difference is discussed below in Sec.\,\ref{subsec:nonaxiDean}.

\subsubsection{Axisymmetric disturbances ($n = 0$)}
In this limit, CS1(a-c) are again vertical (overlapping, for $L \gtrsim 50$) line segments on the imaginary axis, with the solvent CS (CS2) line segment being slightly longer. CS3(a,b) again manifest as symmetric wing-like structures about the imaginary axis (Fig.\,\ref{fig:Dean_CS_schematics_2}a). 
 Figure\,\ref{fig:Effect_of_L_dean_axis} shows the effect of $\Wi/L$ on the analytical and numerical eigenspectra for Dean flow, obtained by varying $L$ for a fixed $\Wi$. 
 As $\Wi/L$ is decreased, the lengths of CS1(a--c) decrease, approaching  a point in the Oldroyd-B limit of $L \rightarrow \infty$. 
 The horizontal and vertical extents of the `wing' CS (CS3(a,b)) and the vertical extent of CS2 also decrease with increasing $L$, but more slowly, and are still finite at $L = 10^4$. In addition, new discrete modes appear from the CS, which are symmetric about the imaginary axis. 
 A comparison of the analytical CS and the full numerical spectra shows a decrease in the number of discrete modes as $\Wi/L$ is increased, with eventually only the CS present for $\Wi/L = 2$. 
 Although not shown, 
for a fixed $L = 100$, as $\Wi$ is decreased, we find that the horizontal extent of the wings shortens and the length of the other vertical CS also decreases. This feature is similar to that exhibited by rectilinear flows discussed previously at the end of Sec.\,\ref{sec:rectilinear_PC} on plane Couette flow. 
All the CS exhibit a  collapse for different pairs of $(\Wi,L)$ for the same $\Wi/L$ ratio for axisymmetric disturbances. However, the wings of CS3(a,b) do not collapse for $n \neq 0$.

\subsubsection{Non-axisymmetric disturbances ($n \neq 0$)}
\label{subsec:nonaxiDean}

The schematic representation of the various CS for $n \neq 0$ is shown in Fig.\,\ref{fig:Dean_CS_schematics_2}b.  Interestingly, and unlike the pressure-driven channel flow case (Fig.\,\ref{fig:PPF_effect_of_k}), we find two distinct curved branches for each CS. This arises due to the lack of mid-plane symmetry in the base-state velocity profile for Dean flow, leading to wall-normal locations (on either side of the maximum) with identical base-flow velocities, but with unequal shear rates. The effective relaxation time is lowest near the inner cylinder due to higher shear rates;  as we proceed from the inner to the outer cylinder, the  base-state shear rate vanishes at a certain radial location which has the highest effective relaxation time, forming the lower branch of the CS. Beyond this radial location, the shear rate increases again, and the effective relaxation time decreases as one proceeds toward the outer cylinder, giving rise to the upper branch of the CS.  That is to say, the wall normal locations with identical base-state velocities have different shear rates, and this leads to each CS having two branches for Dean flow. 
\begin{figure}
  \centering
  \subfigure[$n = 0$ (analytical)]
  {\includegraphics[width=0.35\linewidth]{Dean_axis_L100_ana.eps}\label{fig:L_100_dean_axis_ana_n0}}
  \subfigure[$n = 0$ (numerical)]
  {\includegraphics[width=0.35\linewidth]{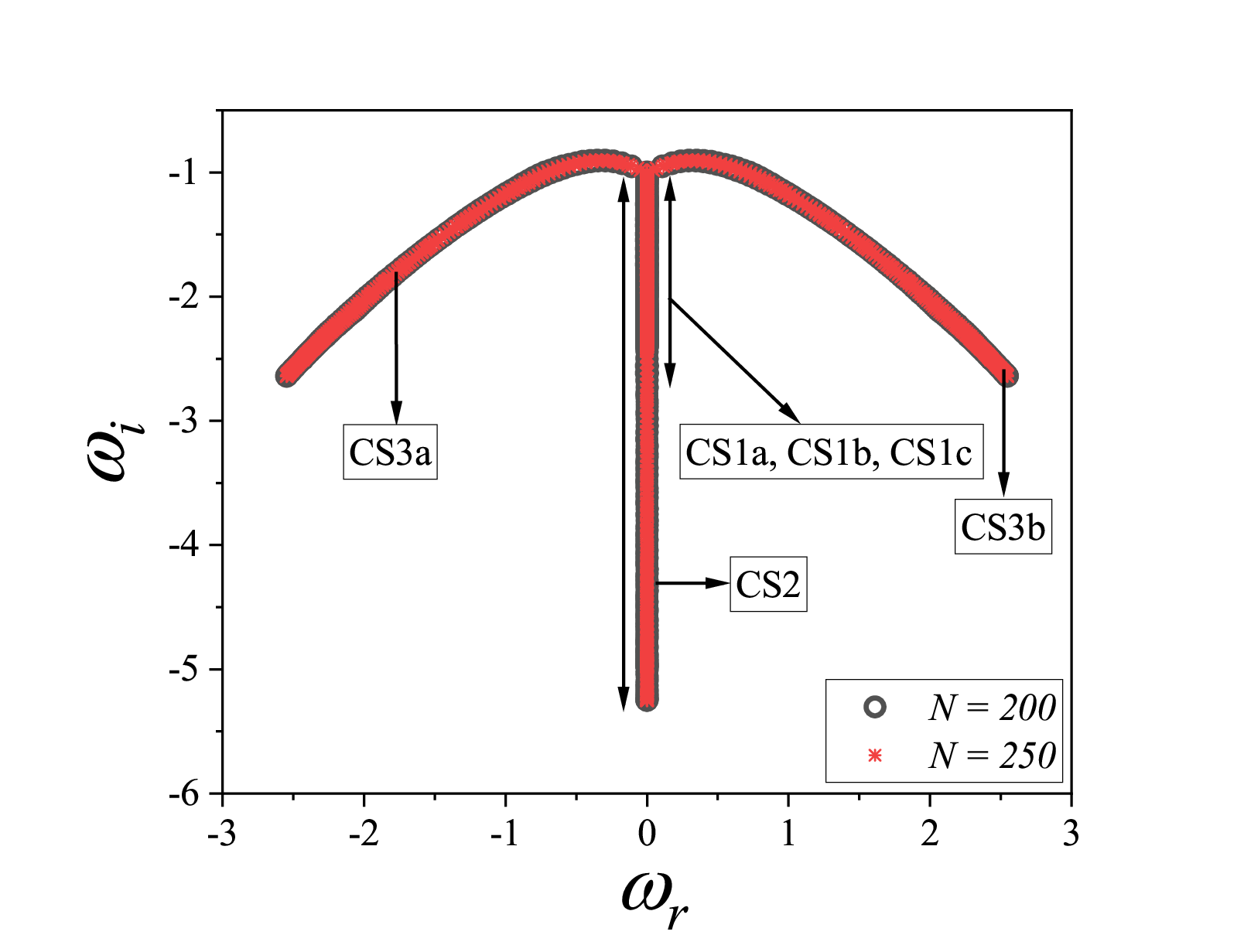}\label{fig:L_100_dean_axis_num_n0}}
  \subfigure[$n = 1$ (analytical)]
  {\includegraphics[width=0.35\linewidth]{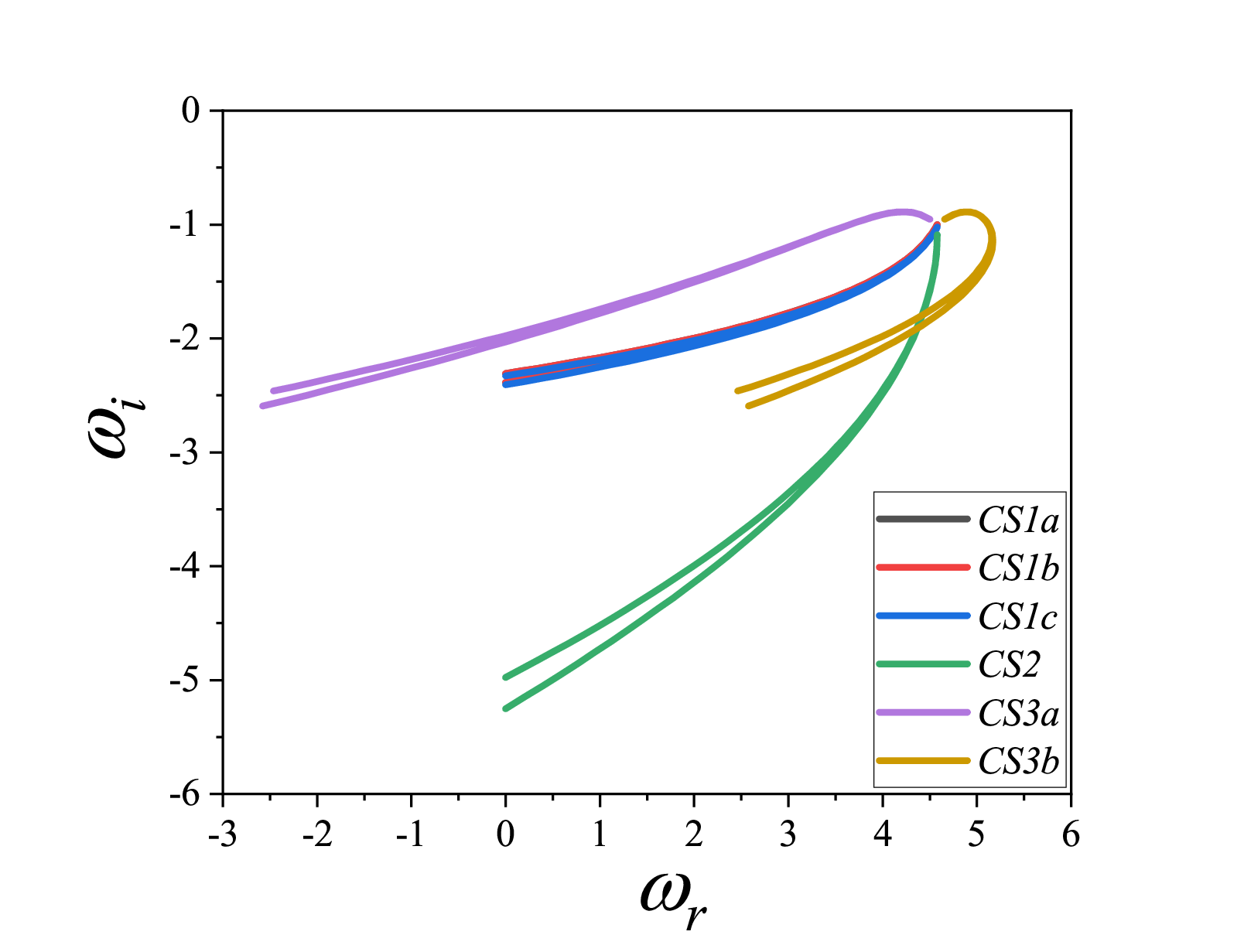}\label{fig:L_100_dean_axis_ana_n1}}
  \subfigure[$n = 1$ (numerical)]
  {\includegraphics[width=0.35\linewidth]{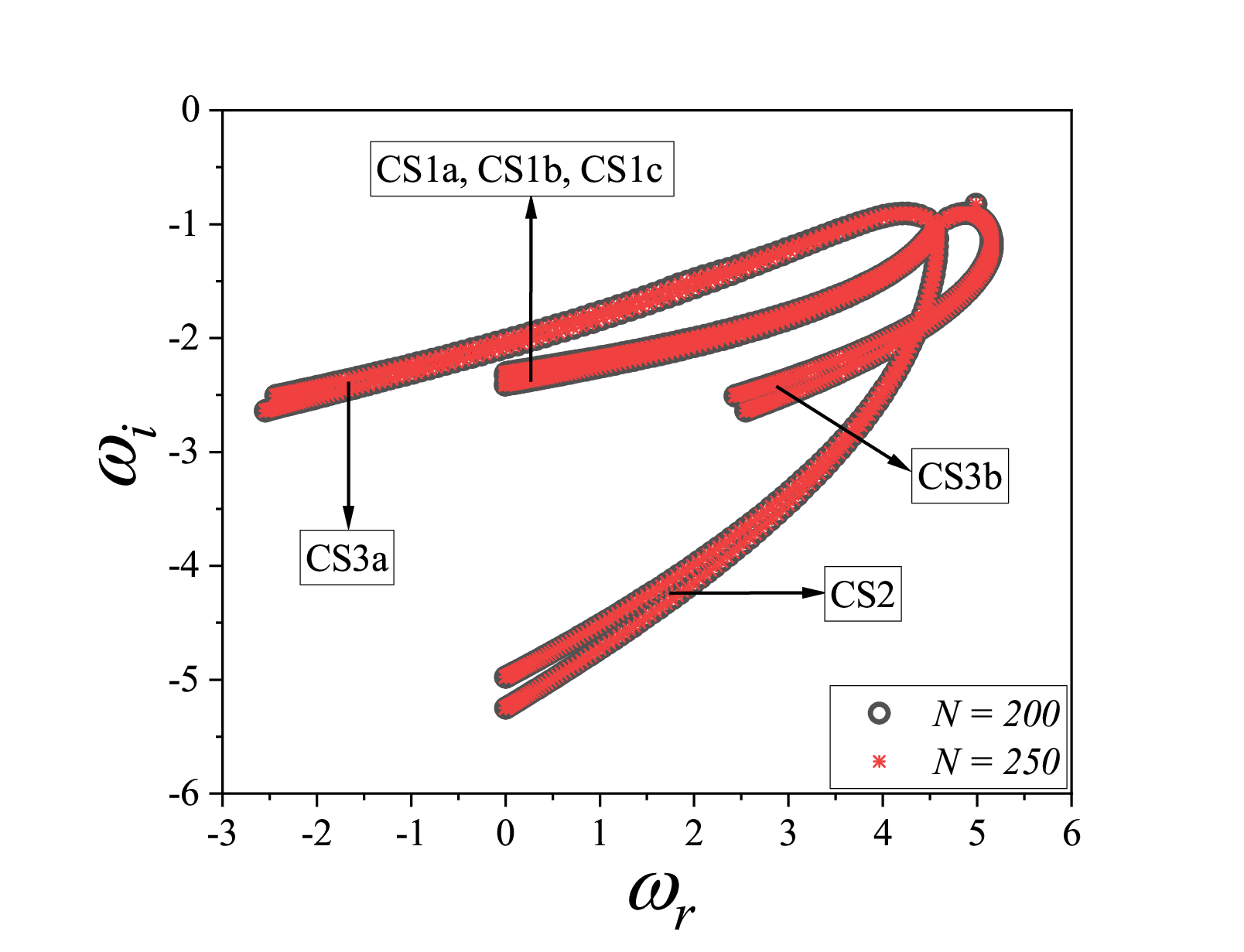}\label{fig:L_100_dean_axis_num_n1}}
  \caption{Effect of $n$ on the  analytical CS and numerical spectra for Dean flow. Data for $\Wi = 200, L = 100, \beta = 0.98, \epsilon = 0.1, \alpha = 7$. (Dashed lines are used to enhance the visibility of overlapping CS).}
  \label{fig:Effect_of_n_Dean}
\end{figure} 
\begin{figure}
  \centering
    \subfigure[$\Wi = 10 $ (analytical)]
  {\includegraphics[width=0.35\linewidth]{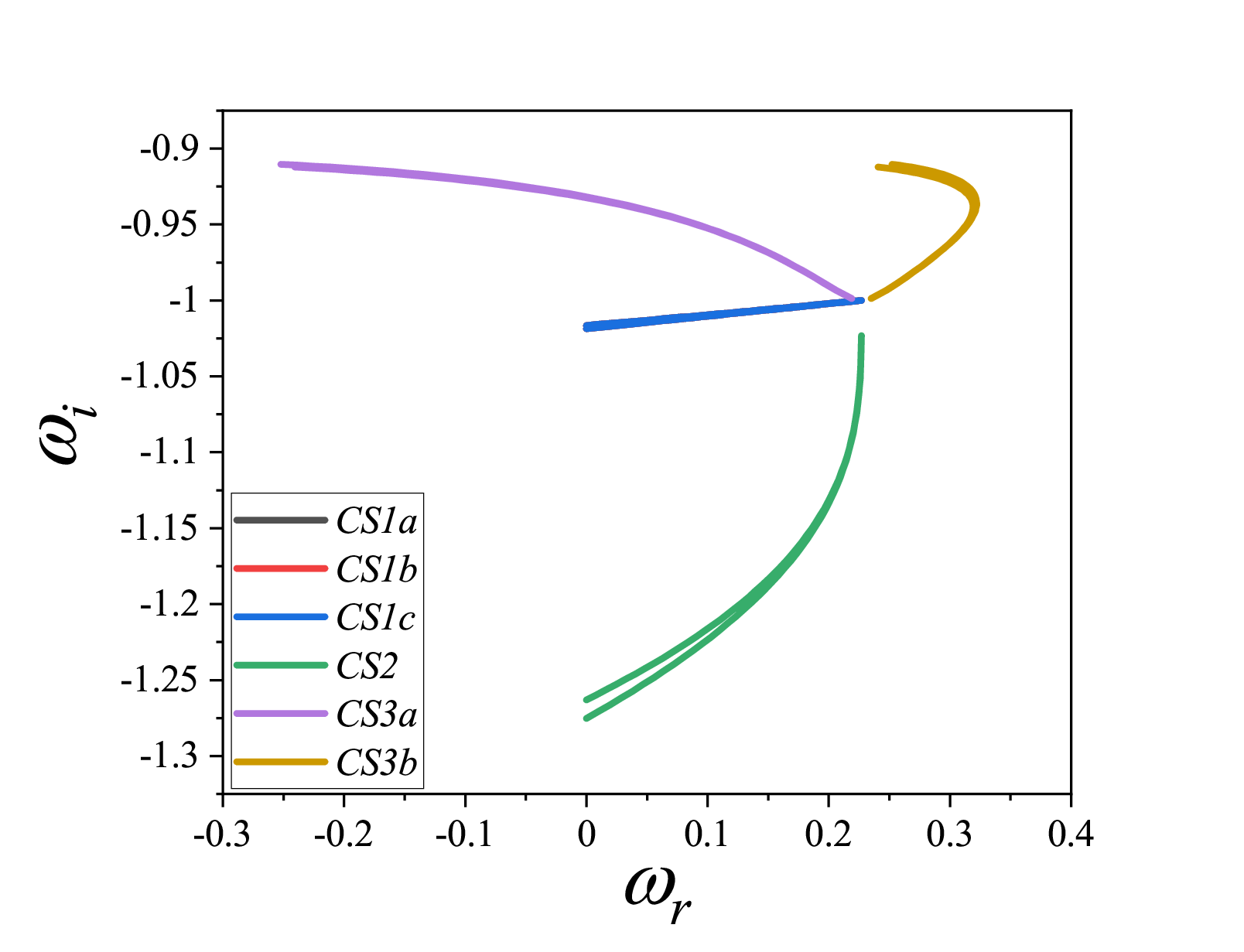}\label{fig:W_10_L_100_n1_dean_nonaxis_ana}}
    \subfigure[$\Wi = 10 $ (numerical)]
  {\includegraphics[width=0.35\linewidth]{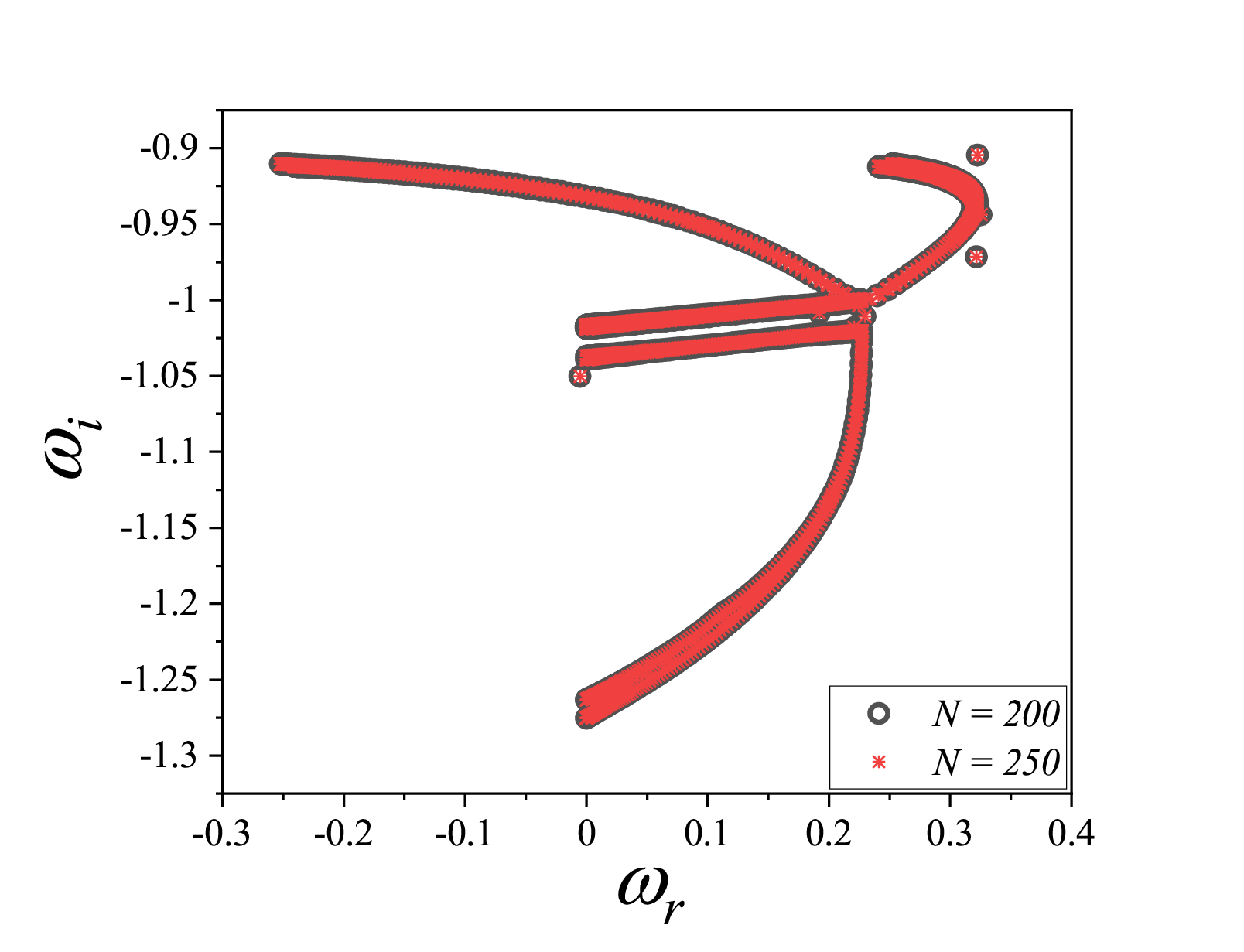}\label{fig:W_10_L_100_n1_dean_nonaxis_num}}
  \subfigure[$\Wi = 50 $ (analytical)]
  {\includegraphics[width=0.35\linewidth]{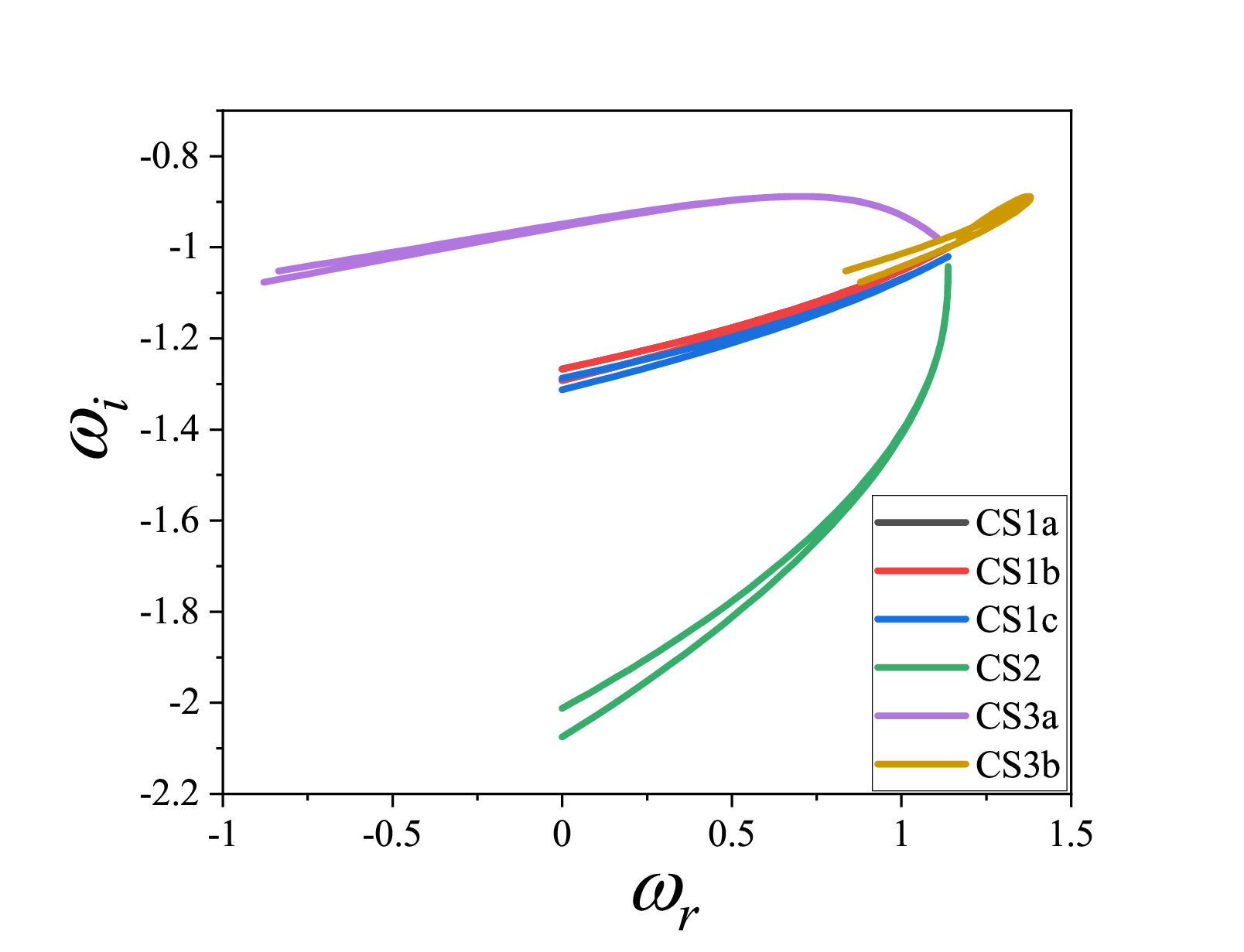}\label{fig:W_50_L_100_n1_dean_nonaxis_ana}}
    \subfigure[$\Wi = 50 $ (numerical)]
  {\includegraphics[width=0.35\linewidth]{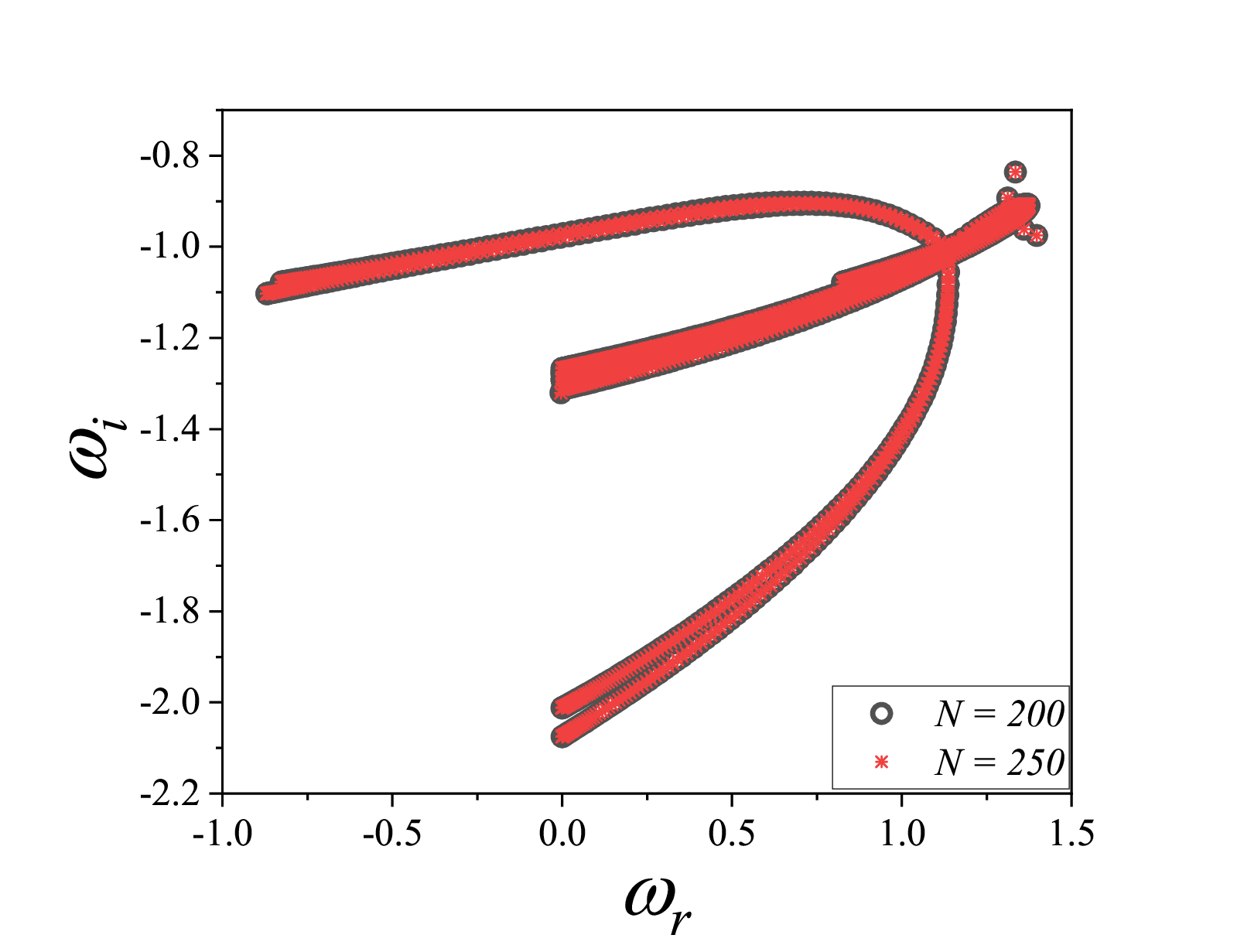}\label{fig:W_50_L_100_n1_dean_nonaxis_num}}
       \subfigure[$\Wi = 100 $ (analytical)]
  {\includegraphics[width=0.35\linewidth]{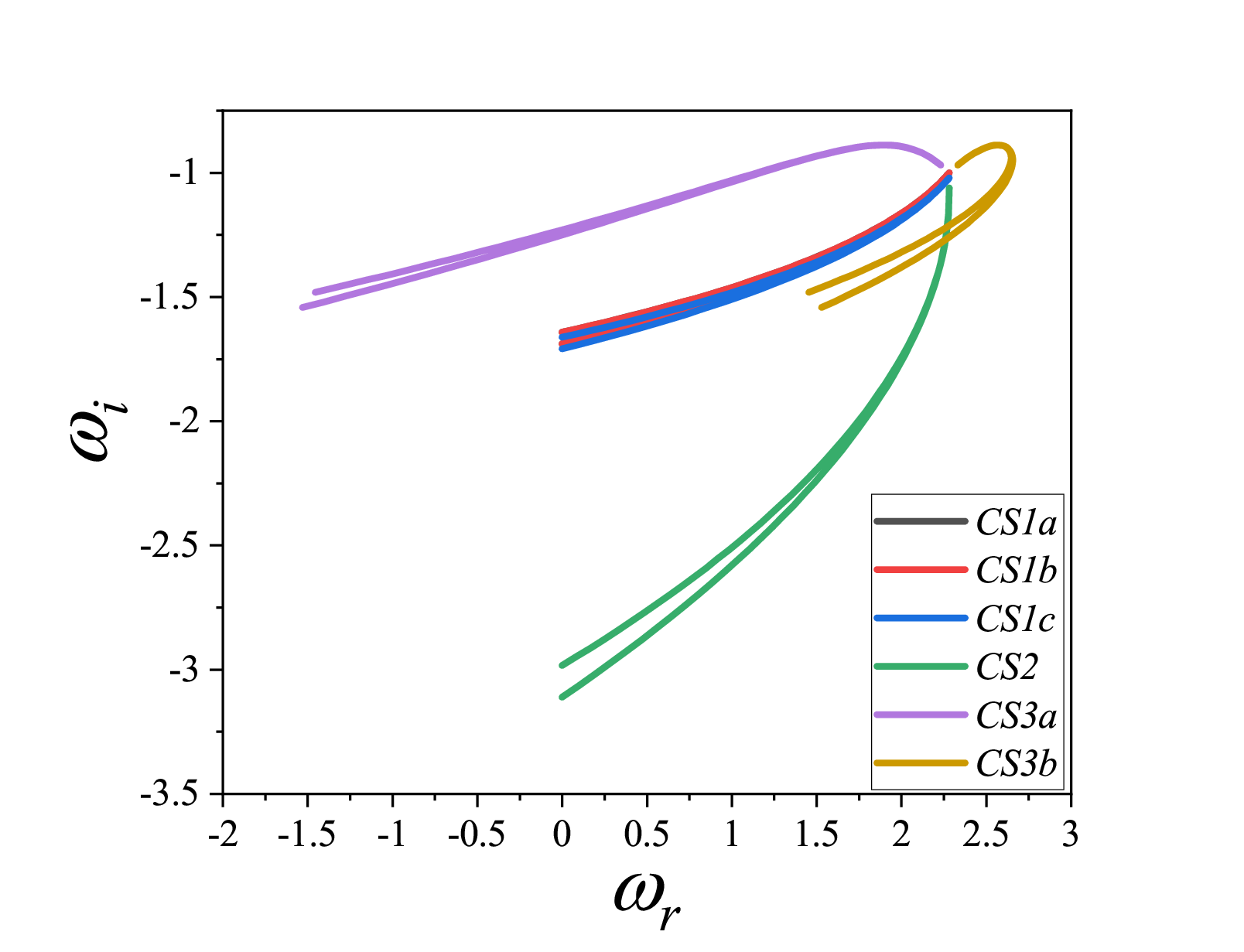}\label{fig:W_100_L_100_n1_dean_nonaxis_ana}}
    \subfigure[$\Wi = 100 $ (numerical)]
  {\includegraphics[width=0.35\linewidth]{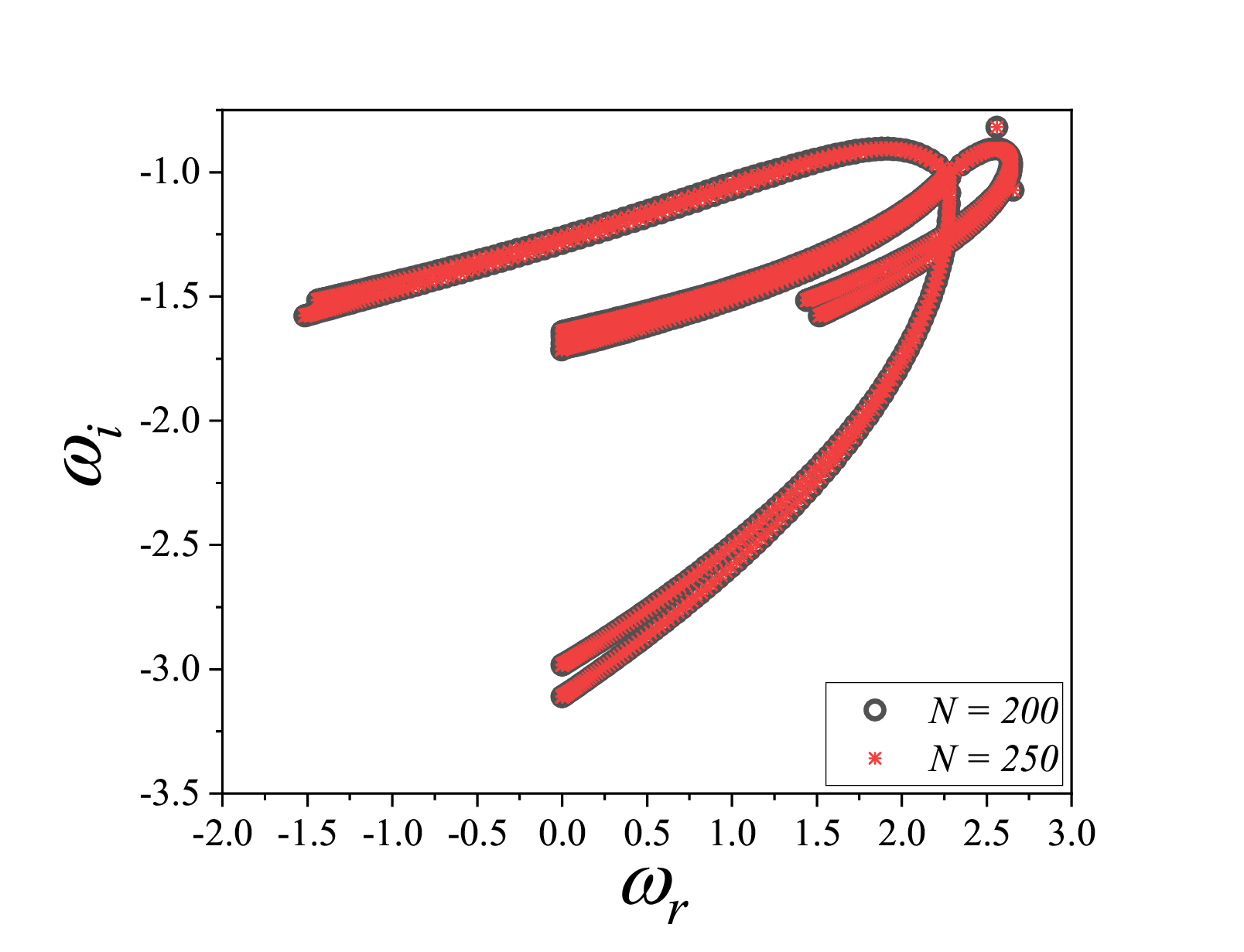}\label{fig:W_100_L_100_n1_dean_nonaxis_num}}
      \subfigure[$\Wi = 1000 $ (analytical)]
  {\includegraphics[width=0.35\linewidth]{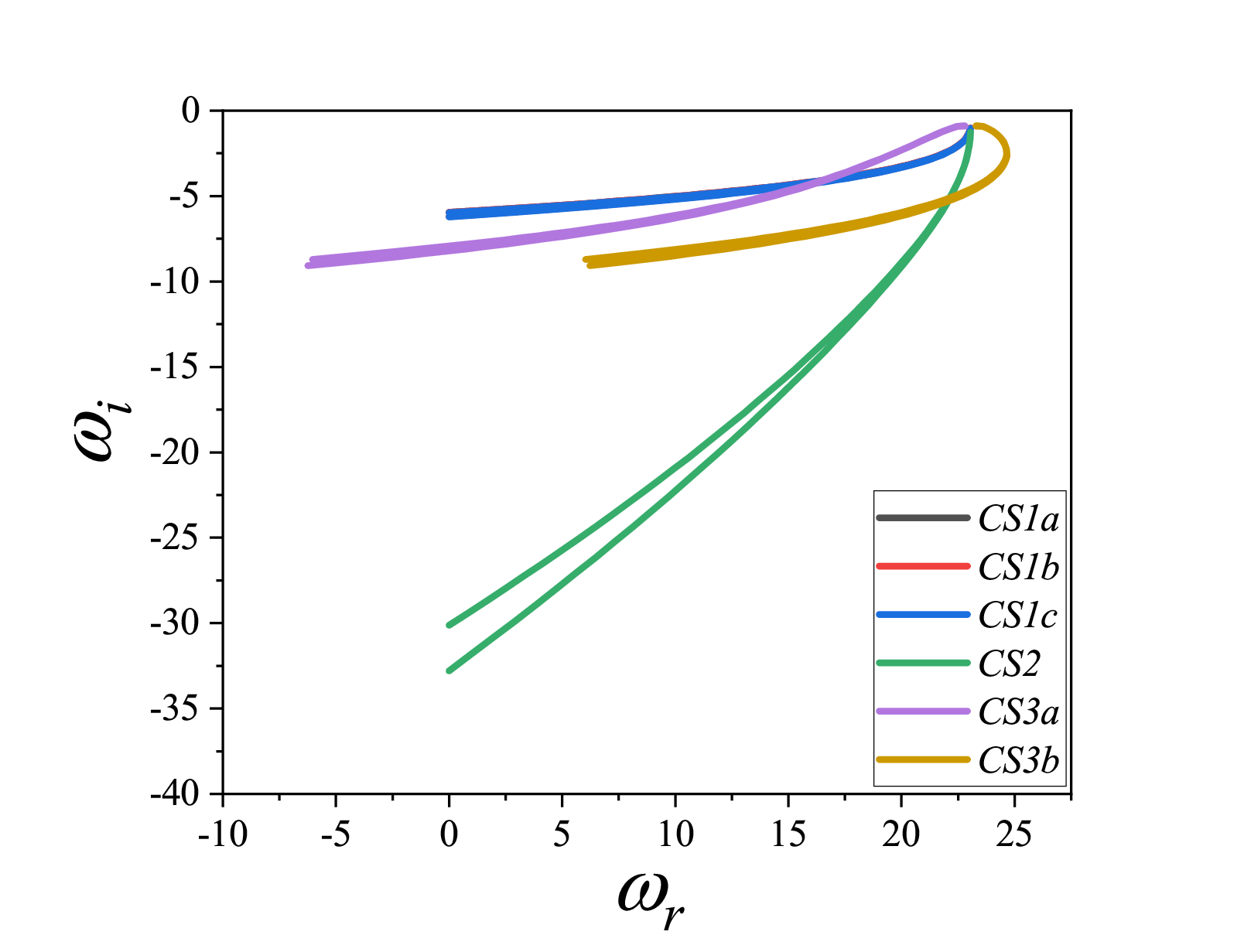}\label{fig:W_1000_L_100_n1_dean_nonaxis_ana}}
    \subfigure[$\Wi = 1000 $ (numerical)]
  {\includegraphics[width=0.35\linewidth]{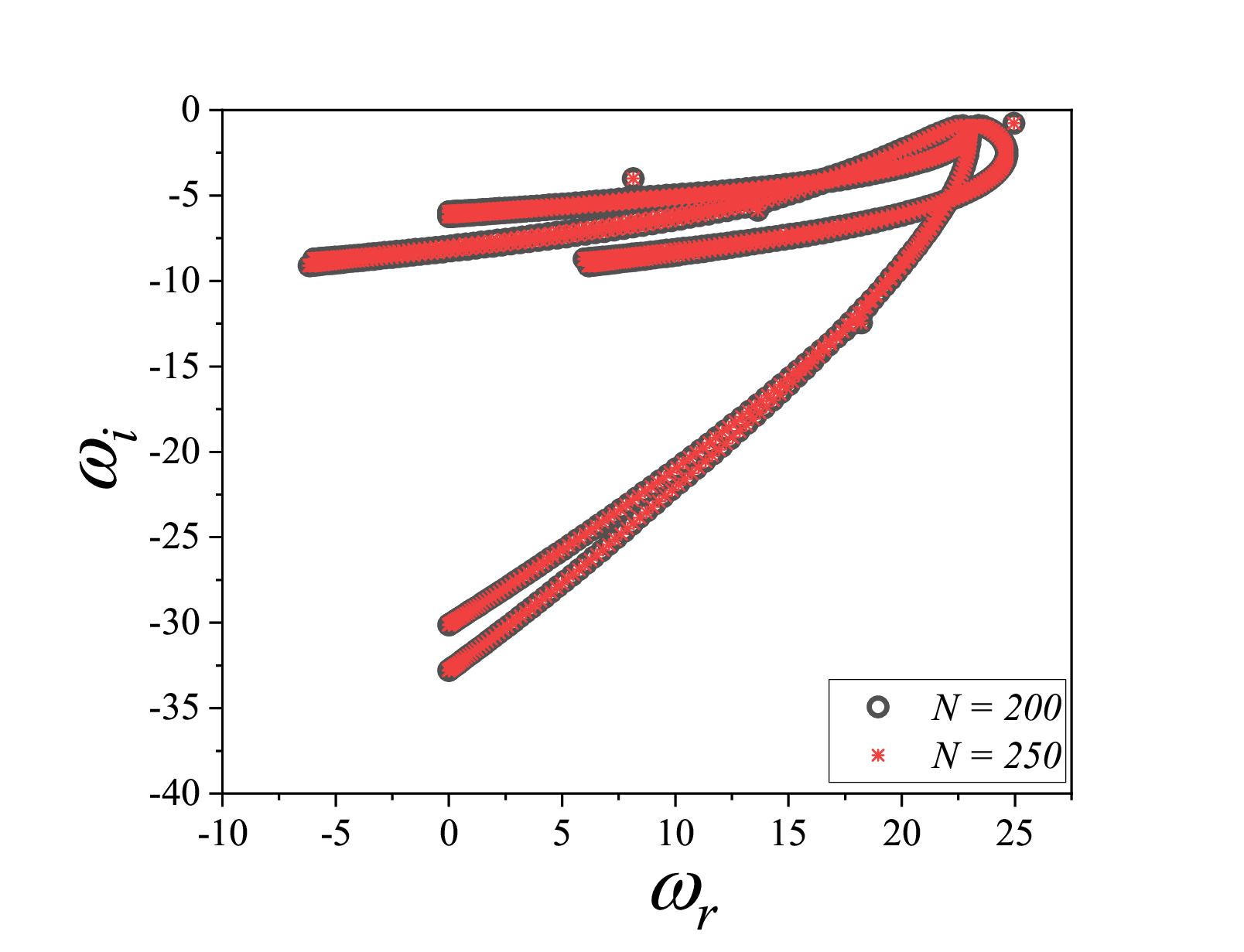}\label{fig:W_1000_L_100_n1_dean_nonaxis_num}}
  \caption{Analytical CS and numerical spectra for Dean flow. Data for $L = 100, \beta = 0.98, n = 1, \epsilon = 0.1, \alpha = 7$, for various $\Wi$.}
  \label{fig:Effect_of_Wi_dean_nonaxis}
\end{figure}
A comparison of  the numerically obtained spectra with the analytical CS locations,  
in Fig.\,\ref{fig:Effect_of_n_Dean}, shows very good agreement between the two, both for $n = 0$ and $n = 1$. 
Figure\,\ref{fig:Effect_of_Wi_dean_nonaxis} shows that the number of discrete modes varies as $\Wi$ is increased at a fixed $L$. 
\begin{figure}
  \centering
  \subfigure[$\beta = 0.5, \epsilon = 1$]
  {\includegraphics[width=0.35\linewidth]{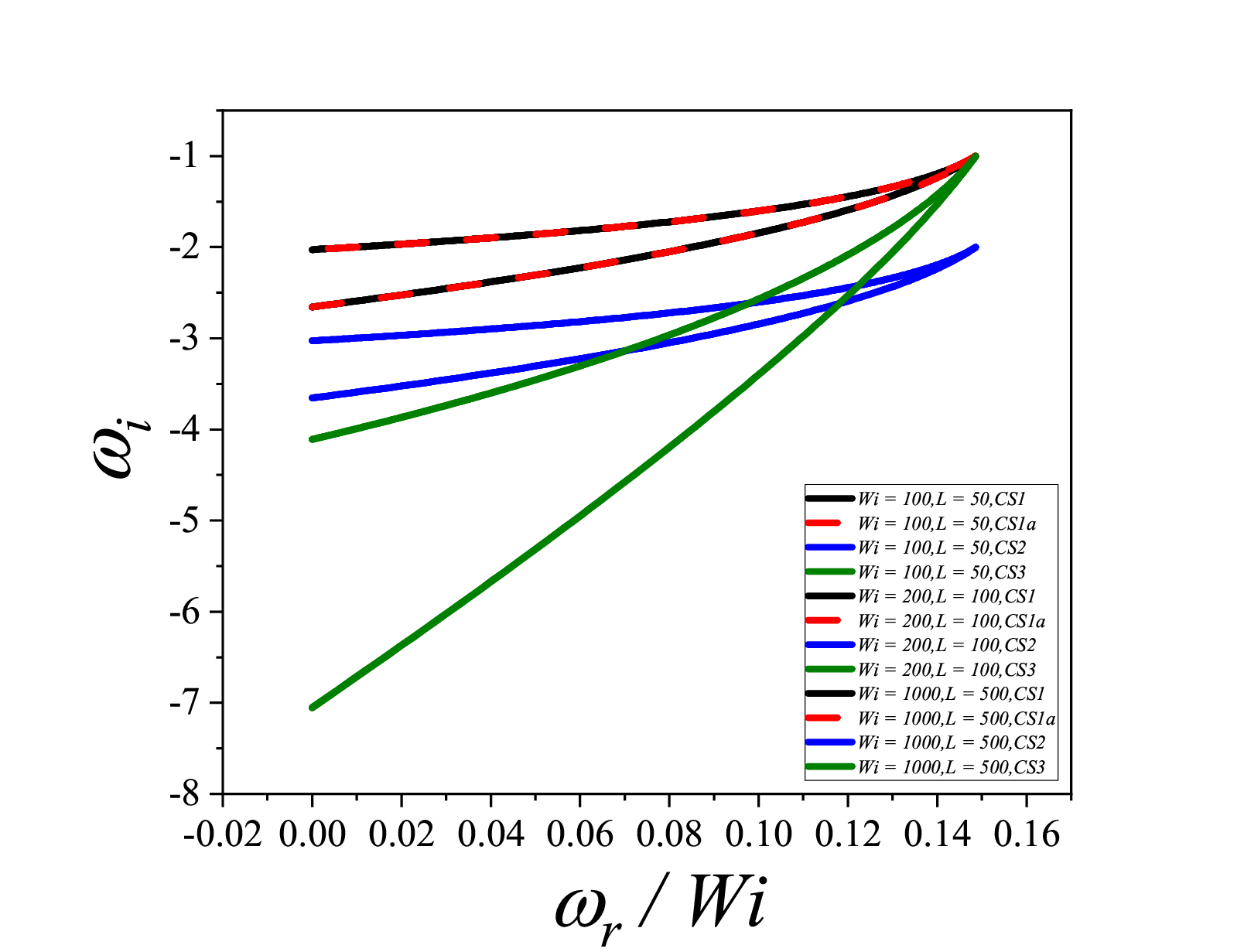}}
    \subfigure[$\beta = 0.98, \epsilon = 0.1$]
  {\includegraphics[width=0.35\linewidth]{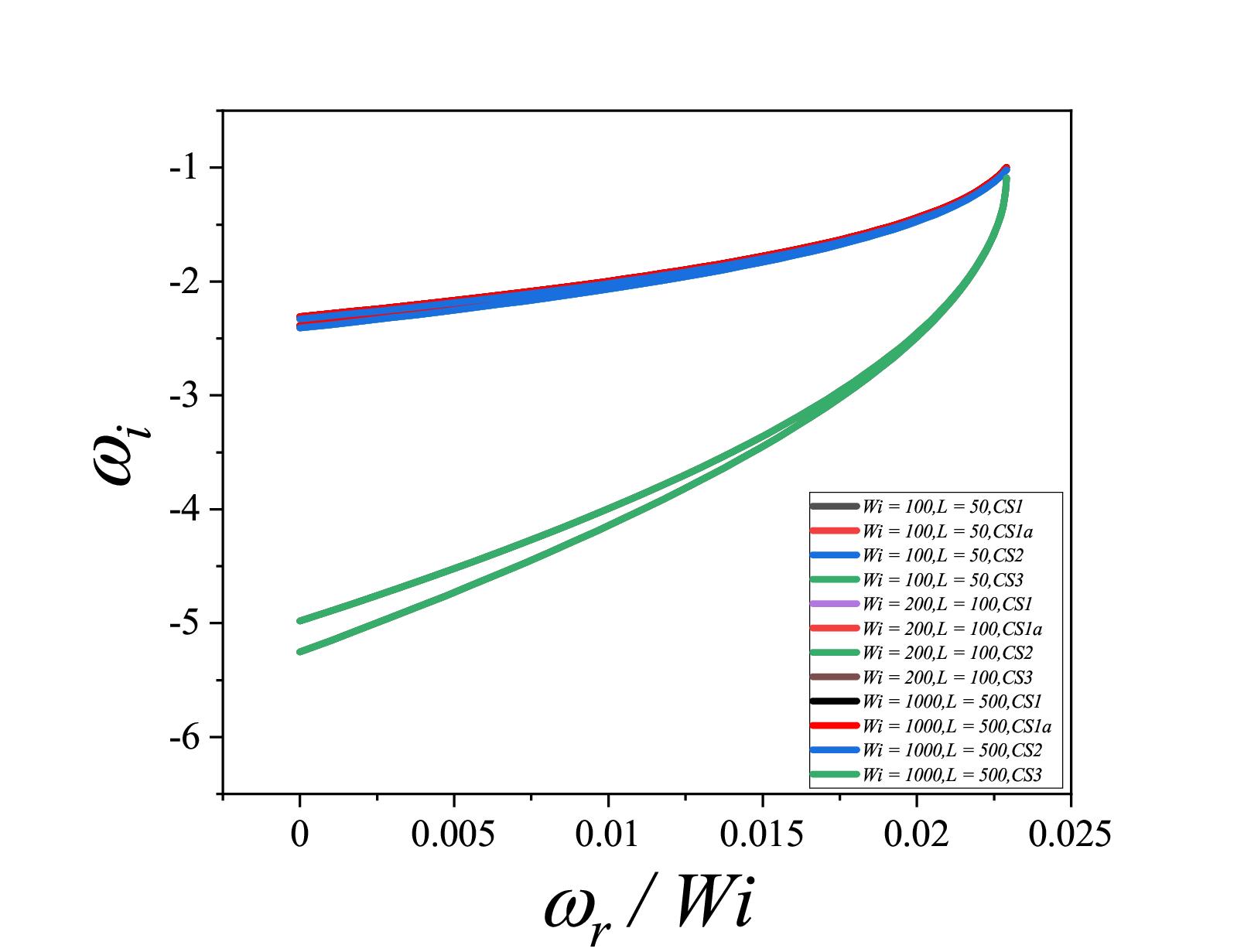}}
  \caption{Collapse of analytical CS for various ($\Wi,L$) pairs at $n = 1$ for Dean flow except CS3 which does not show a collapse. (Dashed lines are used to enhance the visibility of overlapping CS).}
  \label{fig:Collapse_dean_nonaxis}
\end{figure}
\begin{figure}
  \centerline{\includegraphics[width=0.35\linewidth]{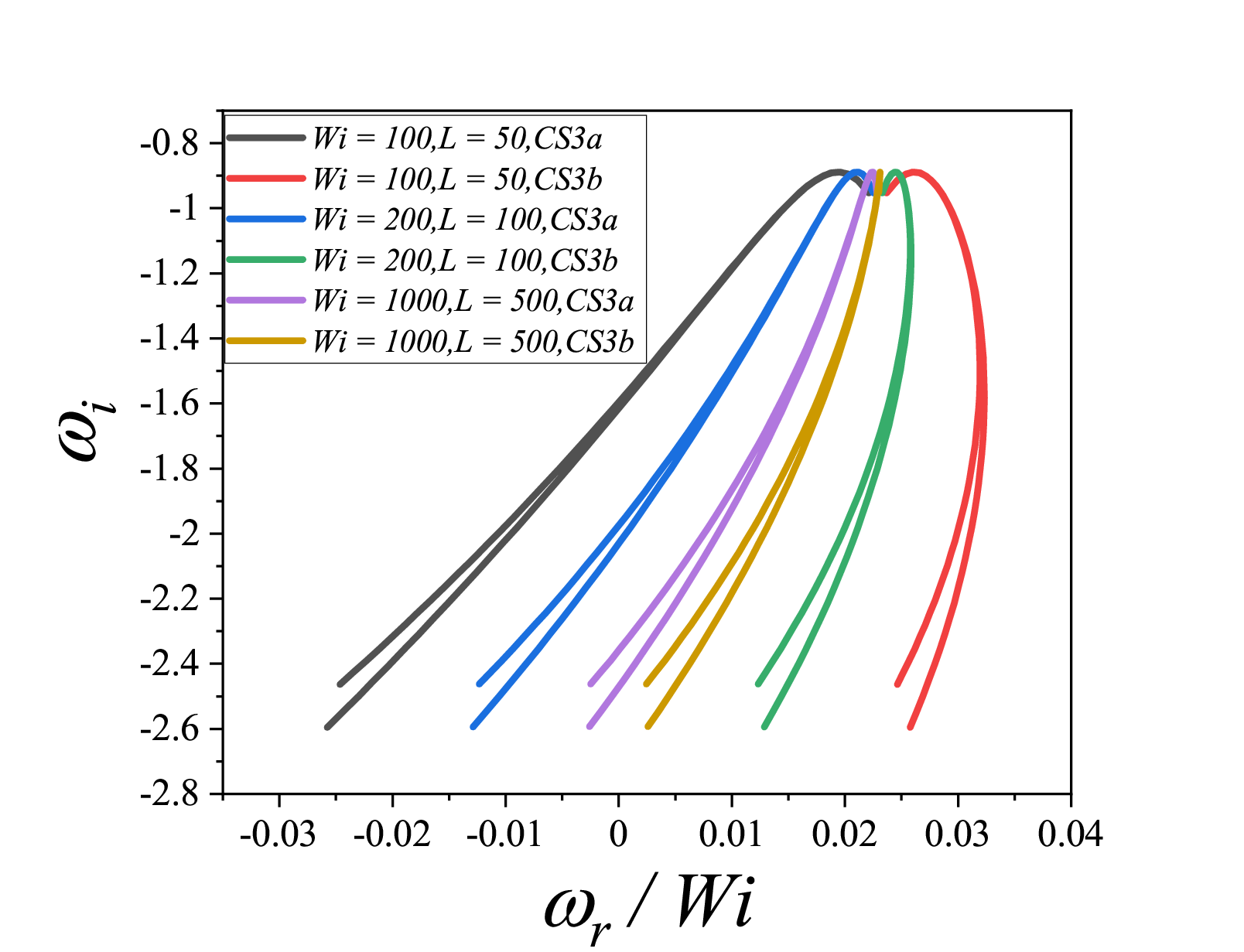}}
  \caption{Analytical predictions for CS3a, CS3b for different $\Wi,L$ pairs (with $\Wi/L = 2$) demonstrating the lack of collapse for CS3(a,b). Data for $\beta = 0.98, \epsilon = 0.1, n = 1, \alpha = 7$ for Dean flow.}
  \label{fig:Collapse_dean_CS4}
\end{figure}
Figure\,\ref{fig:Collapse_dean_nonaxis} shows the collapse of the theoretical CS in Dean flow for various $(\Wi, L)$ pairs, a feature which was demonstrated earlier for rectilinear flows in {Sec.\,\ref{sec:rectilinear_PC} and \ref{sec:rectilinear_PP}.  The collapse is only obtained after rescaling $\omega_r$ with $\Wi$. Here again, the horizontal extent of the CS increases with increase in $\Wi/L$. This is a consequence of using the maximum velocity of the base velocity profile for an Oldroyd-B fluid for the same pressure gradient, and thence the base-state maximum increases with increase in $\Wi/L$ due to shear thinning.
One exception for the collapse is CS3(a,b), as shown in Fig.\,\ref{fig:Collapse_dean_CS4}, which does not follow this trend for $n \neq 0$.

\begin{figure}
\centering
    \subfigure[$n = 0$]
{\includegraphics[width=0.35\textwidth]{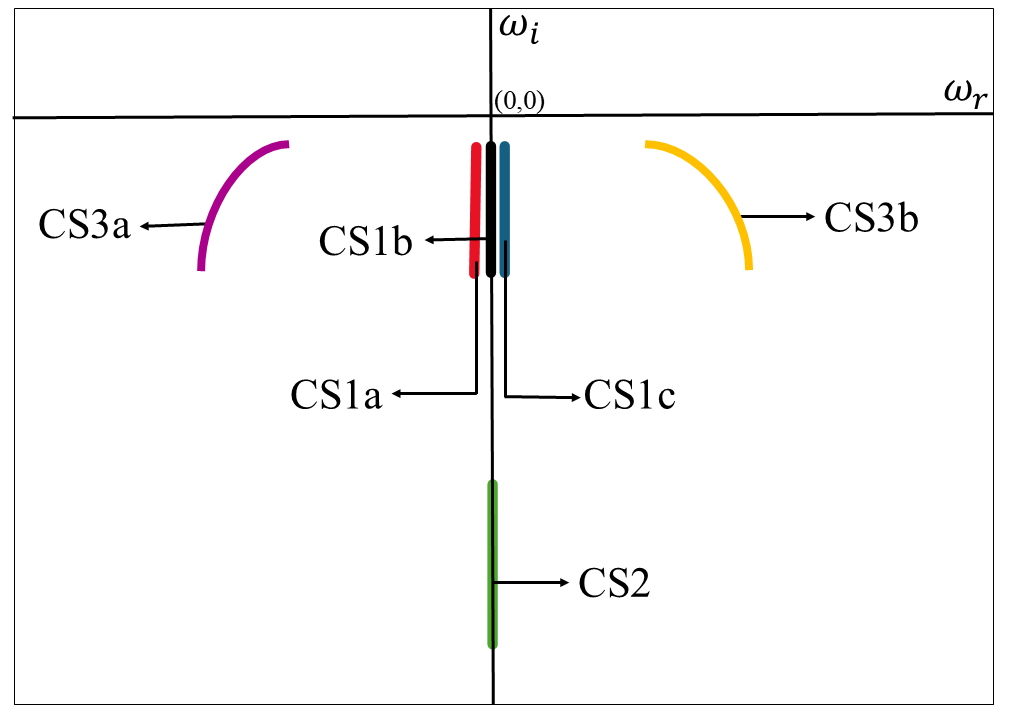}\label{fig:TC_axis_n_0_schematic}}
\quad \quad
\subfigure[$n \neq 0$]
{\includegraphics[width=0.32\textwidth]{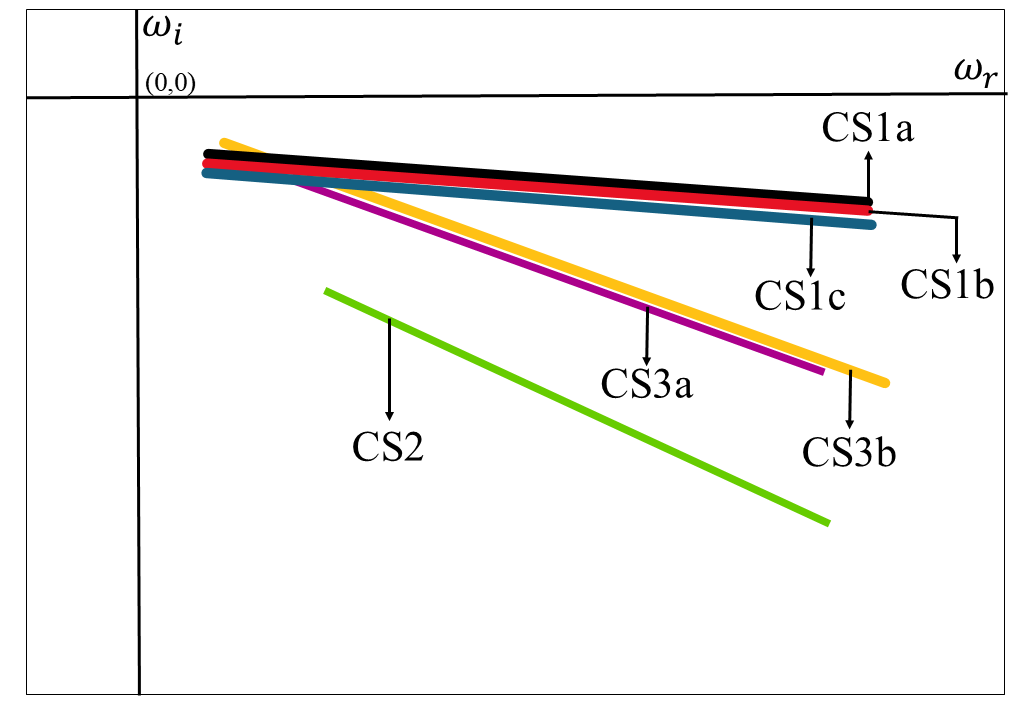}\label{fig:TC_nonaxis_n_1_schematic}}
  \caption{Schematics of the various CS for Taylor Couette flow.}
  \label{fig:TC_CS_schematics}
\end{figure}

 \subsection{Taylor--Couette flow}
 \label{sec:TaylorCouette}

We first show the schematic loci of the various CS in Taylor-Couette flow in Fig.\,\ref{fig:TC_CS_schematics} for both $n = 0$ and $n \neq 0$. In Taylor-Couette flow, again, there are six CS modes, with CS1(a--c) being nearly identical for $L > 50$. The modes corresponding to CS3a and 3b no longer have a complete wing-like structure, but instead appear as `truncated' wings symmetrically placed at a distance about the imaginary axis, while CS1(a-c) and CS2 are line segments on the imaginary axis. It is tempting to conjecture that the gap between the two truncated wings should be related to the absence of a base-state maximum for the Taylor-Couette velocity profile.
For nonzero $n$, all the CS again become extended due to the horizontal shift factor, and stretch out to the right-half plane.  In contrast to the CS for 
$n \neq 0$ in Dean flow (Fig.\,\ref{fig:CS_schematics_Dean_n_1}), each of the non-degenerate CS in Taylor-Couette flow is characterized only by a single branch (Fig.\,\ref{fig:TC_nonaxis_n_1_schematic}), on account of the monotonicity of the velocity profile. In the latter figure, the downward tilt of the CS (from left to right) is related to the configuration wherein the inner cylinder is rotating and the outer one being stationary. The orientation of the CS tilt will reverse if one were to consider the outer cylinder rotating and the inner one stationary.

 \subsubsection{Axissymmetric disturbances}
 \label{subsec:AxiTaylor}

Figures\,\ref{fig:Effect_of_L_TC_axis_e0.1} and \ref{fig:Effect_of_L_TC_axis_e1} show the comparison of the theoretically predicted CS and the numerically obtained spectra for $\epsilon = 0.1$ and $1$ respectively. The various line segments and truncated wings representing the CS are approximated well in the full numerical spectrum. There are many discrete modes that  emanate from CS2 and CS3(a,b) as $L$ is increased from $100$ to $500$ (at a fixed $\Wi = 200$, implying a decrease in $\Wi/L$). For $\epsilon = 0.1$, CS1(a--c) and CS2 are well separated on the imaginary axis (Fig.\,\ref{fig:Effect_of_L_TC_axis_e0.1}), while they are overlapping for $\epsilon = 1$ (Fig.\,\ref{fig:Effect_of_L_TC_axis_e1}). Further, the the line segments and the truncated wings are significantly longer for $\epsilon = 1$. With a further decrease in $\Wi/L$, the lengths of these line segments decrease and eventually reduce to a point in the Oldroyd-B limit (data not shown).

Next, in Fig.\,\ref{fig:TC_effect_of_beta_and_W_by_L_CS4}a,
we demonstrate the effect of $\beta$ and $\Wi/L$ on CS3a and CS3b. 
Similar to plane Couette flow, at higher $\beta$'s ($\beta > 0.4$), CS3a and CS3b are located
symmetrically on either side of the imaginary axis, albeit being truncated wings instead of points. As $\beta$ is decreased gradually from 0.98, the CS3a and CS3b wings shift downward and towards the imaginary axis. At a certain critical $\beta$, they merge and then bifurcate into vertical line segments aligned along the purely imaginary axis. This is analogous to plane Couette flow,  where CS3a and 3b are points, but exhibit a similar bifurcation as $\beta$ is decreased.  Although CS3a and CS3b appear as symmetric wings  for $\Wi/L \sim O(1)$ (Fig.\,\ref{fig:TC_effect_of_beta_and_W_by_L_CS4}b), as we approach the Oldroyd-B limit ($\Wi/L \gg 1$), CS3(a,b) shifts upward toward the imaginary axis and eventually become degenerate with CS1(a--c)  with $\omega_i = -1$, again similar to the behavior observed in plane Couette flow.
 
\begin{figure}
  \centering
  \subfigure[$L = 10^2$ (analytical)\label{fig:L_100_e0.1_TC_axis_ana}]
  {\includegraphics[width=0.35\linewidth]{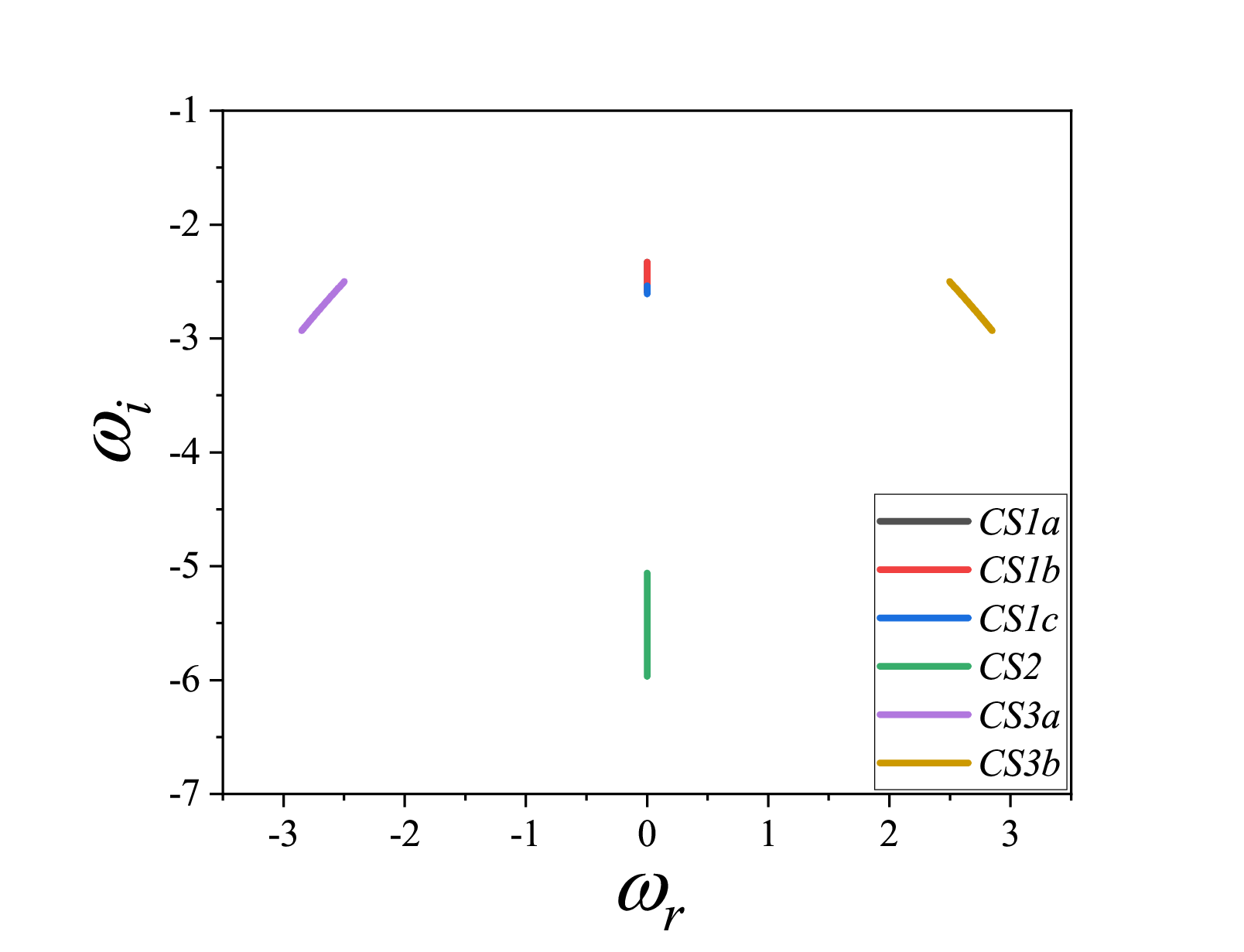}}
  \subfigure[$L = 10^2$ (numerical)\label{fig:L_100_e0.1_TC_axis_num}]
  {\includegraphics[width=0.35\linewidth]{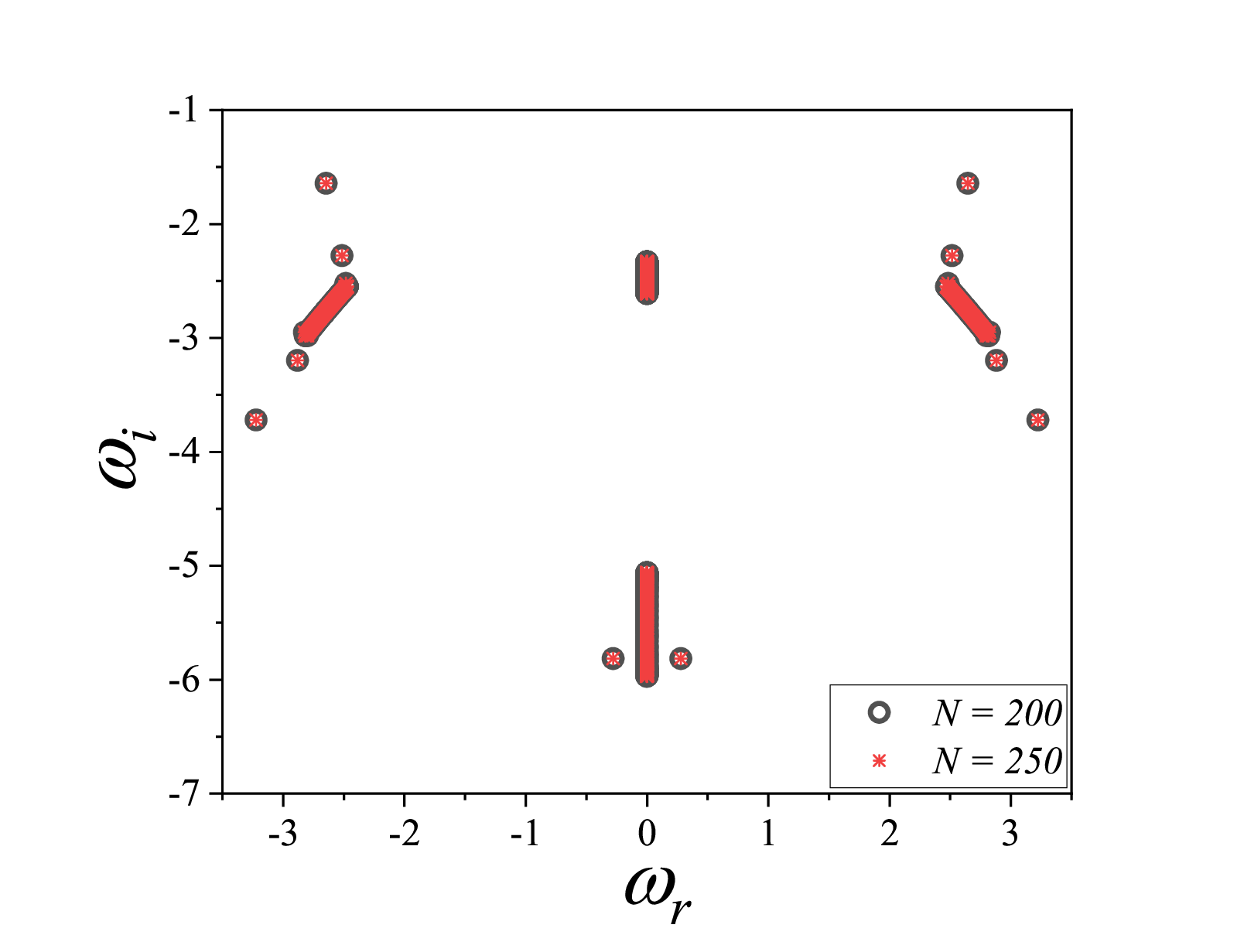}}
  \subfigure[$L = 5 \times 10^2$ (analytical)\label{fig:L_500_e0.1_TC_axis_ana}]
  {\includegraphics[width=0.35\linewidth]{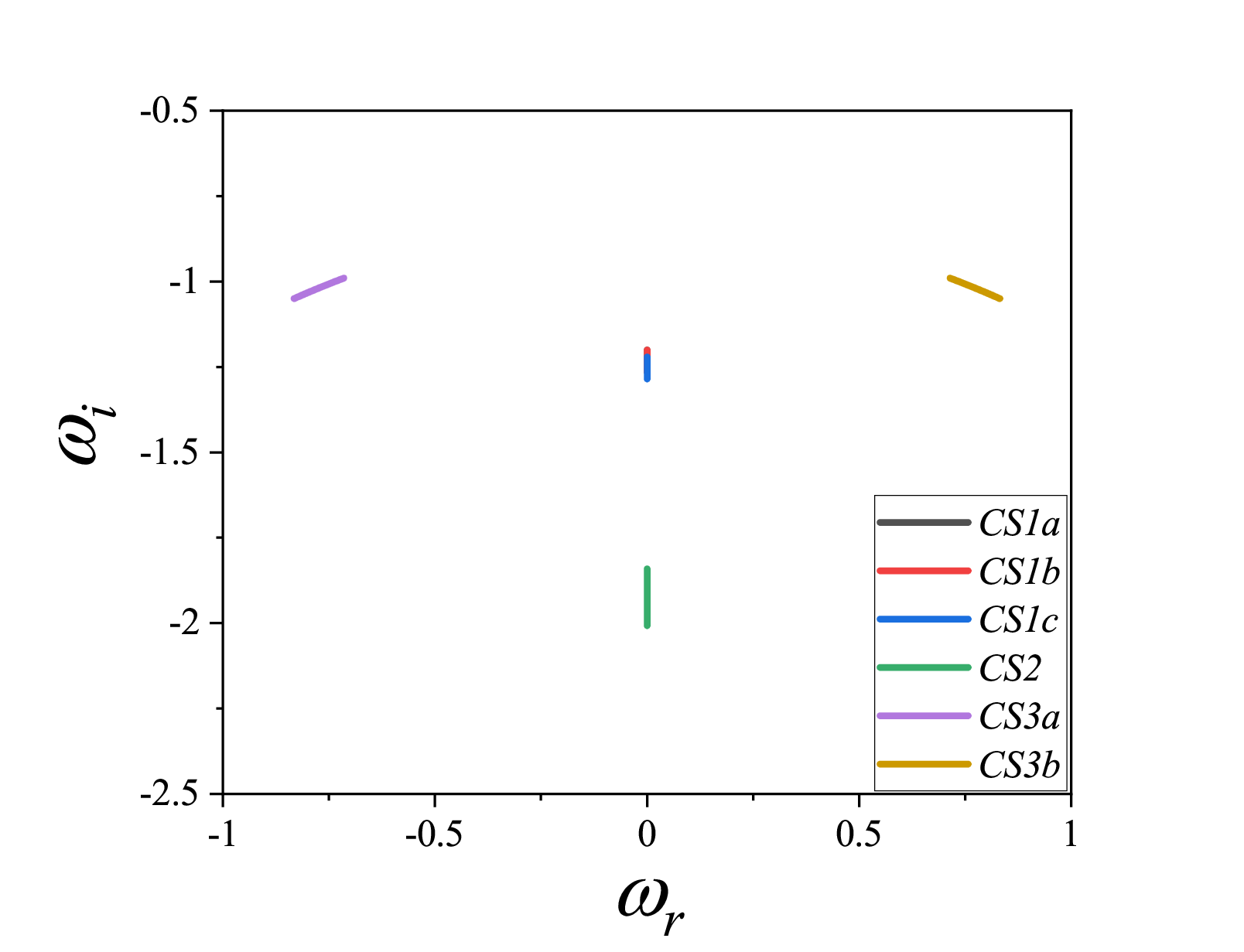}}
  \subfigure[$L = 5 \times 10^2$ (numerical)\label{fig:L_500_e0.1_TC_axis_num}]
  {\includegraphics[width=0.35\linewidth]{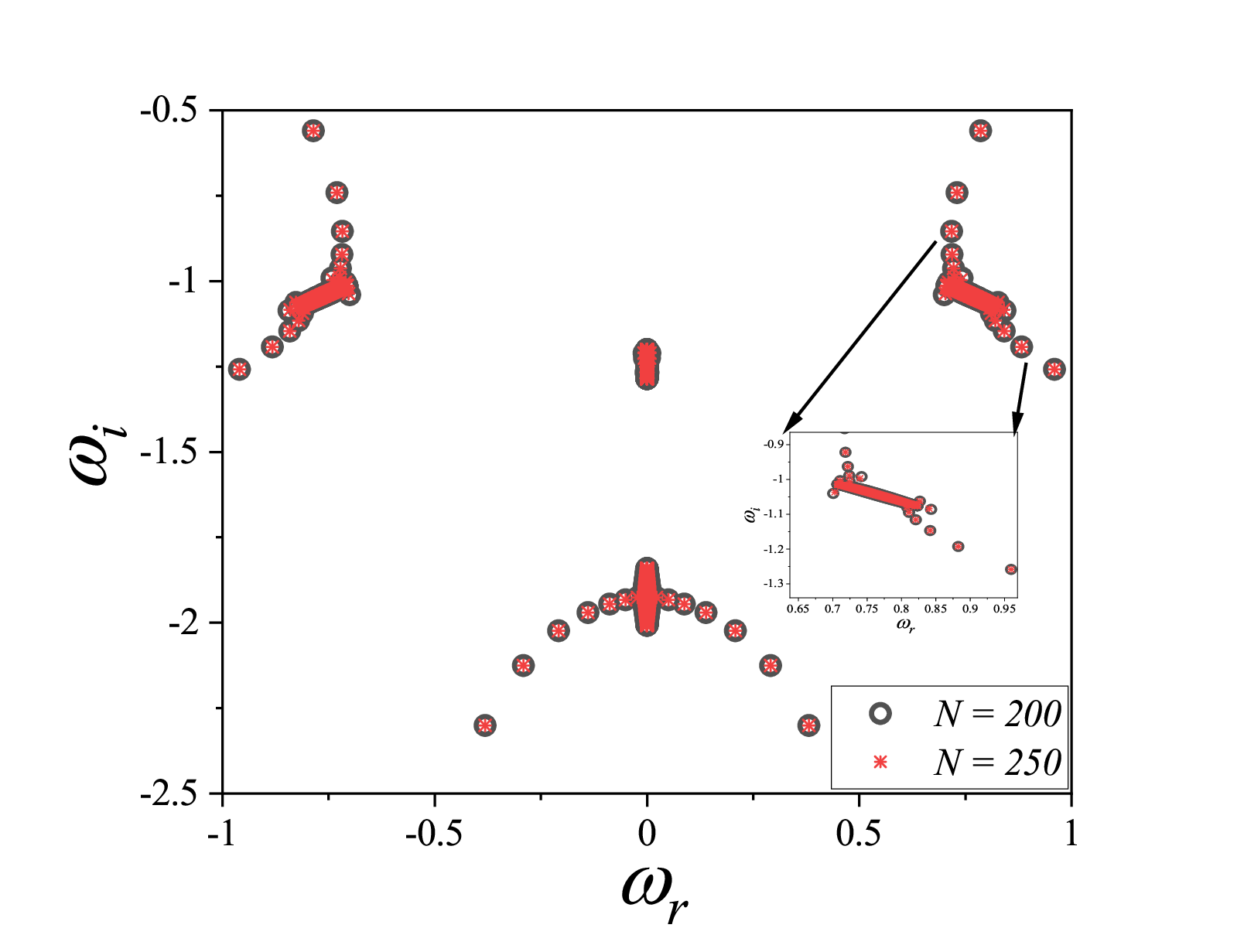}}
  \caption{Analytical CS and numerical spectra for Taylor-Couette flow for $\Wi = 200, \beta = 0.98, \epsilon = 0.1, \alpha = 7$.}
  \label{fig:Effect_of_L_TC_axis_e0.1}
\end{figure}

\begin{figure}
  \centering
  \subfigure[$L = 10^2$ (analytical)\label{fig:L_100_e1_TC_axis_ana}]
  {\includegraphics[width=0.35\linewidth]{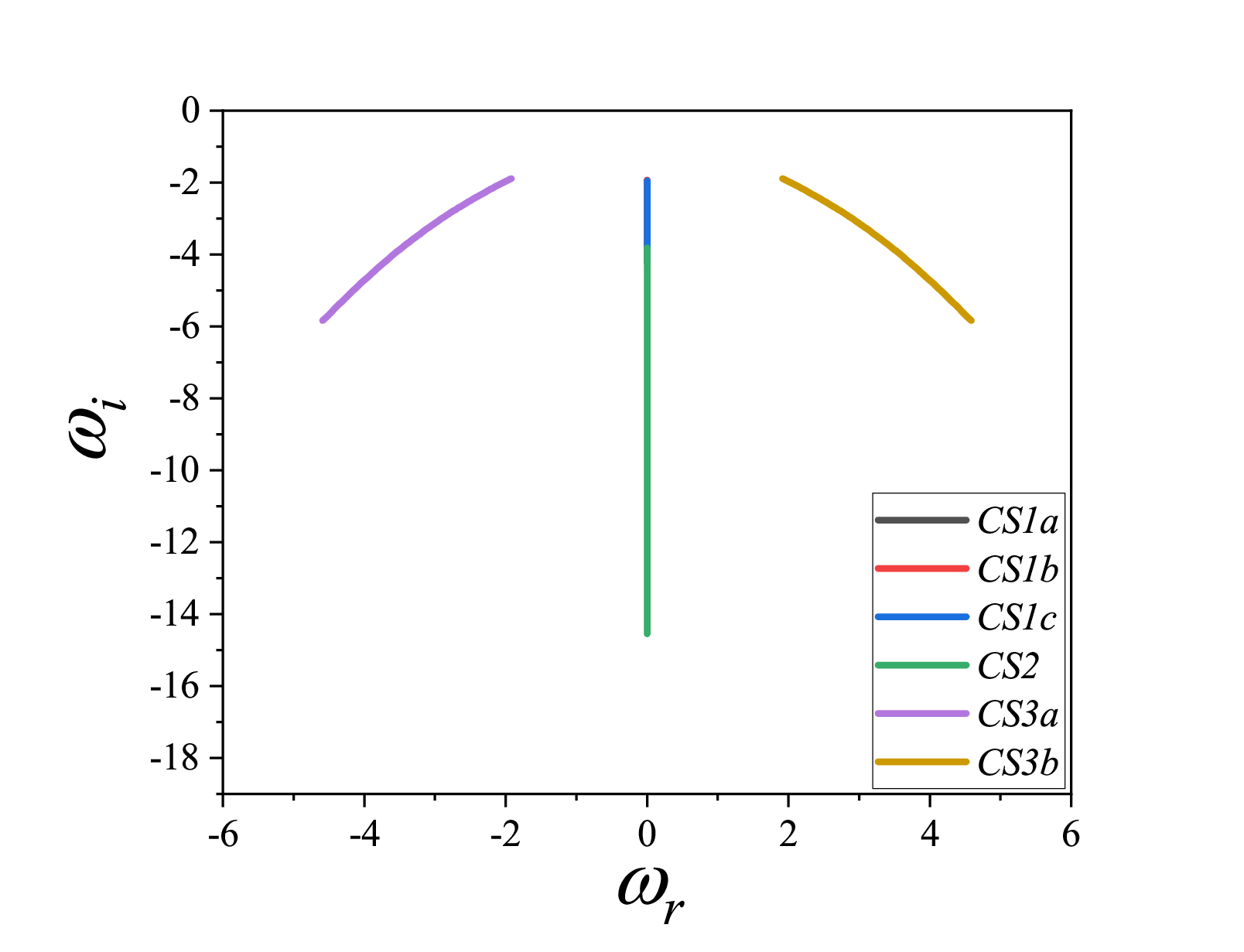}}
  \subfigure[$L = 10^2$ (numerical)\label{fig:L_100_e1_TC_axis_num}]
  {\includegraphics[width=0.35\linewidth]{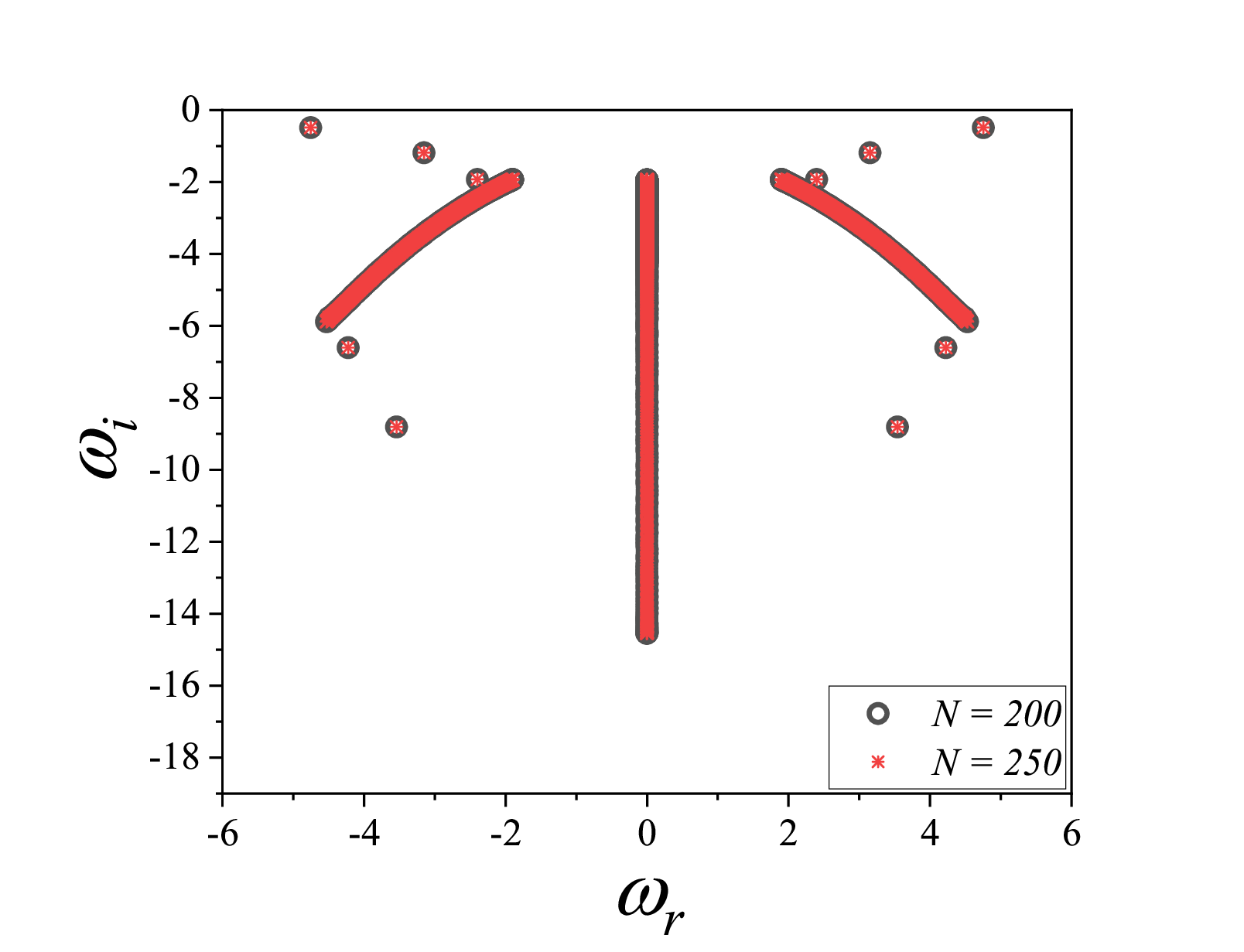}}
  \subfigure[$L = 5 \times 10^2$ (analytical)\label{fig:L_500_e1_TC_axis_ana}]
  {\includegraphics[width=0.35\linewidth]{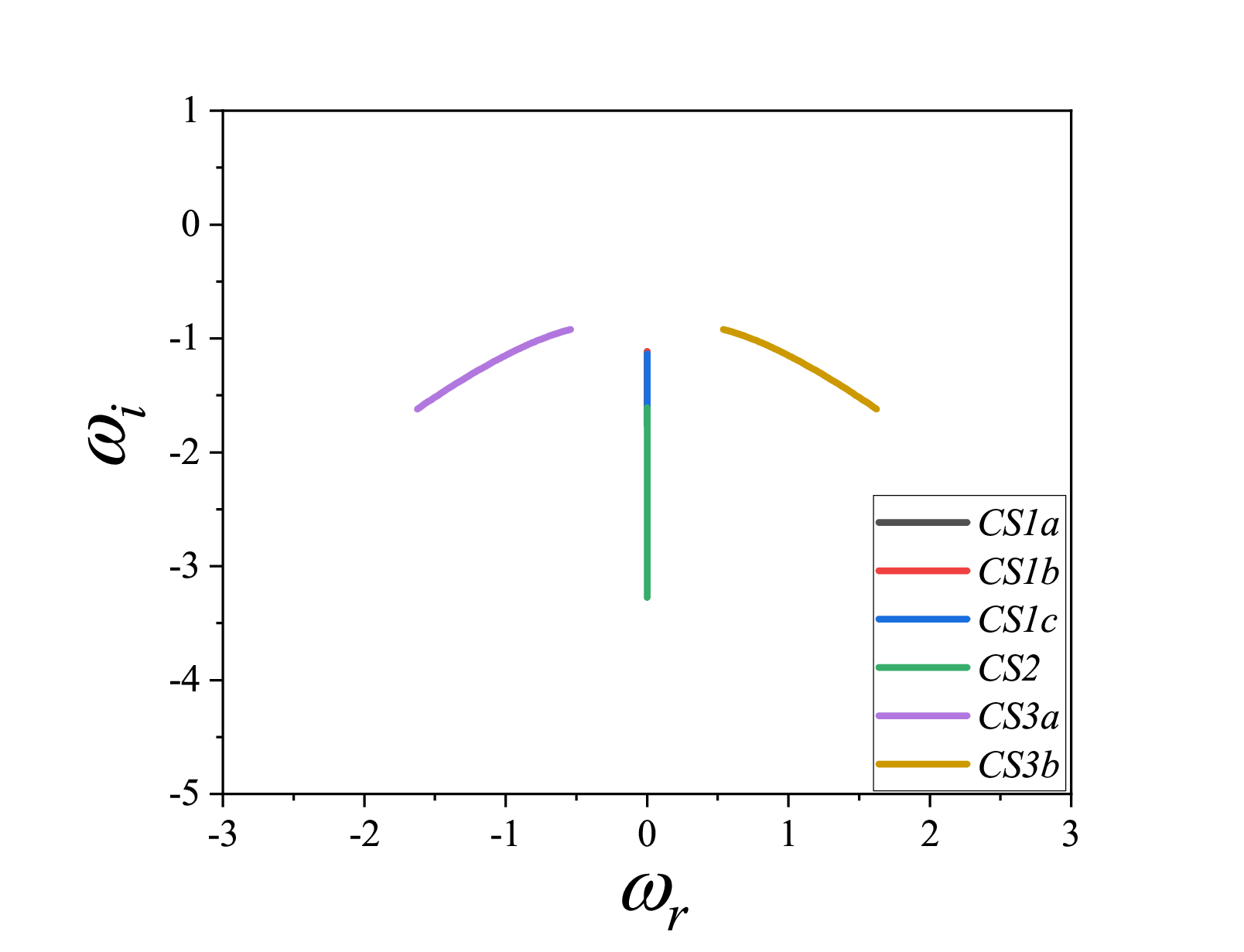}}
  \subfigure[$L = 5 \times 10^2$ (numerical)\label{fig:L_500_e1_TC_axis_num}]
  {\includegraphics[width=0.35\linewidth]{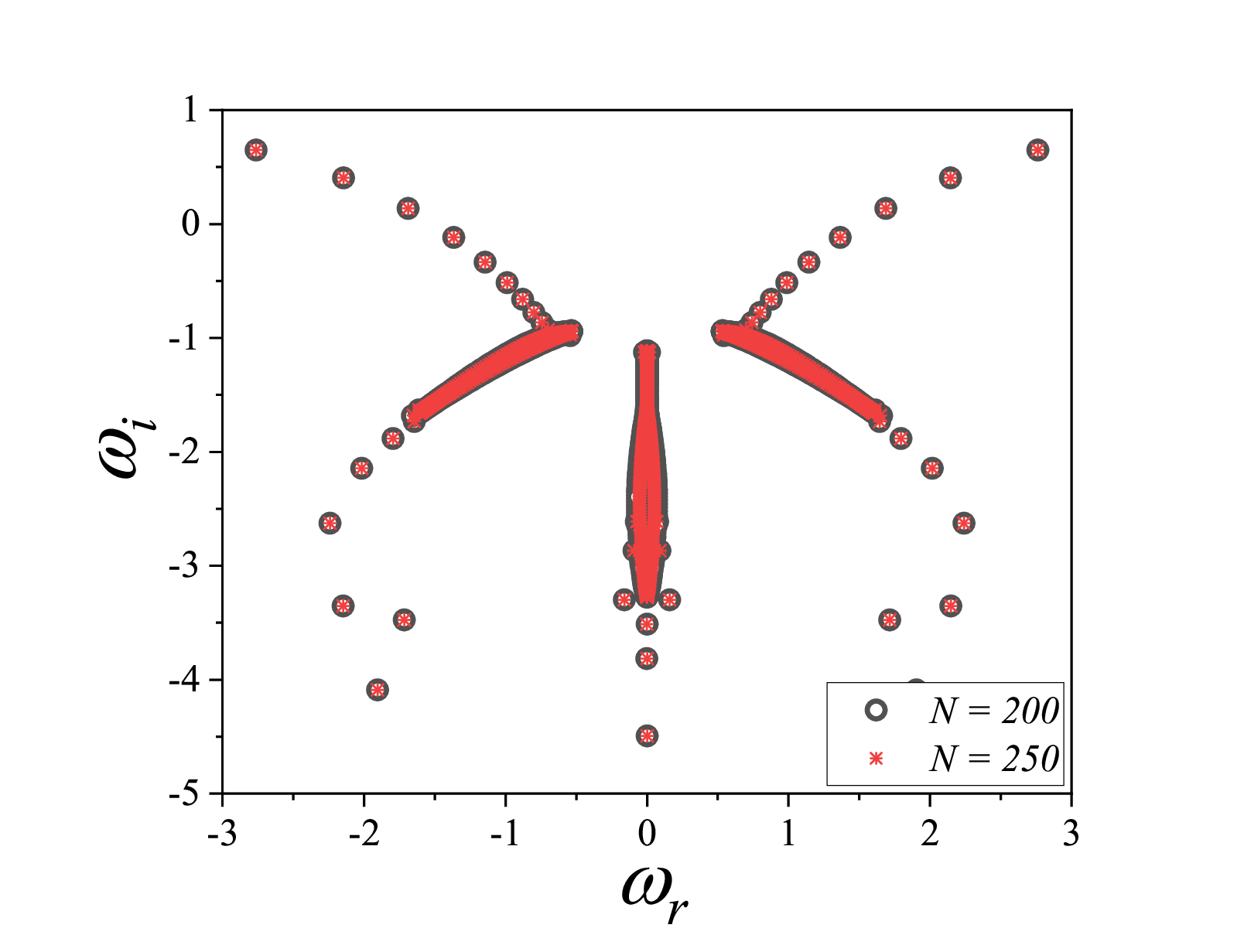}}
  \caption{Analytical CS and numerical spectra of Taylor-Couette flow. Data for $\Wi = 200, \beta = 0.98, \epsilon = 1, \alpha = 7$, $n = 0$.}
  \label{fig:Effect_of_L_TC_axis_e1}
\end{figure}
\begin{figure}
\centering
    \subfigure[Effect of $\beta$]
{\includegraphics[width=0.35\textwidth]{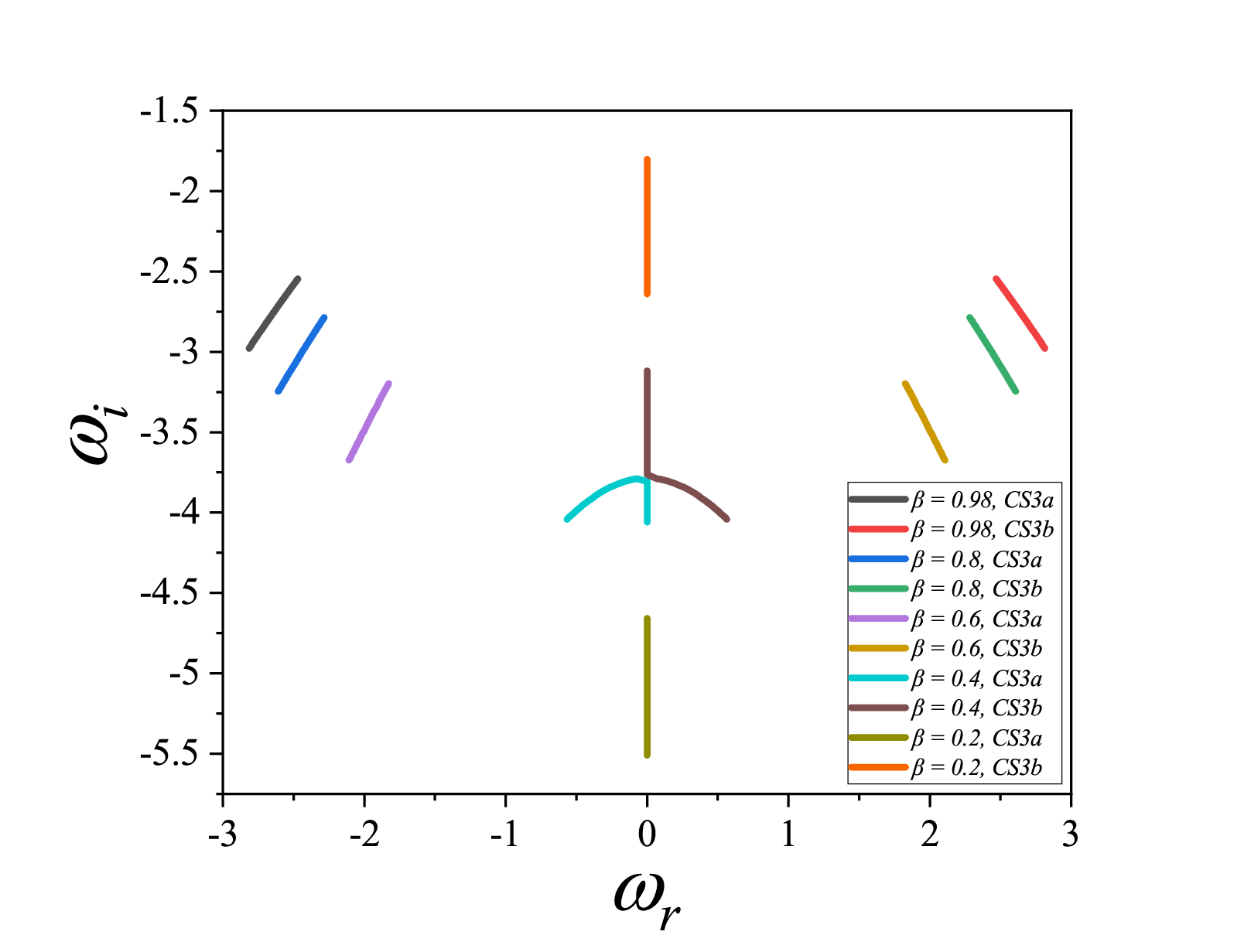}\label{fig:TC_effect_of_beta_CS4_n0}}
\subfigure[Effect of $\Wi/L$]
{\includegraphics[width=0.35\textwidth]{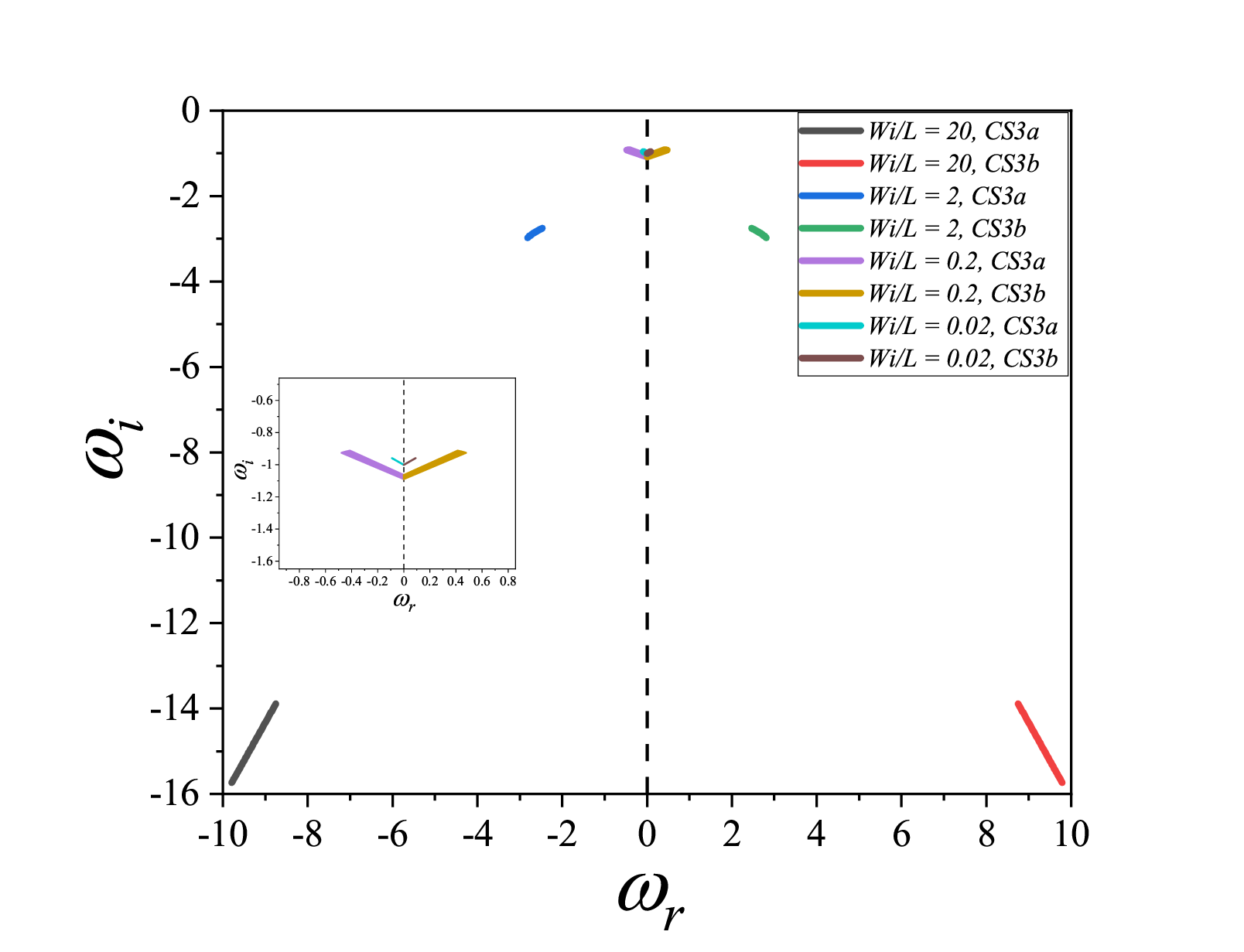}\label{fig:TC_effect_of_Wi_by_L_CS4_n0}}
  \caption{Effect of $\beta$ and $\Wi/L$ on analytically obtained CS3a, CS3b locations for Taylor Couette flow. Data for $\Wi = 200, L= 100, \beta = 0.98, \epsilon = 0.1, Re = 0, n = 0$.}
  \label{fig:TC_effect_of_beta_and_W_by_L_CS4}
\end{figure}

 \subsubsection{Non-axisymmetric disturbances}
 \label{subsec:nonaxiTaylor}
 \begin{figure}
  \centering
  \subfigure[$L = 100 $ (analytical)]
  {\includegraphics[width=0.35\linewidth]{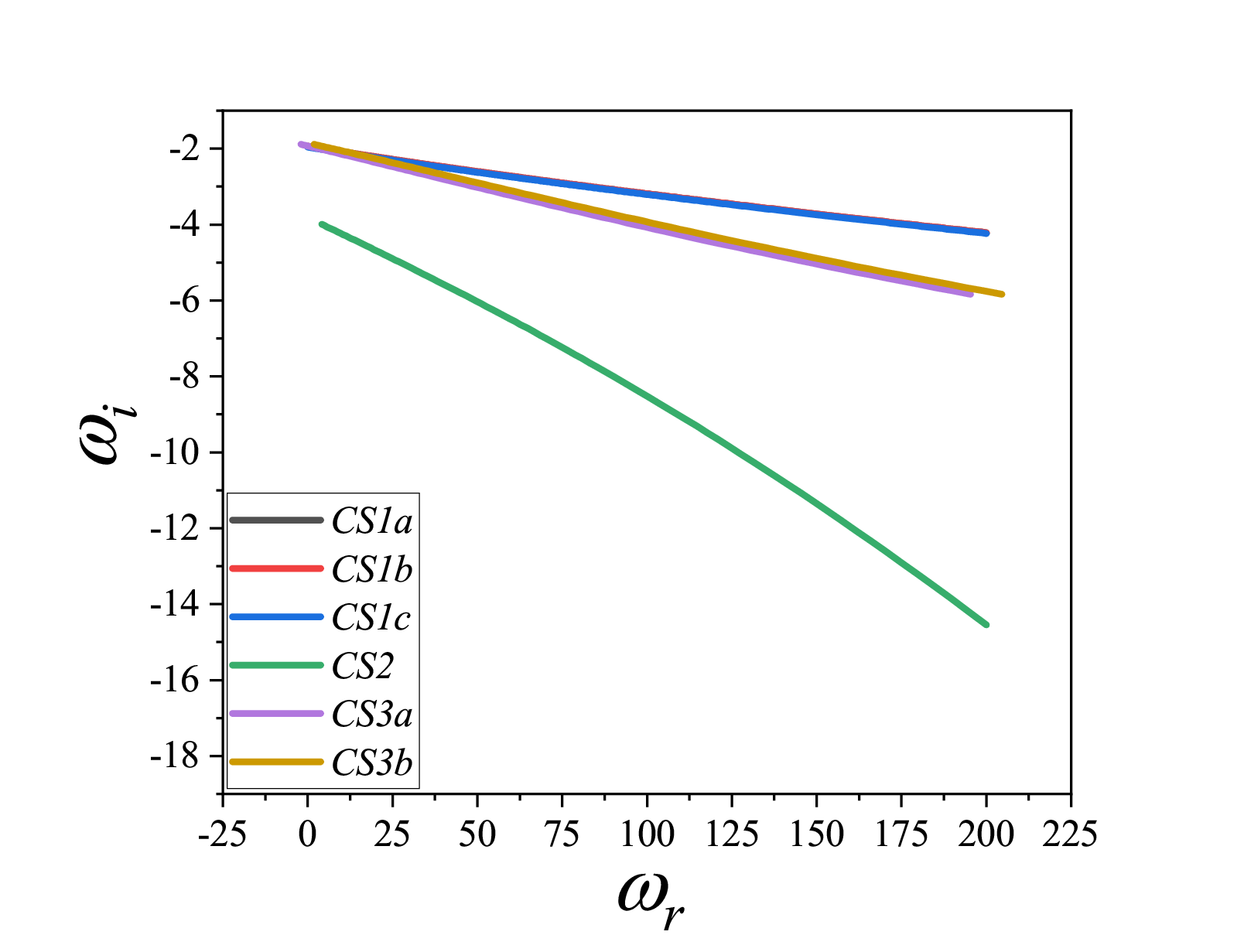}}\label{fig:L_100_e1_TC_nonaxis_ana}
    \subfigure[$L = 100 $ (numerical)]
  {\includegraphics[width=0.35\linewidth]{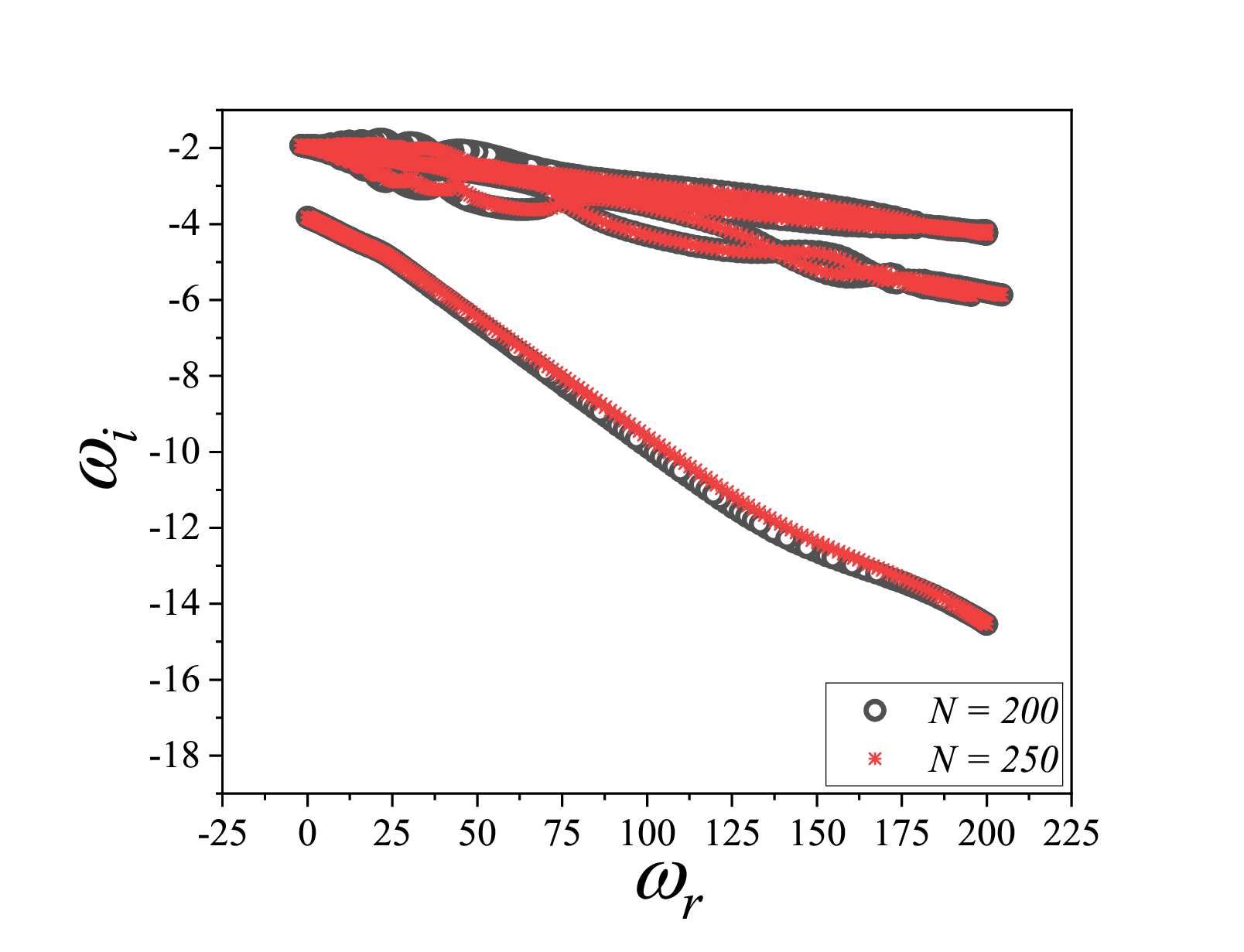}}\label{fig:L_100_e1_TC_nonaxis_num}
     \subfigure[$L = 500 $ (analytical)]
  {\includegraphics[width=0.35\linewidth]{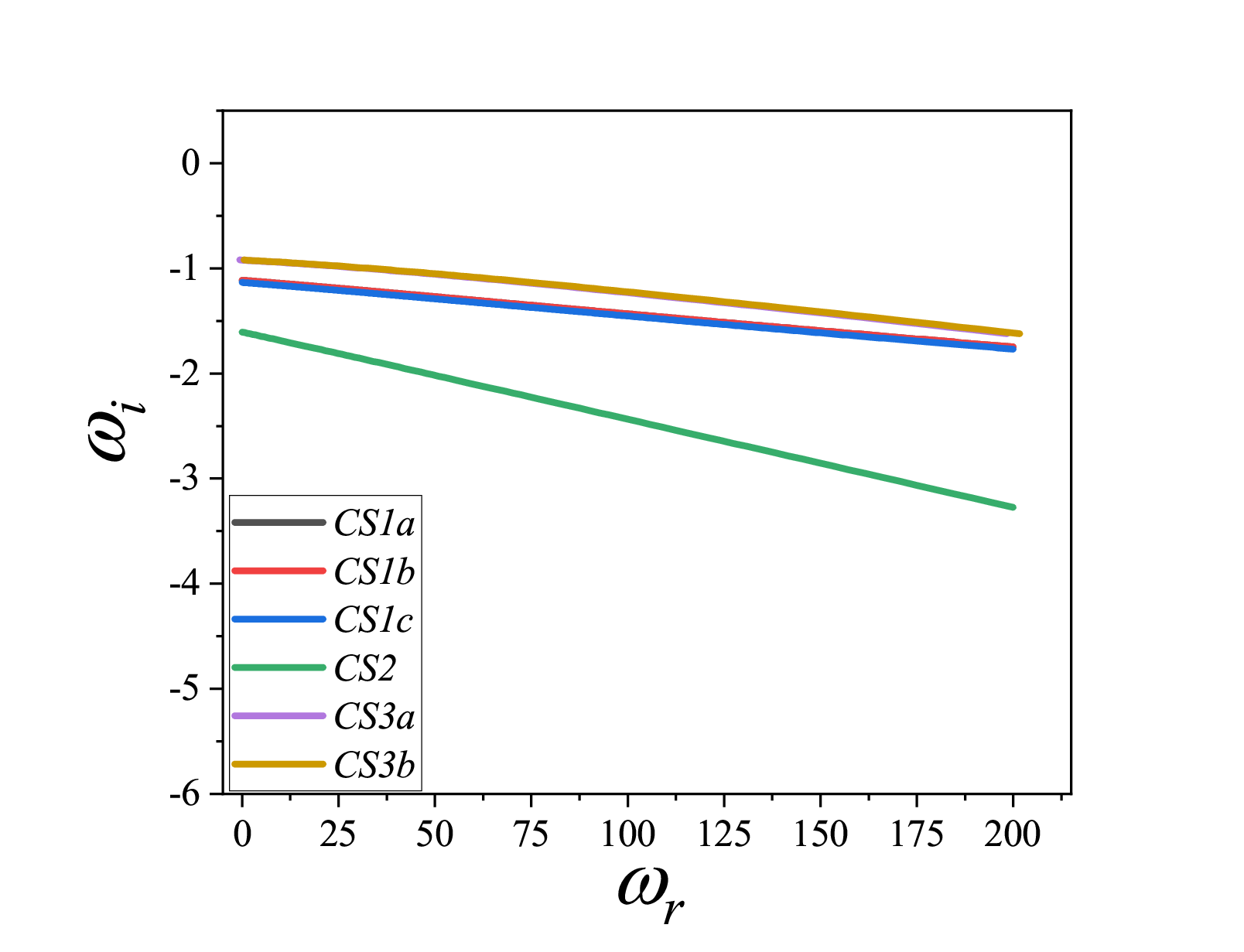}}\label{fig:L_500_e1_TC_nonaxis_ana}
    \subfigure[$L = 500 $ (numerical)]
  {\includegraphics[width=0.35\linewidth]{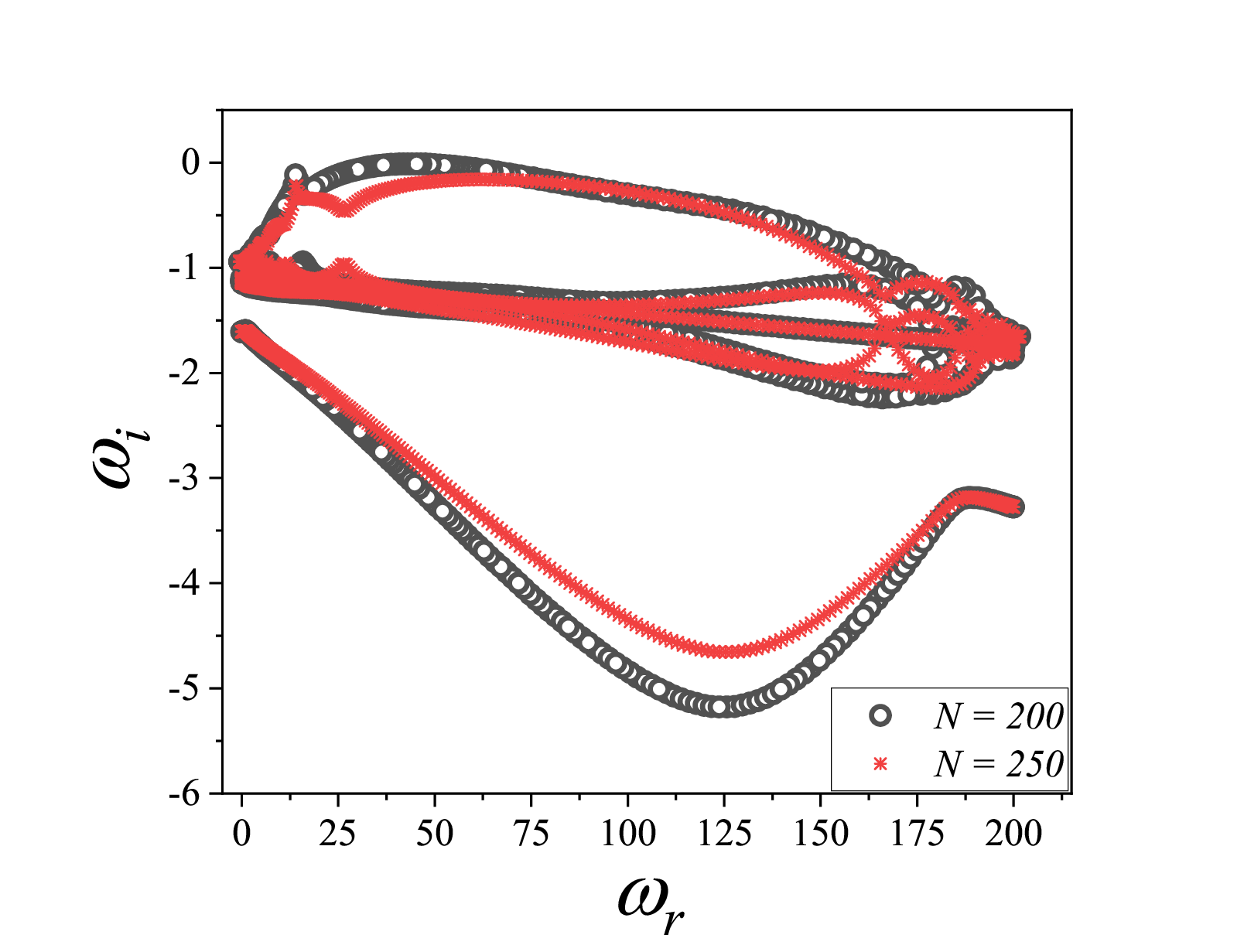}}\label{fig:L_500_e1_TC_nonaxis_num}
  \caption{Analytical CS and numerical spectra for Taylor-Couette flow for $\Wi = 200, \beta = 0.98, \epsilon = 1, n = 1, \alpha = 7$.}
  \label{fig:Effect_of_L_TC_nonaxis}
\end{figure}
Figure\,\ref{fig:Effect_of_L_TC_nonaxis} presents the numerical and analytical CS for non-axisymmetric disturbances in Taylor Couette flow.  Here, the CS are curved, tilting down from left to right, and extend into the right half of the complex plane. 
Regardless of whether the inner or outer cylinder is rotating, the shear rate near the inner wall is higher. Thus, in Fig.\,\ref{fig:Effect_of_L_TC_nonaxis}, with the inner cylinder rotating, the effective relaxation time is lowest near the inner cylinder. Consequently, this results in higher decay rates near the inner wall compared to the CS modes near the outer cylinder, thereby giving rise to the curved nature of the CS that tilt downwards from left to right, a feature alluded to at the beginning of this Section.
Similar to the other flows considered before, the CS for Taylor-Couette flow also exhibit a collapse 
for various $(\Wi, L)$ pairs, but with a fixed $\Wi/L$.
For $\Wi/L \gg 1$, CS2 is  the bottom-most curve, as it scales as $(\Wi/L)^{4/3}$, while the other CS scale as $(\Wi/L)^{2/3}$, consistent with the scaling behavior discussed earlier for rectilinear flows.
 
\section{Conclusions}
\label{sec:concl}
We have provided the first comprehensive account of the nature of the continuous spectra (CS) in both rectilinear and curvilinear shearing flows of FENE-P fluids. The structure of the CS in FENE-P fluids is shown to be much more complex compared to their Oldroyd-B counterparts, with up to six different CS present in general for shearing flows subjected to two-dimensional disturbances; the CS are independent of variations in the spanwise direction. Except in the case of plane Couette flow (where the base-state shear rate is constant in the wall-normal direction), the CS are curves in the complex plane.  For the first time, our study reports the presence of a novel `wing'-shaped CS (labeled CS3) in pressure-driven channel and Dean flows, which is symmetric about the imaginary axis for streamwise constant disturbances. 
The eigenvalues belonging to this wing CS can have phase speeds outside of the base-state range of velocities, in a major departure from the conventional wisdom on viscoelastic continuous spectra.
A variant of the wing-like CS is also present in Taylor-Couette flow subjected to axisymmetric disturbances. Since the hoop-stress induced instability is primarily an axisymmetric one, the presence of the wing-like CS for axisymmetric disturbances has important implications
in the analysis of elastic instabilities, using the FENE-P model, in both Taylor-Couette and Dean flows.
On inclusion of streamwise variation, the wings get distorted, and become asymmetric about the imaginary axis, and the structure of CS3 becomes quite complicated. We also demonstrated that all the CS barring CS3a and 3b collapse for different $(\Wi, L)$ pairs with the same $Wi/L$.

The present theoretical predictions will be  useful in deciphering and discriminating physically genuine discrete modes from the poorly approximated CS eigenvalues in stability analyses of shearing flows of FENE-P fluids. Equipped with the knowledge of the theoretical CS, we find that in many instances, when $L \sim 100$, there are no discrete modes in the numerically obtained spectra in the shearing flows considered here, implying that it is the CS that will govern the linearized dynamics. The absence of discrete modes in some regions of the parametric space, coupled with the fact that the CS modes are stable, implies the absence of a modal (exponential) instability, and thereby suggesting the importance of non-modal/algebraic growth \cite{jovanovic_kumar_2010,jovanovic_kumar_2011,CastilloSanchez2022}. Indeed, Roy and Subramanian \cite{Roy_Subramanian_2014} provided an alternative interpretation of the so-called `lift-up' effect in inviscid parallel shearing flows by an appropriate superposition of the modes belonging to continuous spectrum. Note that in the inviscid limit, there are no discrete solutions to the Rayleigh equation for bounded (inviscid) shearing flows, but it allows for singular solutions that belong to the CS \cite{Case1960}. 
Even in curvilinear shearing flows, such as that induced by a vortex column, the continuous spectrum has been shown to be indispensable in the representation of vortical initial conditions outside the core \cite{Roy_subramanian_vortex_2014}.
In an analogous manner, for parametric regimes where no discrete modes are present in bounded shearing flows of a FENE-P fluid, it may be anticipated that the CS identified in this study will aid in constructing the underlying building blocks for a nonmodal description of instabilities in such regimes. In this regard, it would be instructive to deduce the mathematical nature of the singular eigenfunctions corresponding to the CS reported here, similar to  the earlier efforts of Graham \cite{Graham_1998} and Wilson \textit{et al.} \cite{Wilson1999} for the Oldroyd-B model. Further, many nonlinear constitutive relations for concentrated polymer solutions and melts, such as the Rolie-Poly \cite{LikhtmanGraham2003}, Pom-Pom \cite{McLeish_Larson_PomPom_JOR_1998}, and the extended Pom-Pom \cite{XPPmodel} models also exhibit shear thinning, and the qualitative features reported in this study for the continuous spectra are expected to carry over to those nonlinear models as well.

\appendix
\section{Linearized stability equations for plane Couette/pressure-driven channel flow}
\label{FENEP_PCF_Appendix}
In this Appendix, we provide the linearized governing equations for the rectilinear goemetries considered in this work. Here, $x$, $y$, and $z$, are respectively the base-flow, vorticity, and gradient directions. For rectilinear flows, in our spectral method, we solved only the continuity,  $x$ and $z$ momentum equations along with relevant components of the constitutive equation. 
The situation where rectilinear flows are subjected  to three-dimensional perturbations are obtained in the narrow-gap limit from  curvilinear configurations subjected to three-dimensional perturbations.
Thus, only the relevant equations are given below. \\
Continuity equation:
\begin{equation}
    i k \widetilde{v}_x + \frac{d \widetilde{v}_z}{d z} = 0
    \label{eq:conty_rectilinear}
\end{equation}
$x$-momentum equation:
\begin{equation}
\begin{aligned}
    Re i  (k\overline{V}_x - \frac{\omega}{\Wi}) \widetilde{v}_x + Re \frac{d \overline{V}_x}{d z} \widetilde{v}_z = -i k \widetilde{p} + \beta \left( -k^2 + \frac{d^2}{d z^2} \right) \widetilde{v}_x + (1 - \beta)\left(i k \widetilde{\tau}_{xx} + \frac{d \widetilde{\tau}_{xz}}{d z} \right)
\end{aligned}
\end{equation}
$z$-momentum equation:
\begin{equation}
\begin{aligned}
    Re i  (k\overline{V}_x - \frac{\omega}{\Wi}) \widetilde{v}_z = -\frac{d \widetilde{p}}{d z} + \beta \left( -k^2 + \frac{d^2}{d z^2} \right) \widetilde{v}_z + (1 - \beta)\left(i k \widetilde{\tau}_{xz} + \frac{d \widetilde{\tau}_{zz}}{d z} \right)    
\end{aligned}
\end{equation}
$C_{xx}$ equation:
\begin{equation}
\begin{aligned}
    i  (k\overline{V}_x - \frac{\omega}{\Wi}) \widetilde{C}_{xx} - 2 \frac{d \overline{V}_x}{d z} \widetilde{C}_{xz} - \left( 2 i k \overline{C}_{xx} + 2 \overline{C}_{xz} \frac{d}{d z}\right) \widetilde{v}_x + \frac{d \overline{C}_{xx}}{d z} \widetilde{v}_z = -\widetilde{\tau}_{xx}
\end{aligned}
\label{Cxx_rect}
\end{equation}
$C_{xz}$ equation:
\begin{equation}
\begin{aligned}
    i  (k\overline{V}_x - \frac{\omega}{\Wi}) \widetilde{C}_{xz} - \frac{d \overline{V}_x}{d z} \widetilde{C}_{zz} - \overline{C}_{zz}\frac{d \widetilde{v}_x}{d z} + \left(\frac{d \overline{C}_{xz}}{d z} - i k \overline{C}_{xx} \right) \widetilde{v}_z = -\widetilde{\tau}_{xz}
\end{aligned}
\label{Cxz_rect}
\end{equation}
$C_{yy}$ equation:
\begin{equation}
\begin{aligned}
    i  (k\overline{V}_x - \frac{\omega}{\Wi}) \widetilde{C}_{yy} + \frac{d \overline{C}_{yy}}{d z} \widetilde{v}_z = - \widetilde{\tau}_{yy} 
\end{aligned}
\label{Cyy_rect}
\end{equation}
$C_{zz}$ equation:
\begin{equation}
\begin{aligned}
    i  (k\overline{V}_x - \frac{\omega}{\Wi}) \widetilde{C}_{zz} + \left(\frac{d \overline{C}_{zz}}{d z} - 2 \overline{C}_{xz} i k - 2 \overline{C}_{zz}\frac{d}{d z} \right)\widetilde{v}_z = -\widetilde{\tau}_{zz}
\end{aligned}
\label{Czz_rect}
\end{equation}
\section{Linearized stability equations for Taylor-Couette/Dean flow}
\label{FENEP_TC_Appendix}
This appendix contains the linearized stability equation for continuity, momentum ($r, \theta, z$) and the constitutive equation of the FENE-P model for the curvilinear geometries considered in this work. \\ \\
Continuity equation:
\begin{equation}
    \left( \frac{\epsilon}{1 + \epsilon \zeta} + \frac{d}{d \zeta} \right) \widetilde{v}_r + \frac{\epsilon i n}{1 + \epsilon \zeta}\widetilde{v}_\theta + i \alpha \widetilde{v}_z = 0
\end{equation}

$C_{rr}$ equation:
\begin{equation}
    \begin{aligned}
    \left( \frac{d \overline{C}_{rr}}{d \zeta} - 2 \overline{C}_{rr}\frac{d}{d \zeta } - \frac{2 \overline{C}_{r \theta} \epsilon i n}{1 + \epsilon \zeta} \right) \widetilde{v}_r + 
    \left( \frac{\overline{V}_{\theta} \epsilon i n}{1 + \epsilon \zeta} - \frac{i \omega}{\Wi} \right) \widetilde{C}_{rr} = -\widetilde{\tau}_{rr}
    \end{aligned}
\end{equation}

$C_{r \theta}$ equation:
\begin{equation}
\begin{aligned}
    \left( \frac{d \overline{C}_{r \theta}}{d \zeta} - \frac{\overline{C}_{\theta \theta} \epsilon i n}{1 + \epsilon \zeta} \right) \widetilde{v}_r 
    + \left( \frac{\epsilon \overline{C}_{rr}}{1 + \epsilon \zeta} - \overline{C}_{rr}\frac{d}{d \zeta} \right) \widetilde{v}_\theta 
    + i \alpha \overline{C}_{r \theta} \widetilde{v}_z + \left( \frac{\overline{V}_\theta \epsilon}{1 + \epsilon \zeta} - \frac{d \overline{V}_\theta}{d \zeta} \right) \widetilde{C}_{rr} 
    + \left( \frac{\overline{V}_\theta \epsilon i n}{1 + \epsilon \zeta} - \frac{i \omega}{\Wi} \right) \widetilde{C}_{r \theta} = - \widetilde{\tau}_{r \theta}
\end{aligned}
\end{equation}

$C_{rz}$ equation:
\begin{equation}
\begin{aligned}
    -i \alpha \overline{C}_{zz} \widetilde{v}_r - \left( \overline{C}_{rr} \frac{d}{d \zeta} + \frac{\overline{C}_{r \theta} \epsilon i n}{1 + \epsilon \zeta} \right)\widetilde{v}_z 
    + \left(\frac{\overline{V}_\theta \epsilon i n}{1 + \epsilon \zeta} - \frac{i \omega}{\Wi} \right)\widetilde{C}_{rz} = - \widetilde{\tau}_{rz}
\end{aligned}
\end{equation}

$C_{\theta \theta}$ equation:
\begin{equation}
\begin{aligned}
    \left( \frac{d \overline{C}_{\theta \theta}}{d \zeta} - \frac{2 \overline{C}_{\theta \theta} \epsilon}{1 + \epsilon \zeta}\right)\widetilde{v}_r 
    + \left( \frac{2 \overline{C}_{r \theta} \epsilon}{1 + \epsilon \zeta} - \frac{2 \overline{C}_{\theta \theta} \epsilon i n}{1 + \epsilon \zeta} - 2 \overline{C}_{r \theta} \frac{d}{d \zeta}\right)\widetilde{v}_\theta 
    + 2\left(\frac{\overline{V}_\theta \epsilon}{1 + \epsilon \zeta} - \frac{d \overline{V}_\theta}{d \zeta}\right)\widetilde{C}_{r \theta} 
    + \left(\frac{\overline{V}_\theta \epsilon i n}{1 + \epsilon \zeta} - \frac{i \omega}{\Wi} \right)\widetilde{C}_{\theta \theta} = - \widetilde{\tau}_{\theta \theta}
\end{aligned}
\end{equation}

$C_{\theta z}$ equation : \\
\begin{equation}
\begin{aligned}
    -i \alpha \overline{C}_{zz} \widetilde{v}_\theta - \left(\overline{C}_{\theta r} \frac{d}{d \zeta} + \frac{\overline{C}_{\theta \theta} \epsilon i n}{1 + \epsilon \zeta}\right)\widetilde{v}_z + \left(\frac{\overline{V}_\theta \epsilon}{1 + \epsilon \zeta} - \frac{d \overline{V}_\theta}{d \zeta}\right)\widetilde{C}_{rz} +  \left(\frac{\overline{V}_\theta \epsilon i n}{1 + \epsilon \zeta} - \frac{i \omega}{\Wi}\right)\widetilde{C}_{\theta z} = - \widetilde{\tau}_{\theta z}
\end{aligned}
\end{equation}
$C_{zz}$ equation : \\
\begin{equation}
    \frac{d \overline{C}_{zz}}{d \zeta}\widetilde{v}_r - 2 i \alpha \overline{C}_{zz} \widetilde{v}_z + \left(\frac{\overline{V}_\theta \epsilon i n}{1 + \epsilon \zeta} - \frac{i \omega}{\Wi} \right) \widetilde{C}_{zz} = - \widetilde{\tau}_{zz}
\end{equation}
Radial momentum equation : \\
\begin{equation}
\begin{aligned}
    \left(-\frac{Re i \omega}{\Wi} + \frac{Re \overline{V}_\theta \epsilon i n}{1 + \epsilon \zeta} \right) \widetilde{v}_r 
    - \frac{2 Re \epsilon \overline{V}_\theta}{1 + \epsilon \zeta} \widetilde{v}_\theta 
    = -\frac{d \widetilde{p}}{d \zeta} 
    + \beta \Bigg( 
    \left( \frac{d^2}{d \zeta^2} 
    - \frac{\epsilon^2}{(1 + \epsilon \zeta)^2} 
    + \frac{\epsilon}{1 + \epsilon \zeta} \frac{d}{d \zeta}  
    - \frac{\epsilon^2 n^2}{(1 + \epsilon \zeta)^2} 
    - \alpha^2 \right) \widetilde{v}_r 
     - \frac{2 \epsilon^2 i n}{(1 + \epsilon \zeta)^2} \widetilde{v}_\theta 
    \Bigg) &\\
    + (1 - \beta) \Bigg( 
    \frac{d \widetilde{\tau}_{rr}}{d \zeta} 
    + \frac{\epsilon}{1 + \epsilon \zeta} \widetilde{\tau}_{rr} 
    + \frac{\epsilon i n}{1 + \epsilon \zeta} \widetilde{\tau}_{\theta r}  
    + i \alpha \widetilde{\tau}_{zr} 
    - \frac{\epsilon}{1 + \epsilon \zeta} \widetilde{\tau}_{\theta \theta} 
    \Bigg)
\end{aligned}
\end{equation}

$\theta$-momentum equation :  \\
\begin{equation}
\begin{aligned}
    \left(Re\frac{d \overline{V}_\theta}{d \zeta} + \frac{Re \overline{V}_\theta \epsilon}{1 + \epsilon \zeta} \right) \widetilde{v}_r 
    + \left(\frac{Re \overline{V}_\theta \epsilon i n}{1 + \epsilon \zeta} - \frac{Re i \omega}{\Wi} \right) \widetilde{v}_\theta 
    = -\frac{\epsilon i n}{1 + \epsilon \zeta} \widetilde{p}  
    + \beta \Bigg( 
    \left( \frac{d^2}{d \zeta^2} 
    - \frac{\epsilon^2}{(1 + \epsilon \zeta)^2} 
    + \frac{\epsilon}{1 + \epsilon \zeta} \frac{d}{d \zeta} 
    - \frac{\epsilon^2 n^2}{(1 + \epsilon \zeta)^2} 
    - \alpha^2 \right) \widetilde{v}_\theta &\\
     + \frac{2 \epsilon^2 i n}{(1 + \epsilon \zeta)^2} \widetilde{v}_r 
    \Bigg) &\\
     + (1 - \beta) \Bigg( 
    \frac{d \widetilde{\tau}_{r \theta}}{d \zeta} 
    + \frac{2 \epsilon}{1 + \epsilon \zeta} \widetilde{\tau}_{r \theta} 
    + \frac{\epsilon i n}{1 + \epsilon \zeta} \widetilde{\tau}_{\theta \theta} 
    + i \alpha \widetilde{\tau}_{z \theta} 
    \Bigg)
\end{aligned}
\end{equation}

$z$-momentum equation : \\
\begin{equation}
\begin{aligned}
    \left(\frac{Re \overline{V}_\theta \epsilon i n}{1 + \epsilon \zeta} - \frac{i \omega Re}{\Wi} \right) \widetilde{v}_z = -i \alpha \widetilde{p} 
     + \beta \left(\frac{d^2}{d \zeta^2} + \frac{\epsilon}{1 + \epsilon \zeta}\frac{d}{d \zeta} - \frac{\epsilon^2 n^2}{(1 + \epsilon \zeta)^2} - \alpha^2 \right)\widetilde{v}_z &\\
     + (1 - \beta)\left(\frac{d \widetilde{\tau}_{rz}}{d \zeta} + \frac{\epsilon \widetilde{\tau}_{rz}}{1 + \epsilon \zeta} 
     + \frac{\epsilon i n}{1 + \epsilon \zeta}\widetilde{\tau}_{\theta z} + i \alpha \widetilde{\tau}_{zz}\right)
\end{aligned}
\end{equation}

\bibliography{References}

\end{document}